\documentclass[12pt]{iopart}
\usepackage{graphicx}
\usepackage{multirow}
\usepackage{iopams}  
\usepackage{bm}

\newcommand{\w}{_{\mathrm{w}}}
\newcommand{\tw}{t\w}
\newcommand{\Chi}{\chi}
\newcommand{\Deff}{D}
\newcommand{\dc}{d_{\rm c}}
\newcommand{\dmin}{d_{\rm min}}
\newcommand{\Ei}{{\rm Ei}}
\newcommand{\rbf}{\boldsymbol{r}}

\renewcommand{\H}{\mathcal{H}}
\newcommand{\E}{\mathcal{E}}
\newcommand{\Qo}{\tilde Q^{(1)}_\theta}
\newcommand{\Qop}{\tilde Q^{(1)}_{\theta'}}
\newcommand{\Qoo}{\tilde Q^{(1)}}
\newcommand{\tot}{_\mathrm{tot}}
\newcommand{\be}{\begin{equation}}
\newcommand{\ee}{\end{equation}}
\newcommand{\bea}{\begin{eqnarray}}
\newcommand{\eea}{\end{eqnarray}}
\newcommand{\bra}[1]{\langle #1 |}
\newcommand{\ket}[1]{| #1 \rangle}
\newcommand{\ivec}{\boldsymbol{i}}
\newcommand{\jvec}{\boldsymbol{j}}
\newcommand{\Erfc}{\Phi}
\newcommand{\Ra}{R^{(\mathrm{a})}} 
\newcommand{\Rs}{R^{(\mathrm{s})}} 
\newcommand{\Xa}{\chi^{(\mathrm{a})}} 
\newcommand{\Xs}{\chi^{(\mathrm{s})}} 

\renewcommand{\rmd}{d}
\renewcommand{\rme}{e}

\begin{document}

\title[Non-equilibrium dynamics of spin facilitated models]
{Non-equilibrium dynamics of spin facilitated glass models}

\author{S\'ebastien L\'eonard$^1$, Peter Mayer$^2$, Peter Sollich$^3$, 
Ludovic Berthier$^1$ and Juan P. Garrahan$^4$}

\address{$^1$ Laboratoire des Collo\"{\i}des, Verres et Nanomat\'eriaux,
UMR 5587 Universit\'e Montpellier II \& CNRS, 34095 Montpellier
Cedex 5, France}

\address{$^2$ Department of Chemistry, 
Columbia University, 3000 Broadway, New York, NY 10027, USA}

\address{$^3$ Department of Mathematics, 
King's College London, London WC2R 2LS, UK}

\address{$^4$ School of Physics and Astronomy, University of
Nottingham, Nottingham, NG7~2RD, UK}

\ead{
\mailto{leonard@lcvn.univ-montp2.fr},
\mailto{pm2214@columbia.edu}, 
\mailto{peter.sollich@kcl.ac.uk},
\mailto{berthier@lcvn.univ-montp2.fr},
\mailto{juan.garrahan@nottingham.ac.uk}
}

\begin{abstract}
We consider the dynamics of spin facilitated models of glasses in the
non-equilibrium aging regime following a sudden quench from high to
low temperatures.  We briefly review known results obtained for the
broad class of kinetically constrained models, and then present new
results for the behaviour of the one-spin facilitated
Fredrickson-Andersen and East models in various spatial dimensions.
The time evolution of one-time quantities, such as the energy density,
and the detailed properties of two-time correlation and response
functions are studied
using a combination of theoretical approaches, including 
exact mappings of master operators and reductions to integrable
quantum spin chains, field theory and 
renormalization group, and independent interval and timescale
separation methods. The resulting analytical predictions
are confirmed by means of detailed numerical simulations.  The models
we consider are characterized by trivial static properties, with no
finite temperature singularities, but they nevertheless display a
surprising variety of dynamic behaviour during aging, which can be
directly related to the existence and growth in time of dynamic
lengthscales. Well-behaved fluctuation-dissipation ratios can be
defined for these models, and we study their properties in detail.  We
confirm in particular the existence of negative
fluctuation-dissipation ratios for a large number of observables.  Our
results suggest that well-defined violations of
fluctuation-dissipation relations, of a purely dynamic origin and
unrelated to the thermodynamic concept of effective temperatures,
could in general be present in non-equilibrium glassy materials.
\end{abstract}



\maketitle

\section{Why study the aging regime of spin
facilitated models?}

\subsection{A brief survey of kinetically constrained models}

This paper is concerned with the dynamics of spin facilitated models
of glasses in the non-equilibrium aging regime following a sudden
quench from high temperature to the very low temperature glassy
regime.  Spin facilitated models belong to the broader family of
kinetically constrained models (KCMs).  These are simple statistical
mechanics models which display many of the dynamical features observed
in real glassy materials, such as supercooled liquids~\cite{Donth01},
spin glasses~\cite{Young}, or soft disordered
materials~\cite{lucaramos}.  KCMs are generically defined from a
simple, usually non-interacting, Hamiltonian.  The complexity of
glasses is encoded in specific local dynamical rules, or kinetic
constraints.  For an extensive review of early results on KCMs
see~\cite{RitSol03}.

In this paper we focus on spin facilitated models, in particular
Fredrickson-Andersen (FA)~\cite{FreAnd84} and East models~\cite{east},
but we expect that similar behaviour to the one we describe here will 
also be found in other KCMs such as constrained lattice gases.  The
main insight of FA~\cite{FreAnd84} was to devise models that are
simplistic, as compared to realistic interacting molecular systems,
but whose macroscopic behaviour was in agreement with the
phenomenology of liquids approaching the glass
transition~\cite{FreAnd84,FreAnd85}, displaying a
super-Arrhenius increase of relaxation timescales on decreasing the
temperature and non-exponential relaxation functions at equilibrium.
Early studies also demonstrated that when suddenly quenched to very
low temperatures the subsequent non-equilibrium aging dynamics of the
models compares well with experimental observations on aging
liquids~\cite{FA86a,FA86b}.  Initially it was suggested that FA
models would display finite temperature dynamic transitions similar to
the one predicted by the mode-coupling theory of supercooled
liquids~\cite{FreAnd84}, but it was soon realized that most KCMs do not
display such singularity, and timescales in fact only diverge in the
limit of zero temperature~\cite{RitSol03,TBF04,TBF05}.  
This implies in particular
that after a quench to any non-zero temperature these systems are
eventually able to reach thermal equilibrium.  The time window of the
aging regime, however, becomes large at low temperatures, and the
detailed study of this far from equilibrium aging time regime will be
the main subject of this paper.

There has been much interest in KCMs recently.  This is
partly due to the
realization~\cite{peter1,peter2,peter3,peter4,jackle1,jackle2,GC1}
that their dynamics is characterized by dynamic heterogeneity, that
is, non-trivial spatio-temporal fluctuations of the local relaxation,
which is also a hallmark of supercooled liquids~\cite{Ediger00}.  Many
of the studies relating to dynamic heterogeneity in KCMs are very
recent indeed, having appeared since the review \cite{RitSol03} was
compiled. These studies
characterize in great detail the heterogeneous dynamics of KCMs. They
include: papers defining and quantifying relevant dynamic lengthscales
in KCMs in equilibrium and their relation to relaxation
timescales~\cite{TBF04,TBF05,SchTri99,Whitelam2004,Szamel04,Whitelam2005,Jack2006,JacGarCha06,JacBerGar05,Tonetal06,Tonetal06b,LongI,LongII};
studies of more qualitative or phenomenological consequences of kinetic
constraints, including  dynamic heterogeneity and activated dynamics,
for the physics of glassy
systems~\cite{BerGar03a,BerGar03b,Jung04,VanKetel05,BCG05,Seb06};
and the definition and analysis of new KCMs where kinetic rules, or
lattice geometry, are tuned to explore in more detail the range of possible
behaviours that can be observed in glass
models~\cite{arrow_model,TBF06,SelBirTon05,BerGar05,GeiRei05,Moreno06}.  These
numerous recent studies have in turn been instrumental in encouraging
numerical and experimental efforts to measure and characterize in more
detail dynamic heterogeneity in supercooled
liquids~\cite{yamamoto,glotzer,lacevic,Berthier04,TWBBB,Szamel06,mark,science},
granular materials~\cite{dauchot,LefBerSti04}, colloidal
systems~\cite{science,Puertas04,Weeks06,ReiRabGei05}, and soft glassy
materials~\cite{mayer,Duri06,Hurtado07,Dibble06}.

\subsection{Aging dynamics in glassy systems: 
early studies and open questions}

The previous section suggests that dynamic heterogeneity and activated
dynamics, which have been well-studied and characterized at thermal
equilibrium in various models and systems approaching the glass
transition, play central roles in glassy dynamics.  However, when
moving deeper into the glass phase, glassy materials cannot be
equilibrated anymore on experimental or numerical timescales. In this
non-equilibrium state, physical properties are not stationary, and the
system displays aging behaviour~\cite{Young}. Experimentally, aging
has been well studied at the macroscopic level, in systems as diverse
as polymers~\cite{Struik}, structural and spin glasses~\cite{Young},
and soft materials~\cite{lucaramos}.  A full understanding of the
non-equilibrium glassy state remains a central theoretical
challenge~\cite{Young}.

Theoretical studies of mean-field models have provided important
insights into the aging dynamics of both structural and spin
glasses~\cite{Young,CugKur1,CugKur2}.  In mean-field models, thermal
equilibrium is never reached, and aging proceeds by downhill motion in
an increasingly flat free energy landscape~\cite{laloux}. Time
translational invariance is broken, and two-time correlation and
response functions depend on both their arguments. The
fluctuation-dissipation theorem (FDT), which relates equilibrium
correlation and response functions, does not apply in the aging
regime, but a generalized form is shown to hold~\cite{CugKurPel97}.
This is defined in terms of the two-time connected correlation
function for some generic observable $A(t)$, \be C(t,\tw) = \langle A(t) A(\tw)
\rangle - \langle A(t)\rangle \langle A(\tw) \rangle, \ee with $t \ge
\tw$, and the corresponding two-time (impulse) response function \be R(t,\tw) =
T \frac{\delta \langle A(t) \rangle}{\delta h (\tw)}\Bigg|_{h=0}.  \ee
Here $h$ denotes the thermodynamically conjugate field to the
observable $A$ so that the perturbation to the Hamiltonian (or energy
function) is $\delta E =
-hA$, and angled 
brackets indicate an average over initial conditions and any
stochasticity in the dynamics.  Note that we have absorbed the
temperature $T$ in the definition of the response.  The associated
generalized FDT is then \be R(t,\tw) = X(t,\tw)
\frac{\partial}{\partial \tw} C(t,\tw),
\label{fdr_def}
\ee with $X(t,\tw)$ the so-called fluctuation-dissipation ratio (FDR).
At equilibrium, correlation and response functions are time
translation invariant, depending only on $\tau = t - \tw$, and
equilibrium FDT imposes that $X(t,\tw) = 1$ at all times. A parametric
fluctuation-dissipation (FD) plot of the step response or susceptibility
\be
\Chi(t,\tw)=\int_{\tw}^t dt'\,R(t,t'),
\ee
against
\be
\Delta C(t,\tw)=C(t,t)-C(t,\tw),
\ee
is then a straight line with unit slope. These simplifications do
not occur in non-equilibrium systems.  But the definition of an FDR
through Eq.~(\ref{fdr_def}) becomes significant for aging
systems~\cite{CugKur1,CugKur2}.  In mean-field spin glass models the
dependence of the FDR on both time arguments is only through the
correlation function $X(t, \tw ) \sim X (C (t, \tw ))$ at large
times. This led to the idea that aging systems might be characterized
by an effective temperature~\cite{CugKurPel97}, defined in terms of
the FDR, $T_{\rm eff} = T / X$.  Physically, relaxation in glassy
systems occurs in well-separated time sectors~\cite{CugKur2}; it is
then easy to imagine that each sector could be associated with an effective
temperature~\cite{Kurchan}. A thermodynamic interpretation of
effective temperatures has also been put forward, relating them to the
concept of replica symmetry breaking~\cite{FraMezParPel98}.

Taken together, these results make the mean-field description of aging
very appealing, and have set the agenda for a large body of numerical
and experimental work, as reviewed in~\cite{CriRit03}.  The broader
applicability of the mean-field scenario of aging dynamics remains
unclear, however. While some experiments and simulations seem to support the
existence of well-behaved effective
temperatures~\cite{Grigera99,Abou04,Wang06}, other studies also reveal
the limits of the mean-field scenario.  Experiments have for instance
reported anomalously large FDT violations associated with intermittent
dynamics~\cite{Bellon1,Bellon2,Buisson1,Buisson2}, while theoretical
studies of model systems have also found non-monotonic or even
negative response functions~\cite{nicodemi,CorDeCLipZan02,Viot03,kr,DepSti}, and
ill-defined or observable-dependent FDRs~\cite{FieSol02}.  In
principle, these discrepancies with mean-field predictions are to be
expected, since there are many systems of physical interest in which
the dynamics are not of mean-field type, displaying both activated
processes and spatial heterogeneity.  It is thus an important task to
understand from the theoretical point of view when the mean-field
concept of an FDR-related effective temperature remains viable.

\subsection{Aging and dynamic heterogeneity}

Studying theoretically the interplay between relevant dynamic
lengthscales and thermally activated dynamics in the non-equilibrium
regime of disordered materials is clearly a challenging task.  This
problem has been approached in several different ways, as we briefly
summarize in this subsection.

Spin glasses represent a particular class of materials which has been
studied in detail by means of experiment, simulation and
theory~\cite{Young}. For these systems mean-field theory provides
a set of detailed predictions, in particular for the aging
properties~\cite{CugKur2,CugKur3}, without accounting specifically for growing
dynamic lengthscales or thermally activated processes.  More
phenomenological approaches, on the other hand, directly focus on real space
excitations.  The resulting predictions~\cite{FisHus86,FH2,FH3} differ from
those of mean-field, and aging in particular comes to be associated with the
logarithmic growth with time of a dynamic correlation
lengthscale.  This lengthscale is believed to play a crucial role in
memory and rejuvenation processes observed
experimentally~\cite{Vincent98,Dupuis01,Jonsson,Bert04}, a link that has been
confirmed numerically~\cite{BerBou02,BerYou05}.  There exist
theoretical attempts to extend the mean-field framework to include
spatial fluctuations and dynamic
lengthscales~\cite{barrat,2DXY,kennett,kennett2,kennett3,kennett4}.

The situation is somewhat similar for structural glasses.  While
mean-field theories of aging are by now
well-established~\cite{CugKur1}, several alternative perspectives
insist that dynamic lengthscales and fluctuations should play an
important role~\cite{GC1,BerGar03a,Xia01,Tarjus}.  This includes the
aging regime.  Just as in the spin glass case, extensions of the
mean-field framework to finite dimensions are feasible in 
principle~\cite{LongI,LongII,BB04,Franz05,Dzero05,Franz06,MonSem05,MonSem06}
but this remains a very difficult task.

A natural approach is to consider directly systems with local, finite
ranged interactions.  The study of the aging dynamics of KCMs is
therefore very promising, as these models combine the necessary
features of being defined in terms of (effective) microscopic degrees
of freedom, having local dynamical rules, and displaying thermally
activated and heterogeneous dynamics.

Early studies on the aging dynamics of KCMs were mainly dedicated to
exploring the generic behaviour of FA models after a low temperature
quench, or using more complex thermal protocols~\cite{FA86a,FA86b}.
The idea was to establish KCMs as reasonable glass models.  The
Kob-Andersen lattice gas~\cite{KA} was then studied in some detail,
with results that appeared to be in reasonable agreement with
mean-field ideas~\cite{Kurchan97,SellittoEPJB}. The same model was
also used as a kinetic model to study granular
compaction~\cite{Barrat00,Sellitto01,Sellitto02,Arenzon04}. Both FA and East
models were studied numerically in \cite{Crisanti00,Crisanti02}, with
somewhat unclear results. While two-time correlation functions were
shown to exhibit standard scaling properties, non-monotonic response
functions were found, leading to apparently ill-defined FDRs. However,
these studies considered non-connected correlation functions, making
the resulting FDRs of dubious relevance. The one-spin one-dimensional FA
model, which we study in detail below, was reconsidered in
~\cite{BuhGar02,buhot} with a different conclusion, this time
indicating that FDT is
satisfied for this model at all times.  The East model was also
reconsidered in~\cite{SolEva1,SolEva2}, with results similar to that
found in a family of spin plaquette models~\cite{newman}, which share
many similarities with KCMs~\cite{JacBerGar05}.  Non-monotonic response
functions were again found at low temperature, which seemed to prevent
the existence of genuine FDRs.

It was later realized that although the non-monotonic response
functions were real and physical~\cite{BuhGar02}, they did not prevent
the possibility of defining FDRs with robust scaling properties.  The
confusion arose from the use of an incorrect definition of the
FDR, whereby the $\tw$-derivative in Eq.~(\ref{fdr_def})
was replaced, for numerical convenience, by a $t$-derivative. In the
graphical representation in terms of an FD plot this corresponds to
fixing $\tw$ and letting $t$ vary along the curve, rather than the
correct reverse procedure where the later time $t$ is fixed and the
earlier time $\tw$
is varied. As has been
emphasized several times~\cite{FieSol02,mayer0,Diezemann,Berthier07}, this
procedure can lead to completely different behaviour or at least
incorrect numerical values of the FDR.  Unfortunately, the $\tw$-derivative
in its definition makes the numerical measurement of the correct FDR a very
demanding task; equivalently, it is normally difficult to obtain
the step response $\Chi(t,\tw)$ as a function of the earlier time
$\tw$ with $t$ fixed. However, novel numerical methods for
accessing linear response in simulations have recently become
available~\cite{Berthier07,Chatelain03,chatelain,Ricci} and have
made it possible to systematically study FDRs in
KCMs~\cite{Mayer2006,Robs_plaquette_FDT,MaySol07}. 
The results are, perhaps surprisingly, much simpler to 
interpret physically, as we
shall show throughout this paper.

\subsection{Summary of main results and plan of paper}
\label{sec:summary}

The models considered in this paper are characterized by trivial
static properties with no finite temperature singularity, yet they
display a surprising variety of dynamic behaviour during the aging
which can be directly related to the existence and growth with time of
a dynamic lengthscale.  We find in particular that, as in mean-field
models, well-behaved fluctuation-dissipation ratios can often be defined
for these models from violations of the FDT.  A major difference
between mean-field models and KCMs is that FDT violations are only
transient in KCMs, and a crossover towards equilibrium is always
expected. Therefore, there can be no asymptotic connection between FDT
violations and thermodynamic properties.  Similarly, it is not obvious
how to connect FDRs to effective temperatures in the case of KCMs. In
particular the presence of negative fluctuation-dissipation ratios for
a large number of observables indicates that the interpretation of FDT
violations in terms of some generalized thermodynamics may not be
possible.  In fact, the aging behaviour of KCMs found in this and
related work~\cite{Mayer2006,Robs_plaquette_FDT,MaySol07} suggests that
well-defined violations of fluctuation-dissipation, of purely dynamic
origin and unrelated to the thermodynamic concept of effective
temperatures, might generically be present in non-equilibrium glassy
materials.

The core material of the paper is divided into the following
Secs.~\ref{sec:FA} and~\ref{sec:East} which deal with the FA and East
models respectively. For both models we first give a qualitative
description of the physical processes governing the aging dynamics.
For the FA model we then recall (Sec.~\ref{sec:FA_mapping}) arguments
based on an exact mapping to a diffusion-annihilation process which
demonstrate that the upper critical dimension is $\dc=2$ (rather than
$\dc=4$). Below this dimension, i.e.\ in $d=1$, it turns out that
exact results can be obtained in an appropriate long-time scaling
regime. These are described in Sec.~\ref{sec:FA1d}, first in general
terms and then applied to the local, global and Fourier mode
correlation and response functions. The latter case
(Sec.~\ref{subsubCCR}) includes the other two as opposite limiting
behaviours, and simulations confirm the predictions across the entire
range. In Sec.~\ref{sec:FA_ft} we turn to dimensions $d>\dc=2$, where
field-theoretic methods can be used to calculate the Fourier mode
correlation and response functions. Also these predictions compare
well with simulation results. We
conclude the section on the FA model with the derivation of a useful
identity for global response functions (Sec.~\ref{sec:nofield}) that
makes the latter much more amenable to numerical simulation. The
identity requires only that the dynamics be given by local spin flips
that are modified by a kinetic constraint and so should be of general
use for numerical studies of KCMs with non-conserved (i.e.\ Glauber
spin-flip rather than Kawasaki spin-exchange) dynamics.

In the analysis of the East model, different techniques come to the
fore, in particular master equations based on the independent interval
nature of the dynamics. These can be specifically adapted to find
correlation and response for local (Sec.~\ref{sec:East_local}) and
global (Sec.~\ref{sec:East_global}) observables. Fourier modes turn
out to be less useful for the East model; instead we consider non-local
(distance dependent) quantities (Sec.~\ref{sec:East_nonlocal}) which
reveal more clearly the effects of the directedness of the constraint,
in particular the very intricate spatial structure of two-time
correlations and responses.

We conclude in Sec.~\ref{sec:conclusion} with a summary and discussion
of our results. Some mathematical background for the FA Fourier mode
analysis in $d=1$, and technical details of the more involved East
model calculations, can be found in the appendices.

\section{Fredrickson-Andersen model}
\label{sec:FA}

The FA model~\cite{FreAnd84,FreAnd85} describes the dynamics of $N$ binary
spin-variables $n_i=0,1$ on a hypercubic lattice in $d$ dimensions. An
up-spin ($n_i=1$) represents a highly mobile region or, because of the
associated energy cost, a ``defect''.  Down-spins ($n_i=0$) model
dense, immobile regions and are energetically preferred. The energy
function is assumed to be non-interacting, $E=\sum_i n_i$. In a
continuous-time Markov dynamics, Glauber rates for flipping spin $i$
are prescribed as
\be
w_i(\bm{n}) = f_i(\bm{n})[(1-c)n_i + c(1-n_i)],
\label{Glauber_rates}
\ee
where $\bm{n}=(n_1,\ldots, n_N)$ and $c=1/(1+e^\beta)$ is the
equilibrium up-spin concentration; $\beta=1/T$ denotes the inverse
temperature as usual. Without the factor $f_i(\bm{n})$ the model would
consist of non-interacting spins, with rates $c$ and $1-c$ for up- and
down-flips, respectively. All the interesting physics is therefore in
the facilitation factor $f_i(\bm{n})$. It is chosen to be independent
of $n_i$ and this ensures that detailed balance with respect to the
energy function $E$ is maintained. Physically, $f_i$ represents the
assumption that a change of state in region $i$ is possible only
if some neighbouring regions are in a mobile state. We will focus
exclusively on one-spin facilitated FA models below, which are rather
simpler to understand than those imposing facilitation by more than
one spin~\cite{RitSol03}.  For simulation purposes it is then simplest
to set $f_i(\bm{n})=0$ if $n_j=0$ for all nearest neighbours $j$ of
site $i$, and otherwise $f_i(\bm{n})=1$~\cite{GrahamPiche}. For
analytical work it is more convenient to use the form of $f_i$
suggested originally~\cite{FreAnd84,FreAnd85},
\be
f_i(\bm{n})=\sum_{j={\rm n.n.}(i)} n_j,
\ee
which is again zero when all n.n.\ sites are in immobile states, but
has nonzero rates increasing in proportion to the number of mobile
neighbours. Physically, the key influence of $f_i$ is to make moves
without mobile neighbours impossible; the difference between the two
above definitions of $f_i$ for the rates of the {\em allowed moves}
therefore has no qualitative effects on the observed behaviour.

From the point of view of glass modelling the interesting parameter
regime of FA models is that of low $c$, where mobile regions are few
and far between. After a quench from high temperature (corresponding
to initial defect concentration $c_0\equiv c(T\to\infty)=1/2$) 
any such defects that are
not isolated from each other will quickly flip down, and a state is
reached where essentially all defects are isolated from each other.
From then on, any further reduction in the defect concentration
\be
n(t) = \frac{1}{N}\sum_i\langle n_i(t)\rangle,
\ee
proceeds via a diffusion-coagulation process. Diffusion of defects can
take place when a defect is created next to an existing one, with rate
$c$. Both defects are now mobile and can flip down quickly (rate
$1-c\approx 1$); if the original defect does so first, it leaves the
other one behind and has effectively moved by one site. The diffusion
constant for this process is $\Deff=c/2$, with the factor $1/2$
accounting for the probability of the newly created defect flipping back
down first. When two defects diffusing by this mechanism meet, one of
them can flip down, leading to defect coagulation. The reverse process
of branching is also possible, by creating two defects next to an
existing one which then flips down.  However, as two new defects are
involved the effective rate is $\sim c^2$ and therefore negligible
compared to those for diffusion and coagulation
at least in the time window $c^{-1}\ll t\ll c^{-2}$. 
For larger times
branching must of course eventually become important since it is the
only process that creates defects and so is able to stabilize the
defect density at its equilibrium value, $n(t\to\infty)=c$.

Conceptually, it is important to note that the effective
diffusion-coagulation dynamics has a dynamic critical point at $c\to
0$, where both timescales and lengthscales
diverge~\cite{Whitelam2004,Whitelam2005}. One can define appropriate
critical exponents; this is simplest from the equilibrium dynamics at
low $c$. The order parameter exponent $n(t\to\infty)=c\sim c^\beta$
is fixed to $\beta=1$ because of detailed balance. The lengthscale
$\xi$ and timescale $t_{\rm rel}$ of appropriate relaxation functions
define the other exponents as $\xi\sim c^{-\nu}$ and $\Deff t_{\rm
  rel}\sim \xi^z\sim c^{-z\nu}$. 
We note as an aside that, in the context of dynamic criticality, the
asymptotic FDR $X^\infty$, which is defined as the limit of $X(t,\tw)$ for
widely separated times $t\gg\tw$, appears as a universal amplitude
ratio that can distinguish different dynamic universality
classes~\cite{GodLuc00b}.

\subsection{Mapping to diffusion-annihilation}
\label{sec:FA_mapping}

Initial field-theoretical studies~\cite{Whitelam2005} suggested
that the FA model was in the dynamic universality class of directed
percolation, with an upper critical dimension $\dc=4$. Closer
inspection reveals, however, that there is a hidden symmetry which
reduces this value to $\dc=2$~\cite{Jack2006}. This can be deduced from
an exact mapping between the FA model and another defect-diffusion
model where defects or ``particles'' annihilate in pairs ($A+A\to 0$, where $A$
symbolizes a defect) and are similarly created ($0\to A+A$) in pairs.

The mapping is most easily derived in a quantum mechanical
representation of the Markov
dynamics~\cite{StinchcombeReview,Krebs95,Henkel97}. One associates to
each possible state $\bm{n}$ of the system a distinct unit vector
$|\bm{n}\rangle$ in a vector space, and to the time-dependent
probability distribution over states $p_t(\bm{n})$ the vector
$|p_t\rangle = \sum_{\bm{n}} p_t(\bm{n})|\bm{n}\rangle$. (The basis
vectors $|\bm{n}\rangle$ have unit length and are mutually
orthogonal.) It is then easy to check that the dynamical evolution of
this vector takes the form $\partial_t |p_t\rangle = W|p_t\rangle$
with the master operator
\be
W = \sum_i (F_i-1)\hat{f}_i[(1-c)\hat{n}_i+c(1-\hat{n}_i)].
\label{equ:FAmaster}
\ee
Here, $\hat{n}_i=\sum_{\bm{n}} n_i |\bm{n}\rangle\langle\bm{n}|$ is the
operator measuring the value of the local spin, $\hat{f}_i=\sum_{j={\rm
n.n.}(i)} \hat{n}_j$, and $F_i$ is the operator that flips spin
$i$, $F_i|n_1,\ldots,n_i,\ldots,n_N\rangle =
|n_1,\ldots,1-n_i,\ldots,n_N\rangle$. Averages are obtained by
multiplying with the uniform state
$\langle\bm{e}|=\sum_{\bm{n}}\langle\bm{n}|$ from the left, e.g.\
$\langle n_i(t)\rangle = \langle \bm{e}|\hat{n}_i|p_t\rangle$.

The key mathematical statement that establishes the mapping mentioned
above is that there is an operator $V$ such that the similarity
transformation $VWV^{-1}$ of the FA master operator produces the
master operator of an annihilation-pair creation process with
appropriate rates~\cite{Jack2006}. One can then verify directly that
the equilibrium correlation functions (more precisely those involving
only two separate times) of the two models are identical up to trivial
factors. As a consequence, also the dynamic exponents and dynamic
universality class of the two models must be the same. In particular,
one has $\dc=2$ and the critical exponents are
$(z,\nu,\beta)=(2,1/d,1)$ in $d<2$ while they take the mean-field
values $(z,\nu,\beta)=(2,1/2,1)$ in higher dimensions.

The feature of the annihilation-pair creation process that is
responsible for the upper critical dimension being two rather than
four is the conservation of parity: if the number of particles is even
to start with it remains so for all times, and similarly if it is
initially odd. Mathematically, the parity operator,
\be
\hat{P} = (-1)^{\sum_i \hat{n}_i},
\ee
commutes with the master operator, and maps two different steady
states onto each other (which differ only in the sign of the
probability amplitude of states with an odd number of particles). The
similarity relation between annihilation-pair creation and the FA
model implies that the latter also has a symmetry; this is the
``hidden'' symmetry mentioned above. Its physical nature is somewhat
difficult to elucidate since the symmetry operator that commutes with
the FA master operator is not diagonal and so does not correspond to a
standard observable obeying a conservation law. Mathematically, the
symmetry swaps terms in the master operator across n.n.\ bonds: the
term describing a flip of spin $i$ facilitated from a neighbouring
site $j$ is mapped to the one encoding flips of spin $j$ facilitated
from site $i$. This insight makes clear that the symmetry also applies
to the East model discussed below, except that the mapping here converts
the East model into the West model, where the direction of facilitation
is reversed. While this does not lead to obvious ways of calculating
the correlation and response functions that we will study below, it does
provide some general restrictions. For example, general multi-point
two-time equilibrium correlations
\be
\langle \prod_{\rho=1}^r [n_{i_\rho}(t)-c]\prod_{\sigma=1}^s
[n_{j_\sigma}(0)-c]\rangle,
\ee
can be shown to vanish unless the rightmost spins appearing at time 0
and $t$ are identical, i.e.\ unless
$\max_\rho(i_\rho)=\max_\sigma(j_\sigma)$. For two-point correlations ($r=s=1$)
this reduces to the well-known fact that only local correlations are
non-vanishing at equilibrium.

The above mapping can be extended in a number of ways, for example to
generic coagulation-branching and annihilation-pair creation models,
and to ``bosonic'' processes where each site can be occupied by more
than one particle~\cite{Jack2006}. For the original ``fermionic''
(hardcore repulsion) versions the mapping also has a very appealing
geometric structure, with all operations being effectively rotations
on a ``spin sphere''.

\subsection{Exact results in $d=1$}
\label{sec:FA1d} 

In this section we present exact long-time scaling forms for FDT violations in 
the $1d$ FA model. More precisely, the dynamics following
a quench from a random  
initial state to low temperature $c \ll 1$ is considered, 
in the nonequilibrium aging regime of times $c^{-1} \ll t \ll c^{-2}$. 
To probe violations of FDT we introduce the connected 
correlation and response functions, 
\bea
  C_r(t,\tw) &=& \langle n_{i+r}(t) n_i(\tw)\rangle - n(t)n(\tw), 
  \label{equ:Crdef} \\
  R_r(t,\tw) &=& T \left. \frac{\delta \langle n_{i+r}(t) \rangle}{\delta h_i(\tw)} 
  \right|_{h_i = 0}, 
  \label{equ:Rrdef} 
\eea
where temperature is again included in the definition of the response 
function. 
The fields $h_i$ are conjugate to the $n_i$, so that in their 
presence the energy becomes $E=\sum_i n_i (1-h_i)$. We will also 
consider the step-response $\chi_r(t,\tw) = \int_{\tw}^t 
\rmd t' \, R_r(t,t')$ to a perturbation applied from time $\tw$ 
onwards. 

Our analysis is based on the fact that defects effectively diffuse and 
coagulate on the $\mathcal{O}(c^{-1})$ timescale. As explained above, the rate 
for diffusion is $D = c/2$ while coagulations occur with rate $\gamma \sim 1$. 
We will, from now on and throughout this section, measure time in units of 
$c^{-1}$ (except when stating simulation parameters) so that $D = \frac{1}{2}$ 
and $\gamma \sim c^{-1} \gg 1$. The coarsening dynamics of this diffusion-coagulation 
process at long times $t \gg 1$ becomes independent of $\gamma$ \cite{ZhoBen95,TauHowLee05}: 
defects typically first have to diffuse for a time $\mathcal{O}(t)$ before they 
occupy adjacent sites where coagulation is possible. The duration of this 
reaction, which depends 
on the rate $\gamma$, is only $\mathcal{O}(1)$ and therefore negligible in 
comparison: the process is said to be in the diffusion controlled regime \cite{ZhoBen95}. 
The irrelevance of the precise value of $\gamma$ at long times allows us 
to adjust its value to $\gamma = \frac{1}{2}$, the reason for this choice 
being that (only) the process with $D = \gamma$ admits an exact solution 
\cite{Krebs95,Henkel97,HenOrlSch95}.  
Our results will then be exact to leading order at large time $t \gg 1$ for 
any process with $\gamma > 0$; this includes the FA model. 
Of course, the diffusion-coagulation picture applies 
to the FA model only so long as $t \ll c^{-1}$ 
is still satisfied. In the FA dynamics, branching events become important 
on this times scale and initiate a crossover to equilibrium. Therefore, 
whenever we talk about long time behaviour for $t \to \infty$ it is understood 
that the low temperature limit $c \to 0$ is taken first (in order to remain 
in the regime $1 \ll t \ll c^{-1}$; recall that our time-unit is now
$c^{-1}$). Throughout this section we will use the symbol ``$\sim$'' to
indicate results which become exact in this long-time limit.

The scaling of the connected two-time correlation $C_r(t,\tw)$ in the effective 
FA process with $D = \gamma = \frac{1}{2}$ was derived in \cite{MaySol07}. The result reads 
\begin{eqnarray}
\fl C_r(t,\tw) \sim 
	\rme^{-2\tw} [I_0 + I_1](2\tw) \sum_{i,j=0}^\infty G_{(r,r+1),(-i,j+1)}(\tau) \, H_{i+j+1}(2\tw) 
  \nonumber \\ 
  - \sum_{i,j=0}^\infty G_{(+r,+r-1),(i,j)}(\tau) \, \rme^{-2\tw} [I_i + I_{i+1}](2\tw) 
  \left[ \delta_{j,0} + H_j(2\tw) \right] 
  \nonumber \\ 
  - \sum_{i,j=0}^\infty G_{(-r,-r-1),(i,j)}(\tau) \, \rme^{-2\tw} [I_i + I_{i+1}](2\tw) 
  \left[ \delta_{j,0} + H_j(2\tw) \right], 
\label{equ:cor2tsums}
\end{eqnarray}
where $I_n(t)$ denotes modified Bessel functions, Eq.~(\ref{equ:I}), 
and the explicit form of the functions $H_n(t)$ is given in Eq.~(\ref{equ:H}). 
To save space we use the short-hand notation $[\, \cdot \,](x)$ to indicate 
that all functions contained in the square brackets have the same argument $x$. 
The $G_{\ivec,\jvec}(\tau)$ with $\ivec = (i_1,i_2)$ and $\jvec = (j_1,j_2)$ are 
Green's functions, 
\begin{equation}
  G_{\ivec,\jvec}(\tau) = \rme^{-2\tau} \left[ I_{i_1-j_1} I_{i_2-j_2} - 
  I_{i_1-j_2} I_{i_2-j_1} \right](\tau). 
  \label{equ:G} 
\end{equation}
The result Eq.~(\ref{equ:cor2tsums}) applies to the FA model to leading order in 
$\tw \gg 1$ but uniformly in $\tau \geq 0$, i.e.\ including small values 
of $\tau$. 

An expression for the response function $R_r(t,\tw)$ is also given in \cite{MaySol07}. 
Before stating the result we consider the effect of the associated perturbation. It 
is convenient to express the response function in the operator formalism from above, 
$R_r(t,\tw) = \langle \bm{e} | \hat{n}_{i+r} \rme^{W \tau} V_i | p_{\tw} \rangle$, 
where $V_i = T \partial W/\partial h_i |_{h_i = 0}$ and $W$ the master operator of 
the effective FA process. The perturbation operator $V_i$ accounts for the fact that 
the field $h_i$ reduces the energy barrier for creation of a defect at site $i$; 
recall that $E=\sum_i n_i (1-h_i)$ in the presence of the perturbation. Consequently 
the diffusion rates for entering site $i$ are increased. The rates for leaving site 
$i$, however, are not affected by $h_i$ because they relate to processes involving the
creation of a defect at sites $i-1$ or $i+1$. This locally directed perturbation 
can be represented as the sum of two mechanisms \cite{MaySol07}: 
(i) an {\em asymmetric} perturbation $V_i^{(\mathrm{a})}$ that increases the rates for 
entering site $i$ but decreases the ones for leaving it and thus {\em traps} diffusing 
defects, and (ii) a {\em symmetric} perturbation $V_i^{(\mathrm{s})}$ which increases 
both the rates for entering and leaving site $i$ and thus {\em accelerates} 
the local dynamics. In combination $V_i = V_i^{(\mathrm{a})} + V_i^{(\mathrm{s})}$ 
reproduces the directed perturbation of the local diffusion rates via the field 
$h_i$; the two contributions can be written explicitly as
\begin{eqnarray}
  V^{(\mathrm{a})}_i &=& \frac{1}{4} (F_{i-1} F_i - 1)
  (\hat{n}_{i-1} - \hat{n}_i) \hspace{1ex} +
   \frac{1}{4} (F_{i+1} F_i - 1)
  (\hat{n}_{i+1} - \hat{n}_i),
  \label{equ:Va} \\
  V^{(\mathrm{s})}_i &=& \frac{1}{4} (F_{i-1} F_i - 1)
  (\hat{n}_{i-1} - \hat{n}_i)^2 +
   \frac{1}{4} (F_{i+1} F_i - 1)
  (\hat{n}_{i+1} - \hat{n}_i)^2.
  \label{equ:Vs}
\end{eqnarray}
Clearly an analogous decomposition then applies to the response function 
\begin{equation}
  R_r(t,\tw) = \Ra_r(t,\tw) + \Rs_r(t,\tw), 
  \label{equ:res2tas}
\end{equation}
with $R_r^{(\mathrm{a}/\mathrm{s})}(t,\tw) = \langle \bm{e} | \hat{n}_{i+r} \rme^{W \tau} 
V_i^{(\mathrm{a}/\mathrm{s})} | p_{\tw} \rangle$ the contributions from the asymmetric and 
symmetric perturbations. We will see below that $\Ra_r(t,\tw)$ and $\Rs_r(t,\tw)$ have 
very different scaling properties, each being dominant in a different time regime. 
For now we simply state the relevant expressions \cite{MaySol07}, 
\begin{eqnarray}
\fl  \Ra_r(t,\tw) \sim \partial_{\tw} \frac{1}{2} \rme^{-2t} I_r(t-\tw) [ I_{r-1} 
  + 2 I_r + I_{r+1} ](t+\tw), 
  \label{equ:res2ta} \\
\fl \Rs_r(t,\tw) \sim - \frac{1}{4} \rme^{-2t} \left\{ I_r(t-\tw) [ -I_{r-2} + 2 I_r - I_{r+2} ](t+\tw) \right. 
  \nonumber \\
  \left. + [I_{r-1} - I_{r+1}](t-\tw) [I_{r-1} - I_{r+1}](t+\tw) \right\}. 
  \label{equ:res2ts} 
\end{eqnarray}
Like Eq.~(\ref{equ:cor2tsums}), Eq.~(\ref{equ:res2ta}) applies to leading order in 
$\tw \gg 1$ and uniformly in $\tau \geq 0$. The symmetric contribution 
$\Rs_r(t,\tw)$, Eq.~(\ref{equ:res2ts}), is sensitive to the value 
of $\gamma$ and for this reason only applies when both $\tau,\tw \gg 1$ \cite{MaySol07}.

\subsubsection{Local correlation and response}

The violation of FDT associated with the local defect observable $n_i$ has 
been considered in all simulation studies of the FA model 
\cite{Crisanti00,BuhGar02,buhot,Mayer2006}. However, as we now explain,
this observable is ill suited to the
measurement of FDT violations. One can derive the scaling of the relevant autocorrelation $C_0(t,\tw)$ 
and response $R_0(t,\tw)$ functions from Eqs.~(\ref{equ:cor2tsums}), (\ref{equ:res2ta}) and 
(\ref{equ:res2ts}) evaluated at $r = 0$. In the quasi-equilibrium regime of 
small $\tau \geq 0$ and large $\tw \gg 1$ the autocorrelation scales as \cite{MaySol07}
\begin{equation}
  C_0(t,\tw) \sim \frac{1}{\sqrt{\pi \tw}}\, \rme^{-\tau} I_0(\tau) \sim n(\tw) p_r(\tau). 
  \label{equ:cordst}
\end{equation}
Here we have identified the defect concentration $n(\tw) = 1/\sqrt{\pi \tw}$ and 
the random walk return probability $p_r(\tau) = \rme^{-\tau} I_0(\tau)$. In the 
quasi-equilibrium regime defects are typically too far from each other to meet 
and therefore may be treated as independent random walkers.
Correspondingly, the average $\langle n_i(t) n_i(\tw) \rangle$
is given by 
the probability $n(\tw)$ of having a defect at site $i$ at time $\tw$ multiplied by 
the probability $p_r(\tau)$ for this defect to return and occupy the same site 
at time $t$. This is then also the leading contribution to the local
autocorrelation function $C_0(t,\tw)$; the subtraction of
$n(t)n(\tw)$ only gives a subleading correction because $n(t)\ll 1$.

In the aging regime of large $\tw \gg 1$ with fixed $\tau/\tw$, an expansion of 
Eq.~(\ref{equ:cor2tsums}) yields the scaling behaviour \cite{MaySol07} 
\begin{eqnarray}
\fl C_0(t,\tw) \sim \frac{4}{\pi^{3/2}} \frac{\tw}{\tau^2} 
  \int_0^\infty \rmd x \int_0^\infty \rmd y \, 
  (x+y) \rme^{-(2\tw/\tau) (x^2+y^2)} \Erfc(x+y) 
  \nonumber \\ 
  - \frac{8}{\pi^{3/2}} \frac{\tw}{\tau^2} 
  \int_0^\infty \rmd x \int_0^\infty \rmd y \, 
  (x-y) \rme^{-(2\tw/\tau) (x^2+y^2)} \rme^{-x^2} \Erfc(y).  
  \label{equ:C0} 
\end{eqnarray}
Here $\Phi(z) = (2/\sqrt{\pi}) \int_z^\infty \rmd u \, \rme^{-u^2}$ denotes the 
complementary error function. The inte\-grals in Eq.~(\ref{equ:C0}) can be solved 
but the result is not particularly simple; we only state the resulting
scaling in the two 
limiting cases 
\begin{eqnarray}
  \tau \ll \tw: \, C_0(t,\tw) \approx \frac{1}{\pi \, \sqrt{2 \, \tau \, \tw}}, 
  \label{equ:corddtlltw} \\ 
  \tau \gg \tw: \, C_0(t,\tw) \approx \frac{3\pi-8}{3\pi^2} \frac{\tw}{\tau^2}. 
  \label{equ:corddtggtw}
\end{eqnarray}
This completely characterizes the defect autocorrelation $C_0(t,\tw)$: it has the 
initial value $C_0(\tw,\tw) \sim n(\tw)$ from which it decreases according to 
Eq.~(\ref{equ:cordst}). When $1 \ll \tau \ll \tw$ this quasi-equilibrium result 
matches with the expansion Eq.~(\ref{equ:corddtlltw}). As $\tau$ is further
increased beyond $\tw$ the scaling of $C_0(t,\tw)$ finally crosses over from Eq.~(\ref{equ:corddtlltw}) 
to Eq.~(\ref{equ:corddtggtw}). 

The scaling of the defect response function $R_0(t,\tw) = \Ra_0(t,\tw) + \Rs_0(t,\tw)$ 
is unusual in several ways. It is instructive to first discuss the step-response 
function $\chi_0(t,\tw)$. The latter is always dominated by contributions from the 
asymmetric perturbation $\chi_0(t,\tw) \sim \Xa_0(t,\tw)$ as can be shown 
\cite{MaySol07} from Eqs.~(\ref{equ:res2ta}, \ref{equ:res2ts}). An expression for 
$\Xa_0(t,\tw) = \int_{\tw}^t \rmd t' \, \Ra_0(t,t')$ 
is straightforwardly obtained from Eq.~(\ref{equ:res2ta}). In the quasi-equilibrium 
regime of small $\tau \geq 0$ and large $\tw \gg 1$ one finds the scaling 
\begin{equation}
  \Xa_0(t,\tw) \sim n(\tw) \left[ 1 - p_r(\tau) \right], 
  \label{equ:resdast} 
\end{equation}
while $\Xa_0(t,\tw) \sim n(t)$ in the aging regime $\tw \gg 1$ 
at fixed $\tau/\tw$. This non-monotonic behaviour of $\Xa_0(t,\tw)$ 
is due to a simple mechanism: when applying the asymmetric perturbation -- which traps 
diffusing defects -- at site $i$ and from time $\tw$ onwards, this increases the 
probability $\langle n_i(t) \rangle$ to find a defect at site $i$. The step response 
increases on a $\tau = \mathcal{O}(1)$ timescale, c.f.\ Eq.~(\ref{equ:resdast}), 
from zero to $n(\tw)$. But in the aging regime and for $\tau \gg \tw$, the density 
of defects $n(t)$ in the system decreases and along with it so does $\Xa_0(t,\tw) 
\sim n(t)$. While this makes perfect sense, one has to be careful when
recovering the response $\Ra_0(t,\tw) = - \partial_{\tw} \Xa_0(t,\tw)$ 
from this scaling result: one would incorrectly conclude that $\Ra_0(t,\tw) = - \partial_{\tw} n(t) 
= 0$. The resolution of the apparent paradox is that $\Xa_0(t,\tw)$ is dominated by 
small $\tau$ contributions of $\Ra_0(t,\tw)$, so that the aging behaviour of $\Ra_0(t,\tw)$ 
only shows up in {\em subdominant} terms in
$\Xa_0(t,\tw)$. Summarizing for the step response,
$\chi_0(t,\tw) \sim \Xa_0(t,\tw)$ is dominated by the 
trapping effect of the asymmetric perturbation and $\Xa_0(t,\tw)$, 
in turn, is dominated by short time contributions from $\Ra_0(t,\tw)$. 
This inconvenient insensitivity to aging effects can be avoided by
considering directly the response function $R_0(t,\tw)$, rather than
the step response $\chi_0(t,\tw)$. From Eqs.~(\ref{equ:res2ta}, \ref{equ:res2ts}) 
its constituent asymmetric and symmetric contributions scale as
\begin{eqnarray}
  \Ra_0(t,\tw) \sim \frac{1}{\pi} \frac{\tw}{(t^2 - \tw^2)^{3/2}}, 
  \label{equ:resdaag} \\ 
  \Rs_0(t,\tw) \sim -\frac{1}{2\pi} \frac{t-\tw}{(t^2-\tw^2)^{3/2}}, 
  \label{equ:resdsag}
\end{eqnarray}
in the aging regime $\tau,\tw \gg 1$ and $\tau/\tw = \mathcal{O}(1)$. Note the 
negative sign of $\Rs_0(t,\tw)$. To understand this, recall that applying the symmetric 
perturbation at site $i$ increases the local diffusion rates. This in itself does 
not affect the defect concentration $\langle n_i(t) \rangle$ at site $i$. 
However, the locally accelerated dynamics increase the likelihood for coagulation 
with neighbouring defects and thus reduce the local concentration $\langle n_i(t) \rangle$. 
Consequently $\Rs_0(t,\tw)$ is negative. Considering more
quantitatively the
asymmetric and symmetric response functions, Eqs.~(\ref{equ:resdaag}, 
\ref{equ:resdsag}), we see that they 
are of the {\em same} order $\mathcal{O}(\tw^{-2})$ when $\tau/\tw = \mathcal{O}(1)$. 
For much smaller $\tau\ll\tw$, the trapping effect encoded in $\Ra_0(t,\tw) = 
\mathcal{O}(\tau^{-3/2} {\tw}^{-1/2})$ dominates over $\Rs_0(t,\tw) \sim \mathcal{O}
(\tau^{-1/2} {\tw}^{-3/2})$, and this is at the origin of the
analogous dominance that we found in the step response.
In the opposite regime 
$\tau \gg \tw$ the situation is reversed, with effects from locally accelerated 
dynamics described by $\Rs_0(t,\tw) = \mathcal{O}(\tau^{-2})$ dominating over 
$\Ra_0(t,\tw) = \mathcal{O}(\tau^{-3} \tw)$. 

Based on the above insights regarding the scaling of $C_0(t,\tw)$ and $R_0(t,\tw)$ 
we are now in a position to discuss the resulting violations of the FDT and the 
usefulness of FD plots in revealing them. In the quasi-equilibrium regime 
of small $\tau \geq 0$ and large $\tw \gg 1$ we have that $C_0(t,\tw) \sim n(\tw) 
p_r(\tau)$, Eq.~(\ref{equ:cordst}), and $\chi_0(t,\tw) \sim \Xa_0(t,\tw) 
\sim n(\tw) [ 1 - p_r(\tau)]$, Eq.~(\ref{equ:resdast}). This satisfies the 
equilibrium FDT $\chi_0(t,\tw) \sim C_0(t,t) - C_0(t,\tw)$ to leading order in 
$\tw \gg 1$. Furthermore, the autocorrelation $C_0(t,\tw)$ decreases like $p_r(\tau) 
= \mathcal{O}(\tau^{-1/2})$ with increasing $\tau$ and thus drops to a fraction 
$\mathcal{O}({\tw}^{-1/2})$ of its equal time value $C_0(\tw,\tw)$ by the time 
we reach the aging regime where $\tau$ and $\tw$ are of the same
order; the susceptibility 
$\chi_0(t,\tw)$ approaches $n(t)$ up to a small deviation of the same order. Consequently FD plots for the local 
defect observable show effectively only the quasi-equilibrium behaviour of $C_0(t,\tw)$ 
and $\chi_0(t,\tw)$, with the nontrivial aging regime essentially
invisible at long times \cite{BuhGar02,buhot,Mayer2006}. This type of scaling makes the local defect observable a poor choice 
for probing FDT violations, especially using FD plots.
Regardless of these caveats, however, there is a well defined FDR $X_0(t,\tw) = 
R_0(t,\tw)/\partial_{\tw} C_0(t,\tw)$. It scales as $X_0 \sim X_0(t/\tw)$ 
in the aging limit \cite{MaySol07,Mayer2006}, crossing over from quasi-equilibrium $X_0 \sim 1$ for 
$\tau \ll \tw$ to the asymptotic FDR $X_0 \sim X_0^\infty$ for $\tau \gg \tw$. 
From the expansions Eq.~(\ref{equ:corddtggtw}) and Eq.~(\ref{equ:resdsag}) 
one obtains the {\em negative} value
\begin{equation}
  X_0^\infty = -\frac{3 \, \pi}{6 \, \pi - 16} \approx -3.307.
  \label{equ:Xdinfty}
\end{equation}

\begin{figure}
  \hspace*{1in} \includegraphics[width=4.5in,clip]{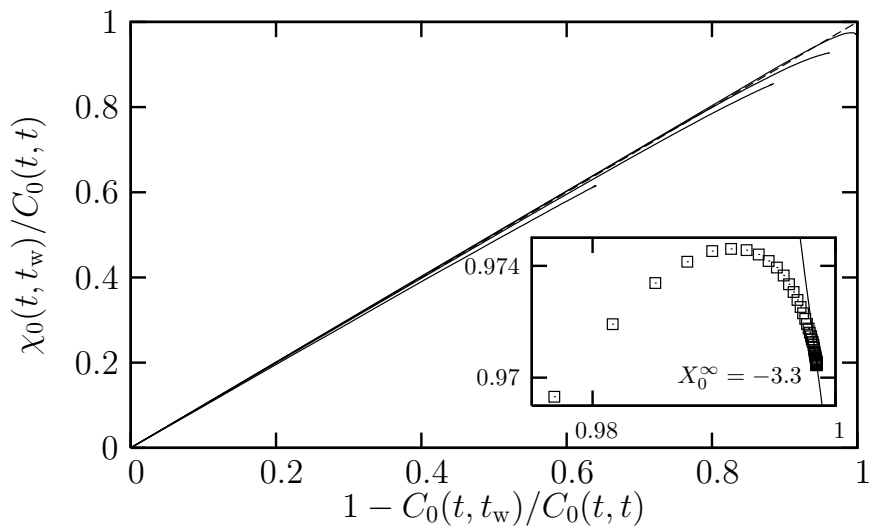} 
  \caption{\label{fig:1dinco} Normalized FD plots for 
    the local observable in the $d=1$ FA model at temperature $T=0.1$
and for times $t=2\times 10^4$, $5\times 10^4$, $10^5$, and $10^6$ 
(from bottom to top).
The dashed diagonal line indicates the equilibrium FDT. Inset:
final part of the $t= 10^6$ data, enlarged to show the nonmonotonic
response and the asymptotic FDR, $X_0^\infty = -3.307$ (solid
line). From Ref.~\cite{Mayer2006}. Copyright American Physical Society.}
\end{figure}

We show simulation results in Fig.~\ref{fig:1dinco} for various times
$t$. These confirm the  
findings of this section. In particular, the FD plot closely follows
the equilibrium FDT except in the regime of
large time differences, and the non-equilibrium part shrinks 
into the top right corner as $t$ increases. Very detailed
measurements in this region, as shown in the inset, are consistent
with the predicted asymptotic value of the FDR from Eq.~(\ref{equ:Xdinfty}).

\subsubsection{Global correlation and response} 
\label{fd_energy1d}

A much clearer picture of the violation of the FDT emerges when considering 
a global observable like the energy \cite{Mayer2006,MaySol07,MayBerGarSol03,
MayBerGarSol04}. We introduce the corresponding 
normalized connected correlation and response functions 
\begin{eqnarray} 
  C_E(t,\tw) = \frac{1}{N} \left( \langle E(t) E(\tw) \rangle - 
  \langle E(t)\rangle\langle E(\tw) \rangle \right), 
  \label{equ:CEdef} \\ 
  R_E(t,\tw) = \frac{T}{N} \left. \frac{ \delta \langle E(t) \rangle}
  {\delta h_E(\tw)} \right|_{h_E = 0}. 
  \label{equ:REdef} 
\end{eqnarray} 
The field $h_E$ uniformly shifts the energy according to $E = (1-h_E) \sum_i n_i$. When 
substituting $E = \sum_i n_i$ in Eq.~(\ref{equ:CEdef}) and using translational 
invariance one readily shows that in fact $C_E(t,\tw) = \sum_r C_r(t,\tw)$ 
and similarly $R_E(t,\tw) = \sum_r R_r(t,\tw)$ from Eq.~(\ref{equ:REdef}). 
The scalings of energy correlation and response functions can thus be derived 
by summing Eqs.~(\ref{equ:cor2tsums}), (\ref{equ:res2ta}) and (\ref{equ:res2ts}) 
over $r$. 

It can be shown \cite{MaySol07} from Eq.~(\ref{equ:cor2tsums}) that the scaling of energy 
correlations in the aging regime $\tw \gg 1$ with fixed $\tau/\tw$ is given by 
\begin{eqnarray}
\fl C_E(t,\tw) \sim \frac{4}{\pi} \frac{\tw}{\tau^{3/2}} 
  \int_0^\infty \rmd x \int_0^\infty \rmd y \, 
  (x+y) \rme^{-(\tw/\tau) (x+y)^2} \Erfc(x+y) 
  \nonumber \\ 
  - \frac{8}{\pi} \frac{\tw}{\tau^{3/2}} 
  \int_0^\infty \rmd x \int_0^\infty \rmd y \, 
  (x-y) \rme^{-(\tw/\tau) (x-y)^2} \rme^{-x^2} \Erfc(y). 
  \label{equ:CE}
\end{eqnarray}
Again these integrals can be solved exactly. The result is bulky but
would be required in full only if we wanted to inquire into the
precise form of the crossover between the simple limit behaviours
\begin{eqnarray}
  \tau \ll \tw: \, C_E(t,\tw) \approx \frac{3-2\sqrt{2}}{\sqrt{\pi \, \tw}}, 
  \label{equ:coreagdtlltw} \\ 
  \tau \gg \tw: \, C_E(t,\tw) \approx \frac{3\pi-8}{3\pi^{3/2}} \frac{\tw}{\tau^{3/2}}. 
  \label{equ:coreagdtggtw}
\end{eqnarray}
Energy correlations have the plateau value Eq.~(\ref{equ:coreagdtlltw}) for $0 \leq 
\tau \ll \tw$. This is because defects typically do not meet when
$\tau \ll \tw$ and therefore  
$C_E(t,\tw) \approx C_E(\tw,\tw)$. Only once $\tau$ increases to
become comparable to $\tw$ do 
coagulation events start to decorrelate $E(t)$ from $E(\tw)$, leading
to the crossover from Eq.~(\ref{equ:coreagdtlltw}) to (\ref{equ:coreagdtggtw}).

The scaling of the energy response function $R_E(t,\tw) = \Ra_E(t,\tw) + \Rs_E(t,\tw)$ 
can also be understood in very simple terms. The response $\Ra_E(t,\tw) = 0$ vanishes exactly 
because, when applied to all sites, the asymmetric perturbations of the diffusion rates 
cancel. Therefore $R_E(t,\tw) = \Rs_E(t,\tw) = \sum_r \Rs_r(t,\tw)$. From 
Eq.~(\ref{equ:res2ts}) this gives 
\begin{equation}
  R_E(t,\tw) \sim - \frac{1}{2 \, \sqrt{\pi} \, t^{3/2}} \sim \partial_t n(t), 
  \label{equ:resesag}
\end{equation}
in the aging regime $\tw \gg 1$ with fixed $\tau/\tw$. Due to the
normalisation by $1/N$, $R_E(t,\tw)$ 
measures changes in the defect concentration $n(t) = (1/N) \langle E(t) \rangle$. Now, 
in the presence of the perturbation all diffusion rates are increased. This is equivalent 
to giving the system some additional time $\delta t$ to evolve. The energy response function 
is therefore $R_E(t,\tw) \sim [n(t+\delta t) - n(t)]/\delta t \sim
\partial_t n(t)$. To understand the sign of this, note that the 
field $h_E$ decreases the energy barriers for defect creation. While in equilibrium this 
would increase the density of defects, in the nonequilibrium coarsening dynamics this 
speeds up the relaxation of the system -- it decreases $n(t)$ and hence $R_E(t,\tw) < 0$ 
is {\em negative} throughout. 

From the scalings Eq.~(\ref{equ:CE}) and Eq.~(\ref{equ:resesag}) one obtains the FD plot 
for the energy. Contrary to the case of the local defect observable $n_i$ the energy 
produces a nontrivial FD plot (contained in Fig.~\ref{fig:FD1d} 
below). This can be measured rather easily in simulations as discussed in Sec.~\ref{sec:nofield} 
and makes the energy a good observable for probing violations of FDT in the FA 
model. From the scalings in 
Eqs.~(\ref{equ:coreagdtggtw}, \ref{equ:resesag}) one shows 
that the asymptotic FDR for the energy is 
\begin{equation}
  X_E^\infty = - \frac{3\pi}{6\pi-16}. 
  \label{equ:XE}
\end{equation}
It matches exactly the value $X_0^\infty$, Eq.~(\ref{equ:Xdinfty}), obtained from the local 
defect observable but is much more accessible to measurement \cite{Mayer2006,MayBerGarSol03}. 
We will discuss this and related points further in the following section. 

\subsubsection{Fourier mode correlation and response}
\label{subsubCCR}

The above results describe the two extreme cases of spatially local and global 
observables. We will now consider observables with a finite intrinsic lengthscale 
in order to reveal in more detail the physics underlying the violation 
of FDT in the $1d$ FA model. Specifically, we consider the Fourier
modes of the arrangement of defects, $n_q = \sum_j n_j \, \rme^{-\rmi q j}$. The associated connected 
correlation and response functions are 
\begin{eqnarray}
  C_q(t,\tw) = \frac{1}{N} \left[ \langle n_q(t) n_{-q}(\tw) \rangle - 
  \langle n_q(t) \rangle \langle n_{-q}(\tw) \rangle \right], 
  \label{equ:Cqdef} \\ 
  R_q(t,\tw) = \frac{T}{N} \left. \frac{ \delta \langle n_q(t) \rangle}{\delta h_{-q}(\tw)} 
  \right|_{h_{-q} = 0}. 
  \label{equ:Rqdef} 
\end{eqnarray}
The fields $h_q$ are conjugate to $n_q$, resulting in a perturbation $\delta E = - h_q n_q$. 
Clearly, $C_q(t,\tw)$ is the dynamic structure factor and $R_q(t,\tw)$ the conjugate 
response function. Using translational invariance one readily verifies that $C_q(t,\tw) = 
\sum_r \rme^{-\rmi q r} C_r(t,\tw)$ is the Fourier transform of the spatial 
correlations, Eq.~(\ref{equ:Crdef}), and likewise for $R_q(t,\tw)$. The 
general results Eqs.~(\ref{equ:cor2tsums}), (\ref{equ:res2ta}) and (\ref{equ:res2ts}) are 
therefore once again the key for analysing the functions $C_q(t,\tw)$ and $R_q(t,\tw)$. 

In order to derive $C_q(t,\tw)$ from Eq.~(\ref{equ:cor2tsums}) the Fourier transforms 
of the Green's functions are needed; these contain all $r$ dependence in $C_r(t,\tw)$. 
Using the identity Eq.~(\ref{equ:S1}) one immediately has 
\begin{eqnarray*} 
\fl G_{q}^{(+)}(\tau;i,j) = \sum_r \rme^{-\rmi q r} G_{(r,r+1),(-i,j+1)}(\tau) = \rme^{\rmi \frac{1}{2}(i-j)q} 
  \rme^{-2\tau} [I_{i+j} - I_{i+j+2}]\left(2 \tau \cos \frac{q}{2} \right), 
  \\
\fl G_{q}^{(-)}(\tau;i,j) = \sum_r \rme^{-\rmi q r} G_{(r,r-1),(i,j)}(\tau) = \rme^{-\rmi \frac{1}{2}(i+j+1)q} 
  \rme^{-2\tau} [I_{i-j-1} - I_{i-j+1}]\left(2 \tau \cos \frac{q}{2} \right), 
\end{eqnarray*}
and in terms of these functions
\begin{eqnarray}
\fl C_q(t,\tw) \sim 
	\rme^{-2\tw} [I_0 + I_1](2\tw) \sum_{i,j=0}^\infty 
	\mathrm{Re} \left[ G_{q}^{(+)}(\tau;i,j) \right] H_{i+j+1}(2\tw) 
  \nonumber \\ 
  - \sum_{i,j=0}^\infty 2 \, \mathrm{Re}\left[ G_{q}^{(-)}(\tau;i,j) \right] 
  \rme^{-2\tw} [I_i + I_{i+1}](2\tw) \left[ \delta_{j,0} + H_j(2\tw) \right]. 
\label{equ:Cq}
\end{eqnarray}
In the first line of Eq.~(\ref{equ:Cq}) the imaginary part of $G_{q}^{(+)}(\tau;i,j)$ 
drops out since it is odd in $i-j$. The second line, on the other hand, is a sum of 
two contributions $G_{q}^{(-)}(\tau;i,j) + 
G_{-q}^{(-)}(\tau;i,j)$ from Eq.~(\ref{equ:cor2tsums}), which is again real. The scaling behaviour of Eq.~(\ref{equ:Cq}) 
in the aging limit $\tw \to \infty$ at fixed $\tau/\tw$ and $\tw q^2$ follows 
from the asymptotic expansions Eqs.~(\ref{equ:IHx}, \ref{equ:IHxn}) of the functions 
$I_n(t)$ and $H_n(t)$, and is given by 
\begin{eqnarray}
\fl C_q \sim \frac{4}{\pi} \frac{\tw}{\tau^{3/2}} \rme^{-\frac{1}{4} \tau q^2} 
  \left[ 
  \int_0^\infty \rmd x \int_0^\infty \rmd y \, 
  (x+y) \cos\left( \sqrt{\tw} q (x-y) \right) 
  \rme^{-(\tw/\tau) (x+y)^2} \Erfc(x+y) 
  \right. 
  \nonumber \\ 
  - 2 \left. 
  \int_0^\infty \rmd x \int_0^\infty \rmd y \,  
  (x-y) \cos\left( \sqrt{\tw} q (x+y) \right) 
  \rme^{-(\tw/\tau) (x-y)^2} \rme^{-x^2} \Erfc(y) 
  \right]. 
  \label{equ:Cqagaux}
\end{eqnarray}
Time arguments $C_q = C_q(t,\tw)$ were omitted to save space. 
Energy correlations are obtained as the special case $C_E(t,\tw) = C_{q=0}(t,\tw)$ 
as is clear from the definition of $C_q(t,\tw)$,
Eq.~(\ref{equ:Cqdef}), and indeed 
Eq.~(\ref{equ:Cqagaux}) with $q = 0$ reduces to Eq.~(\ref{equ:CE}). Moreover, the 
inverse Fourier transform $C_{r=0}(t,\tw) = \int (\rmd q/2\pi) \, C_q(t,\tw)$ 
yields the autocorrelation. The resulting scaling of $C_{r = 0}(t,\tw)$, Eq.~(\ref{equ:C0}), 
is then also a special case of Eq.~(\ref{equ:Cqagaux}). 

Equation~(\ref{equ:Cqagaux}) can be rearranged into a more useful form. 
We rotate the integration coordinates $u=y+x$ and 
$v = y-x$, whereby the integration domain becomes $u \in [0,\infty]$ and 
$v \in [-u,u]$. In the first line of Eq.~(\ref{equ:Cqagaux}) the integration over 
$v$ can then be carried out. In the second line, we integrate by parts in $u$. 
This leads to some simplifications and produces, after combining the $v \in [-u,0]$ 
and $v \in [0,u]$ integration ranges, 
\begin{eqnarray}
\fl C_q \sim \frac{1}{\sqrt{\pi}} \frac{\tw}{t^{3/2}} 
  \rme^{ - \frac{1}{4} (1 - \tw^2/t^2) t q^2 }
  - \frac{4}{\pi} \frac{\tw}{\tau^{3/2}} \rme^{-\frac{1}{4} \tau q^2}
  \int_0^\infty \rmd u \frac{\sin\left( \sqrt{\tw} q u\right)}{\sqrt{\tw} q} 
  \int_0^u \rmd v \, v \, \rme^{- (\tw/\tau) v^2} f(u,v), \nonumber \\
\fl \mbox{where} 
  \label{equ:Cqag} 
\end{eqnarray}
\begin{equation}
  f(u,v) = \frac{u+v}{2} \rme^{-\frac{1}{4} (u+v)^2} \Erfc\left( \frac{u-v}{2} \right)
  - \frac{u-v}{2} \rme^{-\frac{1}{4} (u-v)^2} \Erfc\left( \frac{u+v}{2} \right). 
  \label{equ:fuv}
\end{equation}
This is our general scaling result for the dynamic structure factor. 
Let us now consider some limit cases of Eq.~(\ref{equ:Cqag}). We first study the 
regime $\tau \ll \tw$. In this case the exponential $\rme^{- (\tw/\tau) v^2}$ peaks 
sharply at $v = 0$. To leading order in $\tau/\tw$ we may replace the upper integration 
limit $u$ of the $v$-integration with $\infty$ and approximate $f(u,v)$ by its expansion at 
$v = 0$. Since $f(u,0) = 0$ we use $f(u,v) \approx v \, \partial_v f(u,v)|_{v=0}$. 
Also noting that $f(u,v)$ satisfies the identity $\partial_v f(u,v)|_{v=0} = \partial_u 
[u \rme^{-u^2/4} \Erfc(u/2) - (2/\sqrt{\pi}) \rme^{-u^2/2}]$ and using integration 
by parts in $u$ then gives 
\begin{equation}
\fl C_q \approx \frac{1}{\sqrt{\pi \tw}} \rme^{-\frac{1}{4} \tau q^2} \left[ 
  \rme^{-\frac{1}{4} \tau q^2} - \sqrt{2} \, \rme^{-\frac{1}{2} \tw q^2} 
  + \int_0^\infty \rmd u \, \cos\left( \sqrt{\tw} q u \right) u \, \rme^{-\frac{1}{4} u^2} 
  \Erfc\left( \frac{u}{2} \right) \right], 
  \label{equ:Cqagdtlltw} 
\end{equation}
in the regime $\tau \ll \tw$ and to leading order in $\tau/\tw$. It turns out that 
Eq.~(\ref{equ:Cqagdtlltw}) reduces correctly to the static structure factor for 
$\tau \to 0$; this can be verified rigorously by setting $\tau = 0$ in the full 
expression Eq.~(\ref{equ:Cq}) and then expanding for large $\tw$ and fixed $\tw q^2$. 
Either way one ends up with the result 
\begin{equation}
\fl C_q(\tw,\tw) \sim \frac{1}{\sqrt{\pi \tw}} \left[ 
  1 - \sqrt{2} \, \rme^{-\frac{1}{2} \tw q^2} 
  + \int_0^\infty \rmd u \, \cos\left( \sqrt{\tw} q u \right) u \, \rme^{-\frac{1}{4} u^2} 
  \Erfc\left( \frac{u}{2} \right) \right]. 
  \label{equ:Cqstat} 
\end{equation}
The static structure factor has a local minimum at $q = 0$, where it measures energy 
fluctuations $C_E(\tw,\tw) = C_{q=0}(\tw,\tw)$. We find the value $C_{q=0}(\tw,\tw) = (3-2\sqrt{2})/
\sqrt{\pi \tw}$ from Eq.~(\ref{equ:Cqstat}). This is, of course, consistent 
with the expansion Eq.~(\ref{equ:coreagdtlltw}) of $C_E(t,\tw)$ from above. For 
$\tw q^2 \gg 1$, $C_q(\tw,\tw)$ quickly approaches $n(\tw) = 1/\sqrt{\pi \tw}$. 
This large $q$ behaviour originates from the scaling of $C_{r=0}(\tw,\tw) = n(\tw) 
- n(\tw)^2 \sim n(\tw)$. The local minimum in the static structure factor is caused 
by an ``effective repulsion" of defects in the diffusion-coagulation dynamics: 
defects that have survived up to time $\tw$ are likely to be far from each other, 
i.e.\ at a distance $\mathcal{O}(\sqrt{\tw})$. 
A similar effect was seen in the aging regime of 
plaquette models~\cite{Robs_plaquette_FDT}.

To understand the scaling of $C_q(t,\tw)$ for $1 \ll \tau \ll \tw$ we return to 
Eq.~(\ref{equ:Cqagdtlltw}). 
The profile of $C_q(t,\tw)$ is as follows: on the scale $\tw q^2 = \mathcal{O}(1)$, 
where $\tau q^2 \ll 1$ since $\tau \ll \tw$, Eq.~(\ref{equ:Cqagdtlltw}) is still 
essentially equal to the static structure factor Eq.~(\ref{equ:Cqstat}). However, 
on the larger scale in $q$ where $\tau q^2 = \mathcal{O}(1)$, and hence $\tw q^2 \gg 1$, 
one has from Eq.~(\ref{equ:Cqagdtlltw}), 
\begin{equation}
  C_q(t,\tw) \approx \frac{1}{\sqrt{\pi \tw}} \rme^{-\frac{1}{2} \tau q^2}. 
  \label{equ:Cqdtlltwdiff}
\end{equation}
This makes sense intuitively. The wavevector scale $\tau q^2 = \mathcal{O}(1)$ corresponds 
to a length scale $\ell(\tau) = \mathcal{O}(\sqrt{\tau})$. Here, since $\tau \ll \tw$, we may 
treat defects as independent random walkers. Correlations in $C_r(t,\tw)$ on the lengthscale 
$\ell(\tau)$ are diffusive, and this is consistent with Eq.~(\ref{equ:Cqdtlltwdiff}). On the 
lengthscale $\ell(\tw) = \mathcal{O}(\sqrt{\tw})$, corresponding to $\tw q^2 = \mathcal{O}(1)$, 
on the other hand, spatial correlations $C_r(t,\tw)$ are still essentially static and 
this translates into $C_q(t,\tw) \approx C_q(\tw,\tw)$, c.f.\ Eq.~(\ref{equ:Cqagdtlltw}) 
and Eq.~(\ref{equ:Cqstat}). 


We now turn to the scaling of $C_q(t,\tw)$ in the opposite regime $\tau \gg \tw$. Here 
an expansion of Eq.~(\ref{equ:Cqag}) follows rather straightforwardly. The exponential 
$\rme^{- (\tw/\tau) v^2}$ becomes flat and may be replaced by unity to leading order in 
$\tw/\tau$; this is justified since $f(u,v)$ vanishes sufficiently fast in $v$. Moreover, 
the only relevant wavevector scale is $\tau q^2 = \mathcal{O}(1)$ so that $\sin(\sqrt{\tw} q u) 
\approx \sqrt{\tw} q u$. Here we have used $\sqrt{\tw} q \ll 1$ since $\tw \ll \tau$ 
and the fact that again only small $u$ contribute to the integral. Thus altogether, 
\begin{eqnarray}
  C_q(t,\tw) \approx \frac{\tw}{\tau^{3/2}} \rme^{-\frac{1}{4} \tau q^2} 
  \left[ \frac{1}{\sqrt{\pi}} - \frac{4}{\pi} 
  \int_0^\infty \rmd u \, u \int_0^u \rmd v \, v f(u,v) \right], 
  \label{equ:Cqdtggtwaux} 
\end{eqnarray}
The expression in the square brackets only produces an overall coefficient. To evaluate 
it one has to change the integration variables back to $x,y$. Then the integrals can 
be factorized and one readily obtains 
\begin{equation}
  C_q(t,\tw) \approx \frac{3\pi - 8}{3\pi^{3/2}} \frac{\tw}{\tau^{3/2}} \rme^{-\frac{1}{4} \tau q^2},  
  \label{equ:Cqdtggtw} 
\end{equation}
in the regime $\tau \gg \tw$ and to leading order in $\tw/\tau$. 
This result contrasts with Eq.~(\ref{equ:Cqdtlltwdiff}) in several
ways. First, the scaling 
of the amplitude of correlations $C_q(t,\tw)$ changes from $\mathcal{O}(\tw^{-1/2})$ 
in the regime $\tau \ll \tw$ to $\mathcal{O}(\tw/\tau^{3/2})$ when $\tau \gg \tw$. Second, 
the shape of $C_q(t,\tw)$ on the wavevector scale $\tau q^2 = \mathcal{O}(1)$ is always 
Gaussian, however, the exponents in Eq.~(\ref{equ:Cqdtlltwdiff}) and Eq.~(\ref{equ:Cqdtggtw}) 
differ by a factor of $\frac{1}{2}$. Third, the overall coefficient in 
Eq.~(\ref{equ:Cqdtggtw}) is {\em not} related to the static structure factor at $q=0$. 
These differences between Eq.~(\ref{equ:Cqdtlltwdiff}) and Eq.~(\ref{equ:Cqdtggtw}) 
arise from the fact that for $\tau \gg \tw$ many-body effects play an important role; defects meet 
and coagulate. This regime is fluctuation-dominated with all loop diagrams contributing 
in a field theoretic framework, see Sec.~\ref{sec:RGdltdc} below. 


Having discussed the dynamic structure factor $C_q(t,\tw)$ in detail we now turn to 
the analysis of response functions. Let us first consider the asymmetric part $\Ra_r(t,\tw)$ 
of the response function. The Fourier transform of Eq.~(\ref{equ:res2ta}) follows 
immediately from the identity Eq.~(\ref{equ:S2}) and is given by 
\begin{equation}
  \Ra_q(t,\tw) \sim \partial_{\tw} \rme^{-2t} \left[ I_0(2A) + \frac{t \cos(q/2)^2 + 
  \tw \sin(q/2)^2}{A} I_1(2A) \right], 
  \label{equ:Raq}
\end{equation}
where $A = \sqrt{ t^2 \cos(q/2)^2 + {\tw}^2 \sin(q/2)^2}$. The scaling of this expression 
in the aging limit $\tw \to \infty$ with $\tau/\tw$ and $\tw q^2$
fixed follows using Eq.~(\ref{equ:IHx}) as
\begin{equation}
  \Ra_q(t,\tw) \sim \partial_{\tw} \frac{1}{\sqrt{\pi t}} \rme^{-\frac{1}{4} (1-\tw^2/t^2) t q^2}.
  \label{equ:Raqag} 
\end{equation}
At $q=0$ this reduces to the asymmetric part of the energy response $\Ra_E(t,\tw) = \Ra_{q=0}(t,\tw)$ 
and vanishes, c.f.\ Eq.~(\ref{equ:Raqag}), since -- as discussed above -- the asymmetric perturbations 
cancel when applied uniformly. Integration over $q$, on the other hand, yields the scaling Eq.~(\ref{equ:resdaag}) 
for the auto-response $\Ra_{r=0}(t,\tw) = \int (\rmd q/2\pi) \, \Ra_q(t,\tw)$. 
The small and large $\tau$ scaling forms of Eq.~(\ref{equ:Raqag}) are 
\begin{eqnarray}
  \tau \ll \tw: \, \Ra_q(t,\tw) \approx \frac{1}{2} \tw q^2 \frac{1}{\sqrt{\pi} \tw^{3/2}} 
  \rme^{-\frac{1}{2} \tau q^2}, 
  \label{equ:Raqagdtlltw} \\ 
  \tau \gg \tw: \, \Ra_q(t,\tw) \approx \frac{1}{2} \tw q^2 \frac{1}{\sqrt{\pi} \tau^{3/2}} 
  \rme^{-\frac{1}{4} \tau q^2}.
  \label{equ:Raqagdtggtw} 
\end{eqnarray}
In analogy with the results for correlations, the exponents in the Gaussians differs by 
a factor of $\frac{1}{2}$ between Eq.~(\ref{equ:Raqagdtlltw}) and Eq.~(\ref{equ:Raqagdtggtw}). 
For the construction of FD plots we need the step response function associated with 
$\Ra_q(t,\tw)$. This is obtained straightforwardly from Eq.~(\ref{equ:Raqag}), 
\begin{equation}
  \Xa_q(t,\tw) = \int_{\tw}^t \rmd t' \Ra_q(t,t') \sim 
  \frac{1}{\sqrt{\pi t}} \left[ 1 - \rme^{- \frac{1}{4} (1 - \tw^2/t^2) t q^2 } \right]. 
  \label{equ:chia}
\end{equation}

It remains to work out the scaling of the symmetric part $\Rs_q(t,\tw)$ of the response 
function. To calculate the Fourier transform of Eq.~(\ref{equ:res2ts}) we use the 
identities Eq.~(\ref{equ:S2}) and Eq.~(\ref{equ:S3}), which produce 
\begin{equation}
\fl \Rs_q(t,\tw) \sim - \rme^{-2t} \cos\left(\frac{q}{2}\right)^2 \left[ I_0(2A) - 
  \frac{t^2 - (t-\tw)^2 \sin(q/2)^2}{A^2} I_2(2A) \right], 
  \label{equ:Rsq} 
\end{equation}
again with $A = \sqrt{ t^2 \cos(q/2)^2 + {\tw}^2 \sin(q/2)^2}$. We now perform the 
usual aging expansion $\tw \to \infty$ with $\tau/\tw$ and $\tw q^2$ fixed. But for 
Eq.~(\ref{equ:Rsq}) some care is needed: the leading terms in the expansion cancel. 
Therefore the first subdominant terms in the modified Bessel functions have to be 
retained. In this way one obtains the scaling 
\begin{equation}
  \Rs_q(t,\tw) \sim - \frac{1}{2\sqrt{\pi}t^{3/2}} \left( 1 - \frac{1}{2} \tw q^2 
  \frac{t-\tw}{t} \right) \rme^{-\frac{1}{4} ( 1 - \tw^2/t^2) t q^2}. 
  \label{equ:Rsqag} 
\end{equation}
%
%
The small and large $\tau$ scaling forms of Eq.~(\ref{equ:Rsqag}) are 
\begin{eqnarray}
  \tau \ll \tw: \, \Rs_q(t,\tw) \approx - \frac{1}{2\sqrt{\pi}\tw^{3/2}} \left( 1 - \frac{1}{2} \tau q^2 
  \right) \rme^{-\frac{1}{2} \tau q^2}, 
  \label{equ:Rsqagdtlltw} \\ 
  \tau \gg \tw: \, \Rs_q(t,\tw) \approx - \frac{1}{2\sqrt{\pi}\tau^{3/2}} 
  \rme^{-\frac{1}{4} \tau q^2}. 
  \label{equ:Rsqagdtggtw} 
\end{eqnarray}
Once again the coefficient in the exponents of the Gaussians crosses over from 
$\frac{1}{2}$ to $\frac{1}{4}$ as we go from the regime $\tau \ll \tw$ to $\tau \gg \tw$. 

Integration of the expression Eq.~(\ref{equ:Rsqag}) yields a simple result for the 
step response function associated with $\Rs_q(t,\tw)$, 
\begin{equation}
  \Xs_q(t,\tw) = \int_{\tw}^t \rmd t' \Rs_q(t,t') \sim 
  - \frac{t-\tw}{2\sqrt{\pi} t^{3/2}} \rme^{- \frac{1}{4} (1 - \tw^2/t^2) t q^2 }. 
  \label{equ:chis}
\end{equation}

\begin{figure}
  \hspace*{1in} \includegraphics[width=4.5in,clip]{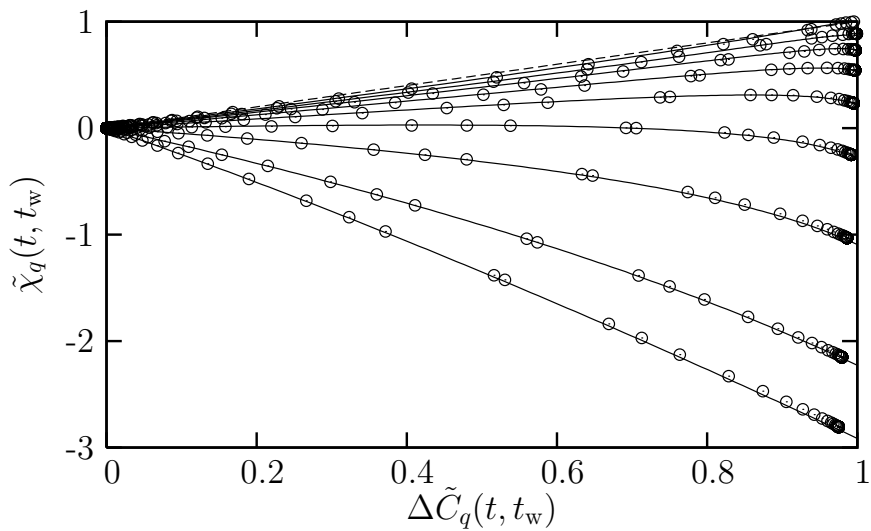} 
  \caption{\label{fig:FD1d} Normalized FD plots for Fourier mode 
    observables $n_q$ in 
    the $d=1$ FA model. Symbols are simulation data at $T = 0.08$ in a 
    system of linear size $L = 700$, at final time $t = 
    4.87\times 10^7$ and wavevector $q = (2\pi/L) j$ 
    with $j = 0$, 3, 6, 9, 12, 15, 18, 21 and 27 (from bottom to top); 
    note that $t$ is given in microscopic 
    units rather than units of $1/c$, as used in the analysis. The dashed line
    represents equilibrium FDT. Full lines show the theoretical 
    scaling predictions. The lowest curve is for
    $q = 0$ and gives the FD plot for the energy. It is slightly curved, 
    with slope increasing (in modulus) from $- (1+\sqrt{2})$ 
    at the origin to $X_q^\infty$, 
    Eq.~(\ref{equ:Xq}), on the right.} 
\end{figure}

Putting together the results from above yields the FD plots 
for the observables $n_q$ in the FA model. In contrast to the case of the 
local observable $n_i$ non-trivial limit plots do exist as long as we
keep $tq^2$ or equivalently $\sqrt{t}q$ constant as we increase $t$. We use normalized 
plots showing $\widetilde{\chi}_q$ versus $\Delta \widetilde{C}_q = 1-\widetilde{C}_q$ where 
\begin{equation}
  \widetilde{\chi}_q(t,\tw) = \frac{\chi_q(t,\tw)}{C_q(t,t)} 
  \quad \mbox{and} \quad 
  \widetilde{C}_q(t,\tw) = \frac{C_q(t,\tw)}{C_q(t,t)}. 
\end{equation}
If parameterized with $\tw$, the slope in the plots directly gives the FDR $X_q$. 
For our wavevector observables we use $C_q(t,t)$, Eq.~(\ref{equ:Cqstat}), to normalize 
the plots. The scaling expression for $C_q(t,\tw)$ is stated in Eq.~(\ref{equ:Cqag}) 
and $\chi_q(t,\tw) = \Xa_q(t,\tw) + \Xs_q(t,\tw)$ 
is given by Eq.~(\ref{equ:chia}) and Eq.~(\ref{equ:chis}), respectively. Numerical 
evaluation of these quantities produces the FD plots shown in Fig.~\ref{fig:FD1d}. 

We first note that observables $n_q$ with $\sqrt{t} q \gg 1$ produce FD plots close to 
the equilibrium line $\widetilde{\chi}_q = 1 - \widetilde{C}_q$, similarly as is the case 
for the local defect observable $n_i$. Very small values of $q$, $\sqrt{t} q \ll 1$, on 
the other hand, yield essentially the same FD plot as the energy ($q=0$ curve in 
Fig.~\ref{fig:FD1d}). For intermediate wavevectors $q$ the FD plots interpolate between 
these two extremes. The wavevectors $q$ have an associated intrinsic lengthscale 
$\ell = 1/q$. Thus the observables $n_q$ allow one to probe violations of FDT on any given 
lengthscale. While the coarsening dynamics equilibrate short length scales $\ell \ll 
\sqrt{t}$, large length scales $\ell \gg \sqrt{t}$ remain far from equilibrium 
with a well defined violation of FDT corresponding to the case $q = 0$. 

The FD plots in Fig.~\ref{fig:FD1d} are unusual in two ways: first, they are non-monotonic 
and second, the step-response functions $\chi_q$ can actually become negative. We remark 
that the non-monotonicity of the plots is genuine and not due to incorrect parameterization 
with $t$ instead of $\tw$ \cite{Crisanti00}. As 
for the local and global observables discussed 
above, the response function $R_q$ comprises an asymmetric and symmetric contribution. 
In the regime $\tau \ll \tw$, $\Ra_q > 0$ dominates over $\Rs_q < 0$, see Eqs.~(\ref{equ:Raqagdtlltw}, \ref{equ:Rsqagdtlltw}). One then verifies from Eqs.~(\ref{equ:Cqdtlltwdiff}, \ref{equ:Raqagdtlltw}) 
that $X_q \sim \Ra_q(t,\tw)/\partial_{\tw} C_q(t,\tw) \sim 1$. Equilibrium FDT is satisfied 
for $\tau \ll \tw$ and all FD plots have slope $X_q \sim 1$ close to
the origin. This applies with the exception of $q=0$ itself;
numerically, already for small but nonzero $q$ the initial equilibrium
regime is difficult to discern from the plots.
In the opposite regime $\tau \gg \tw$, on the other 
hand, $\Rs_q < 0$ dominates over $\Ra_q > 0$, c.f.\ Eqs.~(\ref{equ:Raqagdtggtw}, \ref{equ:Rsqagdtggtw}), and the response $R_q$ becomes negative. This causes the non-monotonic 
FD plots and negative step-response functions $\chi_q$. The resulting asymptotic FDR 
$X_q^\infty$ for $\tau \gg \tw$ is given by $X_q \sim \Rs_q(t,\tw)/\partial_{\tw} C_q(t,\tw)$, 
and from the scalings Eqs.~(\ref{equ:Cqdtggtw}, \ref{equ:Rsqagdtggtw}), takes the value 
\begin{equation}
  X_q^\infty = - \frac{3\pi}{6\pi-16} \approx -3.307. 
  \label{equ:Xq}
\end{equation}
All FD plots in Fig.~\ref{fig:FD1d} thus end with the same slope
$X_q^\infty$ on the right, regardless of $q$. 
This is clearly observable in the plots with $\sqrt{t} q \ll 1$ but becomes 
difficult to detect when $\sqrt{t} q \gg 1$. This reinforces the
general point that observables which probe directly large, 
non-equilibrated length scales are better suited for 
measurement of $X_q^\infty$ 
\cite{Mayer2006,MayBerGarSol03}. 

Negativity of the response function $R_q$ can be traced back to the activated nature 
of the dynamics. Perturbations coupling to the $n_i$ affect the energy barriers 
for creation of defects on these sites. This effect and the associated response functions can be 
decomposed into symmetric $\Rs_q$ and asymmetric $\Ra_q$ contributions, with the former 
trapping diffusing excitations but the latter accelerating their dynamics. 
In the regime $\tau \gg \tw$ the acceleration effect dominates, and the faster relaxation 
towards $n(t) \to 0$ decreases the expectation values $\langle n_i(t) \rangle$, thus 
the negative sign of the response functions $R_q$. We will see below that similar arguments 
apply to FA models in dimensions $d>1$ and also to the East model.

The asymptotic scaling results derived in this section are confronted with numerical 
data in Fig.~\ref{fig:FD1d}. The simulations are based on the continuous time BKL 
algorithm \cite{BKL}. Step response functions $\chi_q$ with $q \neq 0$ are measured using 
the no-field method of Chatelain \cite{Chatelain03, chatelain, Ricci}, while a specialised no-field method 
is employed for the $q=0$ case, see Sec.~\ref{sec:nofield} below. These no-field methods enable 
us to construct properly parameterised FD plots with $\tw \in [0,t]$
as the running parameter. 
It is obvious from Fig.~\ref{fig:FD1d} that 
excellent agreement is obtained between the
scaling results and the 
simulation data.

\subsection{Field theoretical approach}
\label{sec:FA_ft}

In this subsection we describe how to calculate aging properties of
the FA model by means of an effective field theory built from the
Doi-Peliti \cite{Doi,Peliti} formalism for systems with stochastic
dynamics.  This approach to treat the FA model was first introduced in
\cite{Whitelam2004} and developed further in
\cite{Whitelam2005,Jack2006}.  A summary of the results below, in so
far as they relate to aging dynamics for dimensions $d>\dc=2$, can be
found in \cite{Mayer2006}.  The aim of this section is to provide a detailed
discussion of the calculations leading to those results, and to
consider their extension to $d<\dc$ where a comparison with the exact
results from the previous section can be performed.

\subsubsection{Effective theory and dynamical action}

We consider the ``bosonic'' version of the FA model
\cite{Whitelam2004,Whitelam2005,Jack2006}: we allow for each site of
the lattice to contain any number of excitations; that is, $n_i =
0,1,\ldots$ indicates the occupation of site $i$.  Since we are
interested in the regime of low temperatures where the equilibrium
density of excitations is low, removing the hard-core restriction on
site occupation does not change the behaviour of the model
significantly \cite{Whitelam2004,Whitelam2005,Jack2006}.

The derivation of the field theory for this version of the FA model
was described in detail in Ref.\ \cite{Whitelam2005} (see also
\cite{Jack2006}).  The effective theory is defined for excitations on
a lattice of $N$ sites in $d$ dimensions with integer occupancy per
site.  As explained above, excitations can effectively diffuse between
neighbouring sites 
with a rate proportional to $c$; neighbouring excitations can
coalesce with a rate likewise proportional to $c$
\cite{Whitelam2004,Mayer2006,Jack2006}.  Excitations can also branch
into pairs, but the rate for this process goes as $c^2$, and we
disregard it in our analysis of aging at low temperatures.  Using the
Doi-Peliti formalism \cite{Doi,Peliti,Lee,Tauber05} we can represent this
dynamics in terms of complex fields $\phi(\rbf,t)$,
$\bar{\phi}(\rbf,t)$ with an action
\cite{Whitelam2004,Whitelam2005,Jack2006,Mayer2006}
\begin{equation}
  S = \int_{\rbf, t} \bar{\phi} (\partial_t - D \nabla^2)
  \phi + \lambda \bar{\phi} (1+\bar{\phi}) \phi^2 - n_0 \bar{\phi} \delta(t),
\label{action}
\end{equation}
where $D,\lambda \propto c$.  This is the field theory for the $A+A
\to A$ reaction-diffusion process \cite{Lee,Tauber05}.  In this representation
the local occupation number operator $\hat{n}(\rbf,t)$ 
is given by 
\begin{equation}
\hat{n}(\rbf,t) = \left[ 1+\bar{\phi}(\rbf,t) \right] \phi(\rbf,t) .
\label{nphi}
\end{equation}
The last term in (\ref{action}) indicates a Poisson distribution of
random initial conditions of density $n_0$.  Correlation functions are
calculated through the path integral: $\langle \hat{A}(t) \rangle =
\int D \bar{\phi} D\phi \,\hat{A}[\hat{n}(t)] e^{-S}$.

\subsubsection{Tree-level density and two-time correlations}

For $d>\dc=2$ the field theory is finite and a tree level
calculation will give the correct behaviour at long times.  Such a
tree-level analysis
amounts to solving the Euler-Lagrange equations.  From (\ref{action})
we have:
\begin{eqnarray}
\frac{\delta S}{\delta \bar{\phi}(t)} &=& (\partial_t - D \nabla^2)
\phi +  \lambda (1+2 \bar{\phi}) \phi^2 - n_0 \delta(t) = 0 ,
\label{EL1} \\ \frac{\delta S}{\delta \phi(t)} &=& -(\partial_t + D
\nabla^2) \bar{\phi} + 2 \lambda \bar{\phi} (1+\bar{\phi}) \phi  = 0
. \label{EL2}
\end{eqnarray} 

The average density is given by:
\begin{equation}
n(t) \equiv V^{-1} \int_{\rbf, t} \langle \hat{n}(\rbf,t) \rangle = V^{-1} \int_{\rbf, t}
\langle \phi(\rbf,t) \rangle = V^{-1}  \langle \phi_{0}(t) \rangle ,
\end{equation}
where $\phi_{\bm{q}}(t)$ is the Fourier transform of $\phi(\rbf,t)$,
$\phi_{\bm{q}}(t)=\int_{\bm{r}} e^{-i\bm{q}\cdot\bm{r}}\phi(\bm{r},t)$
[similarly for $\bar\phi_{\bm{q}}(t)$], and we have used that averages
with a factor $\bar{\phi}$ on the left (i.e.\ at the latest time)
vanish. Taking the expectation value of Eq.\ (\ref{EL1}) allows us to
calculate $n(t)$ at tree level:
\begin{equation}
\langle \frac{\delta S}{\delta \bar{\phi}(t)} \rangle = 0 \Rightarrow
\partial_t \langle \phi_0 \rangle + \lambda V^{-1}\langle \phi_0 \rangle^2 =
0 \Rightarrow n(t) = \frac{n_0}{1+\lambda n_0 t} \approx
\frac{1}{\lambda t} .
\label{ntl}
\end{equation} 
From (\ref{EL2}) we can get the tree-level propagator $G_q(t,t_{\rm
w}) \equiv V^{-1}\langle \phi_{\bm q}(t) \bar{\phi}_{-\bm q}(t_{\rm
w}) \rangle$.  To $O(\bar{\phi}^2)$, Eq.\ (\ref{EL2}) can be
integrated to give:
\begin{equation}
\bar{\phi}_{\bm k}(\tw) = \frac{n^2(t)}{n^2(t_{\rm w})}\, e^{-D q^2 (t-\tw)}
\bar{\phi}_{\bm k}(t) .
\end{equation}
Making use of the identity $\lim_{\epsilon \to 0} V^{-1}\langle
\phi_{\bm q}(t+\epsilon) \bar{\phi}_{-\bm q}(t) \rangle = 1$, we
obtain:
\begin{equation}
G_q(t,t_{\rm w}) = \frac{n^2(t)}{n^2(t_{\rm w})} e^{-D q^2 (t-t_{\rm
w})} .
\label{Gtl}
\end{equation}

Equations (\ref{EL1}) and (\ref{Gtl}) now allow to compute two-time
density correlations, 
\begin{eqnarray}
C_q(t,\tw) &\equiv& \int_{\bm r} \, e^{i\bm{q}\cdot\bm{r}} [ \langle
\hat{n}(\bm{r},t) \hat{n}(\bm{0},\tw) \rangle - n(t) n(\tw) ]
\\ 
&=& V^{-1}\langle \phi_{\bm q}(t) \phi_{-\bm q}(\tw)  + \phi_{\bm
  q}(t) \bar{\phi}_{-\bm q}(\tw) V^{-1}\phi_{0}(\tw) \rangle
\\
&=& V^{-1}\langle \phi_{\bm q}(t) \phi_{-\bm q}(\tw) \rangle +
G_q(t,\tw) n(\tw) .
\label{Cphi}
\end{eqnarray}
We now use (\ref{EL1}) to obtain the correlations at tree-level from
$\langle \phi_{\bm{q}}(t) \delta S / \delta \bar{\phi}_{\bm q}(\tw)
\rangle = 0$:
\begin{eqnarray} 
\partial_{\tw} C_q(t,\tw) + \lambda n(\tw) (z+2) C_q(t,\tw) 
- \lambda
  n^2(\tw) G_q(t,\tw) (2z+1) &=& 0 ,
\label{Ctleq}
\end{eqnarray}
where 
\begin{equation}
z \equiv \frac{D q^2}{\lambda n(\tw)} \approx D q^2 \tw . 
\label{z}
\end{equation}
The general form of $C_q(t,\tw)$ that results is 
\begin{equation}
C_q(t,\tw) = f(z) n(\tw) G_q(t,\tw) ,
\label{Ctl}
\end{equation}
where $f(z)$ obeys $z f'(z) + (2z+3) f(z) - (2z+1)=0$.   The solution
to this equation is:
\begin{equation}
f(z) = \frac{1}{2 z^3} \left( e^{-2z} - 1 \right) + \frac{1}{z^2} -
\frac{1}{z} + 1 \approx \left\{ \begin{array}{cc} \frac{1}{3} +
  \frac{z}{3} & (z \ll 1) \\ 1 - \frac{1}{z} & (z \gg 1) \end{array}
\right. ,
\end{equation}
with $z$ given by Eq.\ (\ref{z}).

\subsubsection{Tree-level response and FDR}

Consider now a perturbation $h_q$ of wavevector $q$, cf. Eq.\
(\ref{equ:Rqdef}).  The corresponding change in the rate constants is
$D_q \to D \delta_{q0} + D \beta h_q$ and $\lambda_q \to
\lambda \delta_{q0} + \lambda \beta h_q$.  The change in the
average density, to linear order in $h_q$, is given by the
linear order correction $\langle \phi_{\bm{q}}(t) \rangle^{(1)}$ to
the expectation value of $\phi$.  This can be calculated at tree level
by expanding $\langle \delta S / \delta \bar{\phi} \rangle = 0$ to
linear order in $\delta h_q$:
\begin{eqnarray}
\fl 0 =
\partial_t \langle \phi_{\bm{q}}(t) \rangle^{(1)} + \left[ D q^2 + 2
  \lambda n(t) \right] \langle \phi_{\bm{q}}(t) \rangle^{(1)} 
%
- \beta
  h_q(t) \lambda n^2(t) \left[ \frac{D q^2}{\lambda n(t)} - 1
  \right]  .
\end{eqnarray}
Here we have used the fact that when the diffusion constant is not
uniform, as is the case with a perturbation like the one here, the
diffusion term in the action reads $- \int_{\rbf, t} \bar{\phi} \nabla
\cdot \left( D \nabla \phi - \phi \nabla D \right)$ \cite{Mayer2006}.
The response, $R_q(t,\tw)$, is given by $R_q(t,\tw) = T \delta
\langle \phi_{\bm{q}}(t) \rangle^{(1)} / \delta h_{\bm q}(\tw)$, and
obeys the differential equation:
\begin{eqnarray}
\fl \partial_t R_q(t,\tw) + \left[ D q^2 + 2 \lambda n(t) \right] R_q(t,\tw)
{} - \delta(t-\tw) \lambda n^2(t) \left[ \frac{D q^2}{\lambda
  n(t)} - 1 \right] = 0 ,
\end{eqnarray}
which implies the initial condition $R_q(\tw,\tw) = \left( z - 1
\right) \lambda n^2(\tw)$.  By integrating the above equation we get:
\begin{equation}
R_q(t,\tw) = \left( z - 1 \right)
\lambda n^2(t) e^{-D q^2 (t-t_{\rm w})} .
\label{Rtl}
\end{equation} 

\begin{figure}
  \hspace*{1in} \includegraphics[width=4.5in,clip]{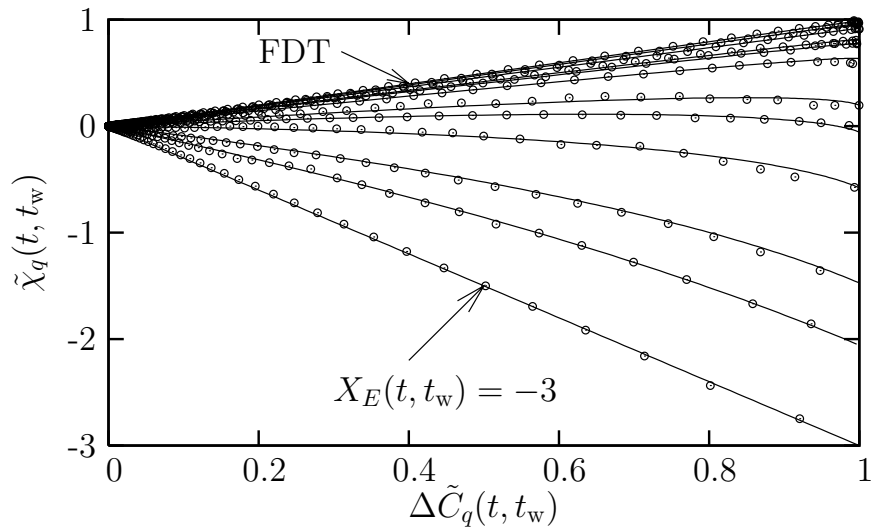} 
  \caption{\label{fig:3dFA} Normalized FD plots for Fourier mode 
    observables $n_q$ in 
    the $d=3$ FA model. Symbols are simulation data at $T = 0.1$
    in a system of linear size $L=32$, for final
    time $t = 
    2\times 10^5$ and wavevectors 
    $q=\pi/x$ with $x=1$, 2, 2.4,
    3, 3.2, 4, 5.33, 6, 8, 12, 16 and $\infty$  (top to bottom).
    Full lines show the field theoretical 
    predictions, accumulating on the equilibrium FDT line at large $q$.
    At the other extreme, the lowest curve
    coincides with the FD plot for the energy at $q=0$, which is
    a straight line with slope $-3$. From
    Ref.~\cite{Mayer2006}. Copyright American Physical Society.}
\end{figure}

From equations (\ref{Ctl}) and (\ref{Rtl}) we obtain the FDR for large
$t$ and $\tw$ and for all momenta,
\begin{equation}
X_q(t,\tw) = \frac{z-1}{\left( 1 + \partial_z \right) z f(z)} = 
\left\{ \begin{array}{cc} -3 + 12 z & (z \ll 1) \\ 1 - 1/z &
  (z \gg 1) \\ \end{array} \right. ,
\label{Xtl}
\end{equation}
where $z$ is given by Eq.\ (\ref{z}) as before. At any waiting time
$\tw$, for wavevectors $q$ larger than $1/\sqrt{D\tw}$, FDT is
recovered: $X_q \approx 1$.  On the other hand, at small enough
wavevectors, $q \ll 1/\sqrt{D\tw}$, the FDR becomes negative. In the
limit $q \to 0$, we get the FDR for energy fluctuations,
\begin{equation}
  X_E(t,\tw) \equiv X_{q=0}(t,\tw) = -3.
  \label{Xhighd}
\end{equation}
From~(\ref{Xtl}) this is also the asymptotic FDR
$X_q^\infty=\lim_{\tw\to\infty} \lim_{t/\tw\to\infty} X_q(t,\tw)$ for
{\em any} wavevector.  
As for the one-dimensional case, a detailed comparison (see
Fig.~\ref{fig:3dFA}) between these analytical
results and direct numerical simulations of the FA model in 
dimension $d=3$ shows very good agreement over the complete
range of wavectors, from the straight line of equilibrium FDT for
large $q$ to the energy FD plot (straight line of slope $X_E=-3$) for
small $q$.

\subsubsection{RG analysis and FDR for $d<\dc$}
\label{sec:RGdltdc}

For dimensions below the critical dimension $\dc$ the field theory
needs to be renormalised.  Our RG analysis here follows that of Ref.\
\cite{Lee,Tauber05}.  The theory only requires coupling constant
renormalization.  The Callan-Symanzik equations for two-time
correlation and response functions, in terms of renormalized
quantities, read:
\begin{eqnarray*}
\fl 0 = \left[ 2 (Dt) \partial_{Dt} + 2 (D \tw) \partial_{D \tw} - q
\partial_q 
\right. 
%
\left. 
{}+ \beta(g_{\rm{R}}) \partial_{g_{\rm{R}}} - d n_0
\partial_{n_0} + d \right] C_q(t,\tw;n_0,g_{\rm{R}},\kappa) ,
\\
\fl 0 = \left[ 2 (Dt) \partial_{Dt} + 2 (D \tw) \partial_{D \tw} - q
\partial_q 
\right. 
\left. 
{}+ \beta(g_{\rm{R}}) \partial_{g_{\rm{R}}} - d n_0
\partial_{n_0} + (d+2) \right] R_q(t,\tw;n_0,g_{\rm{R}},\kappa) ,
\end{eqnarray*}
where we have made explicit the dependence of the functions on the
initial density, the renormalized coupling constant, $g_{\rm R}$, and
the arbitrary scale $\kappa$ which relates the dimensional coupling
$\lambda$ to the adimensional one $g = \left( \lambda / D \right)
\kappa^{d-2}$.  The function $\beta(g_{\rm{R}})$ is the exact
beta-function \cite{Lee,Tauber05}.  These equations are solved by the method of
characteristics \cite{Lee,Tauber05}
\begin{eqnarray}
\fl 
C_q(t,\tw;n_0,g_{\rm{R}},\kappa) =  \left( \frac{t}{t_0}
\right)^{-d/2} 
C_{q \sqrt{t/t_0}} \left( t_0, t_0 \left( \frac{\tw}{t} \right) ; n_0 \left( \frac{t}{t_0}
\right)^{d/2}, g_{\rm{R}}(t), \sqrt{\frac{1}{Dt_0}} \right), 
\label{CZCsol}
\\ 
\fl 
R_q(t,\tw;n_0,g_{\rm{R}},\kappa) = \left( \frac{t}{t_0}
\right)^{-d/2-1} 
R_{q \sqrt{t/t_0}} \left( t_0, t_0 \left( \frac{\tw}{t} \right) ; n_0 \left( \frac{t}{t_0}
\right)^{d/2}, g_{\rm{R}}(t), \sqrt{\frac{1}{Dt_0}} \right),
\label{CZRsol}
\end{eqnarray}
where $g_{\rm{R}}(t)$ is the running coupling \cite{Lee,Tauber05}.  In the
asymptotic limit $t \to \infty$ this approaches its
fixed-point value, $g_{\rm{R}}(t) \to g_{\rm{R}}^*$ ($=2\pi$ for $d=1$).

We can now use the tree-level results (\ref{Ctl}, \ref{Rtl}) and Eqs.\
(\ref{CZCsol}, \ref{CZRsol}) to obtain the renormalized tree-level
correlation and response.  To leading order in $g_{\rm{R}}$ and for
long times we make the replacement $\lambda \to D g_{\rm{R}}^*
\kappa^{2-d}$. Using (\ref{CZCsol}, \ref{CZRsol}) and setting $d=1$
we obtain for large $t$ and $\tw$:
\begin{eqnarray}
C_q(t,\tw) &\approx& f \left( D q^2 \tw \right) \frac{1}{g_{\rm{R}}^*
  \sqrt{D t}} \left( \frac{\tw}{t} \right) e^{-D q^2 (t-t_{\rm w})} ,
\label{CRG} \\
R_q(t,\tw) &\approx& \left( D q^2 \tw - 1 \right) \frac{1}{g_{\rm{R}}^*
  \sqrt{D} t^{3/2}}\, e^{-D q^2 (t-t_{\rm w})} . 
\label{RRG}
\end{eqnarray}
The time dependence of these correlations and responses is different
from that for $d>\dc$, see Eqs.\ (\ref{Ctl}-\ref{Rtl}).  The FDR,
however, is the same as that given in (\ref{Xtl}).  The above
expressions fail to reproduce the exact results obtained above for
$d=1$ (see Sec.~\ref{subsubCCR}).  Eqs.\ (\ref{CRG}, \ref{RRG}) are the
renormalized tree-level functions.  In principle one could calculate
loop corrections which will change the dependence on times and
wavevector.  It is likely, however, that the discrepancy with the exact
results will persist.  The problem is similar to what occurs with
one-time functions \cite{Lee,Tauber05}: the RG analysis is based on a
$d-\dc$ expansion which is not quantitatively accurate at $d=1$.

\begin{figure}
  \hspace*{1in} \includegraphics[width=4.5in,clip]{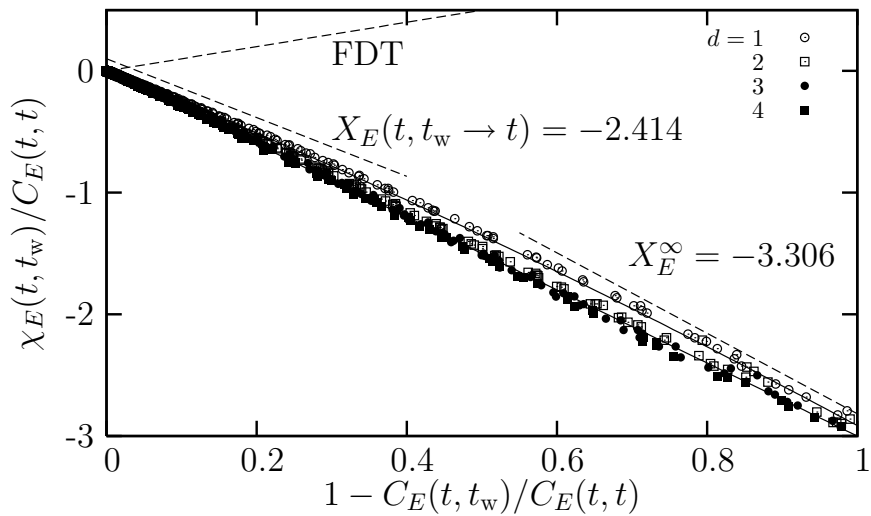} 
  \caption{\label{fig:alldFA} 
Normalized energy FD plots for the FA model in
    dimensions $d = 1$ to $4$.
    Symbols show simulation data, all taken in the aging regime 
    for $T=0.2$ to 0.06 and $t=10^4$ to $10^8$.
    Full curves show
    a  straight line with slope $-3$ 
and the exact limit plot in $d=1$ with 
asymptotic slopes shown as dashed lines. From
Ref.~\cite{Mayer2006}. Copyright American Physical Society.
}
\end{figure}
    
Simulations performed in dimensions $d=1$ to $d=4$
fully confirm our results, see Fig.~\ref{fig:alldFA}. 
While the $d=1$ FD plot for the energy is curved and follows
perfectly the theoretical predictions obtained in Sec.~\ref{fd_energy1d},
the plots in higher dimensions are all compatible with 
the straight line of slope $-3$ calculated in this section.

\subsection{Simulation methodology}
\label{sec:nofield}

Chatelain proposed a very useful no-field method for measuring
response functions in spin systems with Markovian dynamics
\cite{Chatelain03, chatelain, Ricci}.  
The response is re-expressed as a multi-point correlation function,
which has the advantage that it can be calculated from unperturbed
trajectories.  However, the efficiency of this method is poor for
long-ranged perturbations: the statistics can become very noisy,
especially for the collective quantities we want to study here.
It turns out, however, that the 
response of KCMs to energy/temperature perturbations has special properties 
that make it possible to derive a more efficient no-field method. 
Specifically, in any KCM and any dimension~\cite{Mayer2006}, the step
response can be written as
\begin{equation}
\fl \chi_E(t,\tw) = \frac{1}{2} \left[ (1-2c) \, \tau \, \partial_t n(t) 
  + C_E(t,t) - C_E(t,\tw) + C_Y(t,t) - C_Y(t,\tw)\right]. 
  \label{equ:mayer}
\end{equation}
Here $C_Y(t,\tw) = (1/N) ( \langle E(t) Y(\tw) \rangle - \langle E(t) \rangle 
\langle Y(\tw) \rangle)$ is the normalized connected correlation between the 
energy $E$ and the random variable 
\begin{equation}
  Y(t) = \int_0^t \rmd t' \, U(t') 
  \quad \mbox{with} \quad 
  U = \sum_{i=1}^N f_i(\boldsymbol{n}) (n_i - c). 
  \label{equ:Yt}
\end{equation}
In Monte Carlo simulations the integrand in Eq.~(\ref{equ:Yt}) is constant 
between successive spin-flips and hence the integral becomes a sum 
$Y(t) = \sum_{t_i \leq t} (t_i - t_{i-1}) U(t_i)$, with $t_i$ the sequence 
of updating events. Equation (\ref{equ:Yt}) is easily implemented, especially 
in event driven code that keeps track of mobile sites $i$ with $f_i(\boldsymbol{n}) 
\neq 0$. By recording histories of the energy $E(t)$ and the observable $Y(t)$, step 
response functions $\chi_E(t,\tw)$ follow immediately from Eq.~(\ref{equ:mayer}). 
Importantly, one can obtain values of $\chi_E(t,\tw)$ for arbitrary combinations 
of $t,\tw$ by averaging over a stored collection of histories. This is crucial
for the efficient generation of FD plots with fixed $t$ 
and running $\tw$. Without this method, 
it would have been impossible to obtain the collection of data 
shown in Fig.~\ref{fig:alldFA}.

In the remainder of this section we present a derivation of the no-field method 
Eq.~(\ref{equ:mayer}). It is convenient to use the operator formalism introduced 
earlier. We first consider the disconnected correlation $D_E(t,\tw) = 
\langle E(t) E(\tw) \rangle$ which can be expressed in the form 
$D_E(t,\tw) = \bra{\bm{e}} \hat{E} \rme^{W \tau} \hat{E} \rme^{W \tw} \ket{p_0}$, 
with $\ket{\bm{e}} = \sum_{\boldsymbol{n}} \ket{\boldsymbol{n}}$ as
before and $\ket{p_0}$ 
the initial state. Differentiating this with respect to $\tw$ at fixed $t$ 
(note that $\tau = t-\tw$) gives 
\begin{equation}
  \partial_{\tw} D_E(t,\tw) = \bra{\bm{e}} \hat{E} \rme^{W \tau} [\hat{E}, W] 
  \rme^{W \tw} \ket{p_0}, 
\end{equation}
where $[A,B] = A B - B A$ denotes the commutator. For our derivation we will 
also need the response $G_E(t,\tw) = T [\delta \langle E(t) \rangle / \delta h(\tw)]$ 
of the energy to the perturbation $\delta E = - h \sum_i n_i$.
In the operator formalism this is
\begin{equation}
  G_E(t,\tw) = \bra{\bm{e}} \hat{E} \rme^{W \tau} V \rme^{W \tw} \ket{p_0}. 
\end{equation}
The perturbation operator is $V = T \partial_h W(h)|_{h=0}$ where $W(h)$ denotes the 
master operator in the presence of the field $h$. The basic idea for deriving 
a no-field method for the measurement of $G_E(t,\tw)$ is simple: we try to form a 
linear combination $a \, G_E(t,\tw) + b \, \partial_{\tw} D_E(t,\tw) = \bra{\bm{e}} \hat{E} 
\rme^{W \tau} X \rme^{W \tw} \ket{p_0}$ with $X=a V + b \, [\hat{E}, W]$ so that 
off-diagonal components in $X$ match the master operator $W$ (there exist no coefficients 
$a,b$ such that $X$ is purely diagonal). Then this 
quantity can be expressed as a time derivative. It turns out that the relevant 
linear combination is $a = 2$ and $b = -1$. To see this let us work out the explicit 
form of $V$ and $[\hat{E}, W]$. From the master operator Eq.~(\ref{equ:FAmaster}) 
we see that calculation of $V$ only requires $T \partial_h c(h)|_{h=0} = c (1-c)$; 
the Glauber rate for activation in the presence of the field is $c(h) = 
1/[1+\rme^{\beta(1-h)}]$. So, 
\begin{equation}
  V = c (1-c) \sum_{i=1}^N (F_i-1) \hat{f}_i (1-2\hat{n}_i). 
\label{V_FA}
\end{equation}
The calculation of the commutator $[\hat{E},W]$, on the other hand, essentially reduces 
to $[\hat{n}_i, F_j]$ and this is $[\hat{n}_i, F_j] = \delta_{i,j} F_i (1 - 2 \hat{n}_i)$, 
whence 
\begin{equation}
  [\hat{E},W] = \sum_{i=1}^N F_i \hat{f}_i (c - \hat{n}_i). 
  \label{equ:EWcomm} 
\end{equation}
Based on the last two equations one verifies that $X = 2V - [\hat{E},W] = (1-2c) W 
+ \hat{U}$, with $\hat{U} = \sum_i \hat{f}_i (\hat{n}_i - c)$ the (diagonal) operator 
corresponding to the observable $U$, Eq.~(\ref{equ:Yt}). This is now a no-field method 
because 
\begin{eqnarray}
\fl 2 G_E(t,\tw) - \partial_{\tw} D_E(t,\tw) = \bra{\bm{e}} \hat{E} \rme^{W \tau} X \rme^{W \tw} \ket{p_0}
  \nonumber \\ 
  = (1-2c) \bra{\bm{e}} \hat{E} \rme^{W \tau} W \rme^{W \tw} \ket{p_0} 
  + \bra{\bm{e}} \hat{E} \rme^{W \tau} \hat{U} \rme^{W \tw} \ket{p_0} 
  \nonumber \\ 
  =(1-2c) \partial_t \langle E(t) \rangle + \langle E(t) U(\tw) \rangle. 
  \label{equ:nofieldGD}
\end{eqnarray}
Note that this equation applies to {\em any} KCM since we have not
made any assumptions 
on the particular form of the kinetic constraint $f_i(\boldsymbol{n})$. We have 
only used that the unconstrained flip rates $w_i(\boldsymbol{n})$ are Glauber rates 
for $E = \sum_i n_i$ (a similar result can be derived for Metropolis rates). It remains 
to switch from disconnected to connected correlations in Eq.~(\ref{equ:nofieldGD}) 
and to multiply by $1/N$ so that all quantities are intensive. This gives 
\begin{equation}
\fl 2 R_E(t,\tw) = (1-2c) \partial_t n(t) + \partial_{\tw} C_E(t,\tw) + 
  \frac{1}{N} ( \langle E(t) U(\tw) \rangle + \langle E(t) \rangle \partial_{\tw} \langle E(\tw) \rangle). 
  \label{equ:nofieldRC} 
\end{equation}
Obviously $R_E(t,\tw) = (1/N) G_E(t,\tw)$, $C_E(t,\tw) = (1/N) ( D_E(t,\tw) - \langle E(t) 
\rangle \langle E(\tw) \rangle)$ and $n(t) = (1/N) \langle E(t) \rangle$. But the last 
term in Eq.~(\ref{equ:nofieldRC}) is in fact the connected correlation between $E(t)$ 
and $U(\tw)$ since 
\begin{equation} 
  \partial_{\tw} \langle E(\tw) \rangle = \bra{\bm{e}} \hat{E} W \rme^{W \tw} \ket{p_0} 
  = - \bra{\bm{e}} \hat{U} \rme^{W \tw} \ket{p_0} = -\langle U(\tw) \rangle. 
  \label{equ:dEU}
\end{equation} 
The identity $\bra{\bm{e}} \hat{E} W = - \bra{\bm{e}} \hat{U}$ is most easily seen from 
Eq.~(\ref{equ:EWcomm}) by multiplying both sides with $\bra{\bm{e}}$ and using conservation 
of probability $\bra{\bm{e}} W = 0$ as well as completeness of the projection state 
$\bra{\bm{e}} F_i = \bra{\bm{e}}$. Equation (\ref{equ:dEU}) makes intuitive sense: according 
to its definition, Eq.~(\ref{equ:Yt}), $U$ is a measure for how much the concentration 
over {\em mobile} spins (with $f_i(\boldsymbol{n})=1$) differs from the equilibrium 
concentration $c$. Thus, if $\langle U(\tw) \rangle > 0$ there is an excess of mobile 
$n_i = 1$ spins and $\langle E(\tw) \rangle$ will decrease, and vice versa. Our 
no-field method Eq.~(\ref{equ:mayer}), finally, follows by integrating 
Eq.~(\ref{equ:nofieldRC}) to obtain the step response $\chi_E(t,\tw) = \int_{\tw}^t 
\rmd t' R_E(t,t')$. 

For completeness we add that this no-field method of course reduces to the standard 
FDT in equilibrium. This is most easily seen from Eq.~(\ref{equ:nofieldGD}). In 
equilibrium the term $\partial_t \langle E(t) \rangle = 0$ drops out since the energy 
is stationary. Further, $\partial_{\tw} D_E(t,\tw) = - \bra{\bm{e}} \hat{E} W \rme^{W (t-\tw)} 
\hat{E} \ket{p_\mathrm{eq}} = \bra{\bm{e}} \hat{U} \rme^{W (t-\tw)} \hat{E} \ket{p_\mathrm{eq}} 
= \langle U(t) E(\tw) \rangle = \langle E(t) U(\tw) \rangle$, where the last equality 
expresses time-reversal symmetry. Thus Eq.~(\ref{equ:nofieldGD}) retrieves the 
FDT $G_E(t,\tw) = \partial_{\tw} D_E(t,\tw)$ in equilibrium.

\section{East model}
\label{sec:East}

In the second part of this paper we move to the East model, a directed
variant of the FA model. It is defined on a one-dimensional lattice,
i.e.\ a chain of $N$ spins, and differs from the FA model only in the
choice of facilitation factor:
\be
f_i(\bm{n}) = n_{i-1}.
\ee
Compared to $f_i=n_{i-1}+n_{i+1}$ in the FA case, this allows
facilitation only from the left; the spin to the ``East'' of each
defect is mobile. This seemingly harmless change has profound
consequences. In particular, there are no finite groups of defects
that can diffuse unimpeded over a defect-free (all down-spin)
background. This is clear because the leftmost up-spin of any such
group will never be able to change its state, due to the absence of a
facilitating neighbour on the left. The East model therefore has much
more cooperative dynamics, and its relaxation times correspondingly
diverge more quickly as $c$ decreases towards zero. In fact, when
expressed in terms of temperature, relaxation times to equilibrium
after a quench diverge as an exponential of inverse temperature {\em
  squared}, $\ln t_{\rm rel} \propto 1/T^2 \propto
\ln^2(1/c)$~\cite{Tonetal06,Tonetal06b,SolEva1,SolEva2,AldDia02}, compared
to the Arrhenius dependence $\ln t_{\rm rel} \propto \ln (1/c) \approx
1/T$ for the FA case.

We consider again the dynamics after a quench from equilibrium at
infinite temperature, where every spin is up independently with
probability 1/2, to some low final temperature $T$. On timescales of
$O(1)$ the system will reach a state where all up-spins are isolated.
Further reduction in the defect density proceeds via thermally
activated fluctuations, but now in a much more non-local manner than
for the FA model: each up-spin can flip down only when, starting from
the nearest up-spin on the left, a fluctuating ``front'' of up-spins
has extended sufficiently far to the right to mobilize it. To be
specific, consider an up-spin at site $d$, and let the nearest up-spin
on the left be at site 0. To relax this ``domain'' of length $d$, a
front of defects needs to propagate from $n_0$ so that it creates a
defect at site $d-1$, thus mobilizing $n_d$. Naively one might suspect
that the creation of this front takes an energy $d-1$, with spins
$n_1$, $n_2$, \ldots, $n_{d-1}$ simply flipped up in sequence.
However, further reflection~\cite{east,SolEva1,SolEva2}
shows that the most efficient way of creating a front is hierarchical
on lengthscales increasing as powers of two: after flipping up $n_1$
and $n_2$, $n_1$ can be flipped down; then $n_3$ and $n_4$ are flipped
up, $n_3$ is flipped down, and finally $n_2$ is eliminated by
reversing the initial part of the process. Pictorially, one has
\begin{eqnarray*}
10000 \to 11000 \to 11100 \to 10100 \to 10110 \to 10111 \to 10101 \to
\\ \to 11101 \to 11001 \to 10001 \to \ldots
\end{eqnarray*}
Effectively one creates a ``stepping stone'' at site 2, steps from
there to site 4, and removes the stepping stone at site 2; this
process can then be continued by creating a stepping stone at site 8
from the one at site 4 and so on. Keeping track of the maximum number
of extra defects that exist at any one time, one shows that the energy
barrier for flipping down the up-spin at site $d$ (via creation of an
up-spin at site $d-1$) is
\be
k \mbox{\ \ for\ \ } 2^{k-1}<d\leq 2^{k},
\label{East_barriers}
\ee
where $k=0,1,2,\ldots$~\cite{SolEva1,SolEva2}. Such relaxation
processes therefore take place on timescales $t\approx \exp(k/T)\approx
c^{-k}$.  Importantly, in the limit of small $c$ the ratio of any two
of these timescales diverges, and the dynamics separates naturally
into ``stages'' labelled by the value of $k$. We can include in this
picture the initial relaxation after the quench, where domains of
length $d=1$ disappear without activation; this is stage $k=0$.

\begin{figure}
\hspace*{1in} \includegraphics[width=4.5in]{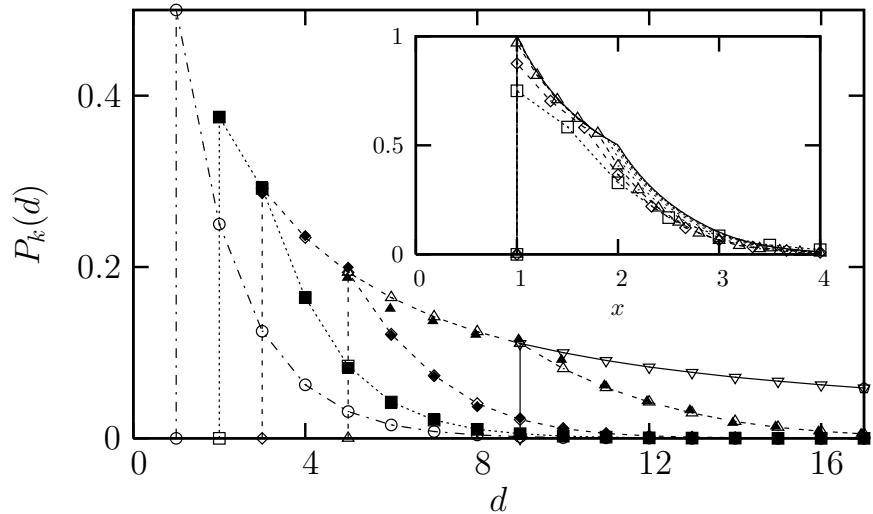}
\caption{Domain size distribution in the East model in the first few
  plateaus of the dynamics. Open symbols are the theoretical prediction for the
  irreversible coarsening regime, filled symbols are from simulations.
Circles: $k=-1$ (initial condition), squares: $k=0$, diamonds: $k=1$,
upward triangles: $k=2$, downward triangles: $k=3$. The inset
shows $\dmin P_k(d)$ versus $x=d/\dmin$ to show the approach to the
predicted scaling form $\tilde{P}(x)$ (solid line) for large $k$.
\label{fig:Pd_East}
}
\end{figure}
To understand the non-equilibrium dynamics in more detail, it is
useful to view it as a coarsening process: when the up-spin at the
right boundary of a domain disappears, this domain merges with the
neighbouring domain on the right. Because the rate for disappearance
of an up-spin depends only on the length $d$ of the domain it bounds,
it is clear that no correlations between neighbouring domain lengths
can ever develop during such a process~\cite{SolEva1,SolEva2}. For
low $c$, the evolution of the domain size distribution, $P(d,t)$, can
therefore be calculated without approximation by an ``independent
interval'' approach. In stage $k$ of the dynamics one finds the
equation of motion (for $d>2^k$)
\be
\frac{\partial}{\partial t}P(d,t) = \sum_{2^{k-1}<d'\leq 2^k}
P(d-d',t)\left[-\frac{\partial}{\partial t}P(d',t)\right].
\label{Pd_dynamics}
\ee
The sum runs over the ``active'' domain sizes $d'$ that are eliminated
during stage $k$. The rate for creating longer ``inactive'' domains is
then the product of the rate $[-(\partial/\partial t)P(d',t)]$ at
which domains of size $d'$ disappear, and the probability of having a
domain of length $d-d'$ on the right. As shown in~\ref{sec:nt}, 
this equation can
be solved to relate the domain size distributions at the beginning and
the end of stage $k$, and this relation can then be iterated
numerically, starting from the initial
$P(d,t=0)=2^{-d}$~\cite{SolEva1}. The results are shown in
Fig.~\ref{fig:Pd_East} for the domain size distributions $P_k(d)$ in
the first few ``plateaus'' of the dynamics. Here we label each plateau
by the stage which it {\em follows}, i.e.\ plateau $0$ refers to times $1\ll
t\ll c^{-1}$. More generally we will use the notation
\be
\nu=T\ln t, \quad k=\lfloor \nu \rfloor, \quad a=\nu-k,
\label{nu_k_a_def}
\ee
so $k$ denotes the plateau that has been reached (and the stage of the
dynamics that has finished) at time $t$. To avoid confusion we note
that plateaus were labelled by $k+1$ in previous
work~\cite{SolEva1,SolEva2}; in our convention, the initial
condition corresponds to plateau $k=-1$.

\begin{figure}
\hspace*{1in} \includegraphics[width=4.5in]{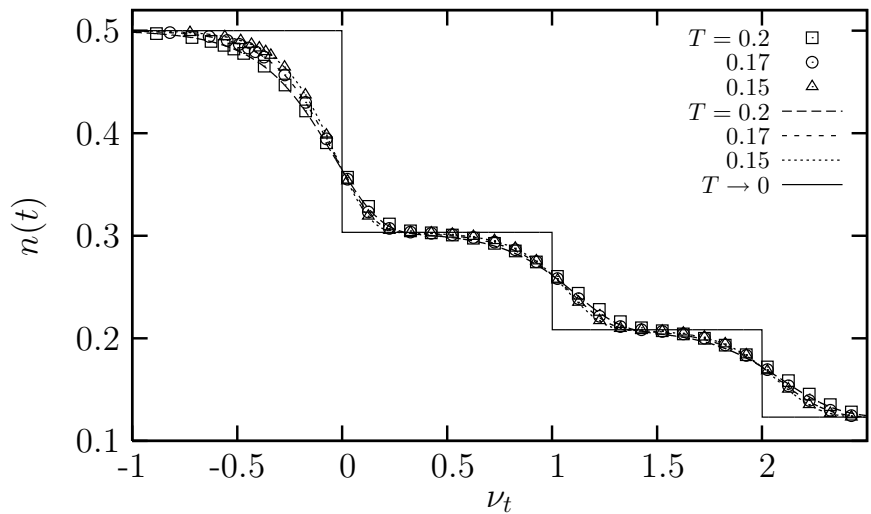}
\caption{Defect density $n(t)$ versus $\nu_t$ for three different
  temperatures. Symbols represent the simulation results. 
The theoretical prediction (\ref{n_t_plateaus}) is shown at finite
$T$ with dotted lines, while its $T \to 0$ limit is shown 
as a full line. 
\label{fig:nt}
}
\end{figure}

As Fig.~\ref{fig:Pd_East} shows, the predicted domain size
distributions agree well with simulation data at low $T$. From the
domain size distribution one can predict in particular the defect
density, giving $n(t)=1/\bar{d}_k$ where $\bar{d}_k=\sum_d d P_k(d)$
is the average domain length in plateau $k$. Again this agrees well
with simulation data (Fig.~\ref{fig:nt}), though the sharp steps that
are predicted for $T\to 0$ in a plot of $n$ vs $\nu$ at the integer values
$\nu=0,1,2,\ldots$ are obviously rounded at nonzero $T$. By analysing
the dynamics in more detail (see~\ref{sec:nt} and
Eq.~(\ref{n_t_plateaus}) below) also this effect can be predicted, and
then gives a very good description of the data.

A further observation from Fig.~\ref{fig:Pd_East} is that the domain
size distributions approach a scaling form for large $k$. This is
obtained by scaling with the minimum domain size present at time $t$,
\be
\dmin(t) = 2^k+1,
\ee
so that $\dmin(t) P_k(d)$ is plotted versus $x=d/\dmin(t)$. The
limiting distribution $\tilde P(x)$ can in fact be worked out
explicitly~\cite{SolEva1,SolEva2}: its Laplace transform is
$1-\exp[-\Ei(s)]$ where $\Ei(s)=\int_1^\infty dz
\exp(-zs)/z$ is the exponential integral. Expanding in powers of
$\Ei(s)$ and inverting the transform gives then
\be
\tilde{P}(x) = \sum_{l=1}^\infty \frac{(-1)^l}{l!} f_l(x),
\label{East_Px}
\ee
where the functions $f_l(x)$ can be defined recursively by successive
convolutions,
\be
f_0(x)=\delta(x), \quad
f_1(x)=\Theta(x-1)\frac{1}{x}, \quad
f_{l+1}(x)=(f_l\ast f_1)(x).
\label{fl_def}
\ee
Here $\Theta(x)$ and $\delta(x)$ are the standard Heaviside step and
Dirac delta functions. Explicitly one has
$f_2(x)=\Theta(x-2)(2/x)\ln(x-1)$, while for $l\geq 3$ the $f_l(x)$
cannot be expressed in any simple closed form. Fortunately each $f_k(x)$
contains a factor $\Theta(x-k)$ so for realistic ranges of $x$ only
the first few functions are needed. The mean value of $x$ across the
distribution~(\ref{East_Px}) is $\exp(\gamma)=1.78\ldots$, with
$\gamma$ being Euler's constant. Converting back to the unscaled
defect density gives then
\be
n(t)=e^{-\gamma}2^{-k},
\label{n_t_large_k}
\ee
for large $k$.  Because of the logarithmic dependence of $k$ on $t$,
this scales as $n(t)\sim t^{-T\ln 2}$: the East model exhibits
anomalously slow coarsening, with the decay exponent $T\ln 2$ that
governs the decay of the defect density decreasing towards zero at low
temperature.

We will refer to the above analysis of the East model dynamics after a
quench to low $c$ (or equivalently low $T$) as ``irreversible
coarsening''. It becomes exact in the limit $c\to 0$ taken at fixed
stage $k$, i.e.\ with typical domain sizes $d$ kept fixed and of order
unity. This way of taking the limit ensures that up-spins do indeed
flip down irreversibly: the probability of observing, at some given
point in time and within a domain of length $d$,
an up-spin as part of a fluctuating
up-spin front is $O(dc)$ and can be neglected in the limit.  Also,
even though within a given stage $k$ there can and will in general be
multiple relaxation timescales because active domains cover a range of
lengths and can be eliminated by passing through different sequences
of intermediate configurations, the timescales for different stages do
always separate by much more than this spread for low $c$.

Both of the above simplifying properties no longer apply if we wish to
look at typical domain sizes of the order of the equilibrium domain
length $\bar{d}_{\rm eq}=1/c$. We would then take $c$ small at fixed
$\delta=dc$, rather than at fixed $d$. For $\delta$ of order unity and
larger, this gives information on the eventual crossover to
equilibrium behaviour. For the purposes of this paper we will only be
concerned with the earliest stages of this crossover, $\delta \ll 1$.
This should connect appropriately with the large-$d$ limit of the
irreversible coarsening regime, where from~(\ref{East_barriers}) the
timescale for relaxation of a domain of length $d$ is $t\sim c^{-k}
\sim c^{-\ln d/\ln 2} \approx d^{1/(T\ln2)}$.  A simple scaling
hypothesis, supported by numerical evaluation of first passage times
for the elimination of domains of large $d$, then gives the following
picture~\cite{SolEva2}. The splitting into discrete stages of the
dynamics is lost because the spectrum of relaxation times within the
stages becomes so broad that a clear separation no longer exists.
However, there is a now a {\em continuous} form of timescale
separation: the time $t$ for eliminating a domain of scaled size
$\delta$ is again $t\sim (\delta/c)^{1/(T\ln 2)}$.  Comparing two
different domain sizes $\delta_1$ and $\delta_2$ then shows that one
always has $t_2/t_1\gg 1$ in the limit $T\to 0$, as long as
$\delta_2/\delta_1>1$. As a consequence, each up-spin has a sharply
(on the spatial scale $d\sim 1/c$) defined ``equilibration zone'' to
its right, of size $\delta\sim ct^{-T\ln 2}$. As $t$ increases these
zones grow; when an equilibration zone reaches the next up-spin to the
right the latter is eliminated and the two domains either side merge.
Put differently, the dynamics consists of always eliminating the
shortest domain and ``pasting'' all of its length on to the domain on
the right. This ``paste-all'' model has been studied independently in
the literature~\cite{paste_all}. Remarkably, the domain size
distribution it produces in the scaling limit is exactly
identical~\cite{SolEva2} to the one we found in Eq.~(\ref{East_Px}).
This shows that the irreversible coarsening regime of the East model
connects smoothly to the paste-all regime just discussed, as it
should: in the large-$k$ limit of irreversible coarsening the domain
size distribution approaches~(\ref{East_Px}), and this form is
preserved -- apart from a trivial scaling with the minimum domain
length $\dmin$ -- in the paste-all dynamics. The paste-all regime has
two advantages for theoretical analysis: only the smallest domain
length is ``active'' and hence the temporal order in which domains
disappear is deterministic; this is not the case in the irreversible
coarsening regime, where two neighbouring domains of different lengths
can be simultaneously active so that either can disappear first.
Secondly, the role of the clock is played by $\dmin$, so that
effectively diverging timescales,
\be
t\sim \dmin^{1/(T\ln 2)},
\label{t_paste_all}
\ee
can be considered. Even where predictions cannot be extracted
analytically, they are then relatively easy to obtain by numerical
simulation.

After this overview of the theoretical tools used to analyse the
low-$T$ out-of-equilibrium dynamics of the East model we turn as in
the FA case to
specific correlation and response function pairs, first for local
observables, then for the global one (which is just the energy), and
finally for non-local observables. Because we need to deal with
two-time functions in all cases it will be useful to refine our
notation from~(\ref{nu_k_a_def}) to
\be
\nu_t=T\ln t, \quad k_t=\lfloor \nu_t \rfloor, \quad a_t=\nu_t-k_t,
\ee
with analogous quantities defined for $\tw$ (giving $\nu\w$, $k\w$,
$a\w$) and the time difference $\tau=t-\tw$ (giving $\nu_\tau$,
$k_\tau$, $a_\tau$). Simulations are carried out for system sizes
$N=250$ and for temperatures $T=0.2$, 0.17, 0.15, with corresponding
equilibrium up-spin concentrations of $c=6.69\times 10^{-3}$,
$2.78\times 10^{-3}$ and $1.27\times 10^{-3}$. For numerical
convenience we mainly restrict ourselves to waiting times $\tw\geq 1$
when simulating two-time quantities.

\subsection{Local correlation and response}
\label{sec:East_local}

\begin{figure}
  \hspace*{1in} \includegraphics[width=4.5in]{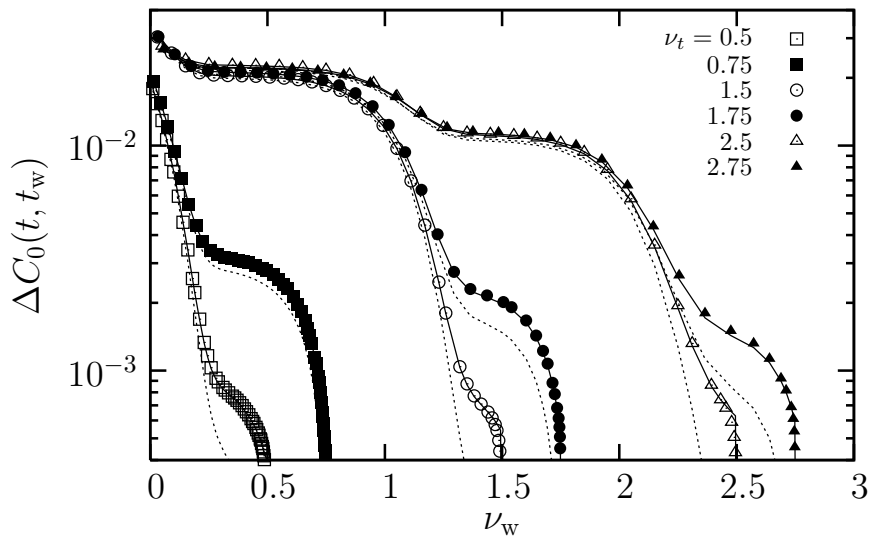}
  \caption{Local correlation function difference from simulations, for
    $\nu_t$ as indicated and $T=0.15$ (symbols). Dotted lines: leading
    order prediction~(\ref{Delta_C0_simple}). Solid lines:
    theory including quasi-equilibrium
    corrections~(\ref{Delta_C0_improved}). Note the logarithmic
    $y$-axis; on a linear axis the correction terms would be difficult
    to discern.
    \label{fig:C0}
}
\end{figure}

We begin with the local correlation function $C_0(t,\tw)=\langle
n_i(\tw)n_i(t)\rangle - n(\tw)n(t)$. Within the irreversible
coarsening regime the time-dependence of this quantity is easy to
determine: any spin that is up at time $t$ must also have been up at
time $\tw$ since to leading order in $c$ spins never flip back up.
This implies $\langle n_i(\tw)n_i(t)\rangle = \langle n_i(t)\rangle =
n(t)$ and so
\be
C_0(t,\tw) = n(t)[1-n(\tw)].
\label{C0_simple}
\ee
Simulation data are compared with this prediction -- which is,
trivially, exact for $\tw=t$ -- in Fig.~\ref{fig:C0}. To make our
later understanding of the FD behaviour easier we plot the data in the
same form as required in an FD plot, showing the correlation function
difference $\Delta C_0(t,\tw) = C_0(t,t)-C_0(t,\tw)$ for which the
prediction~(\ref{C0_simple}) takes the form
\be
\Delta C_0(t,\tw) = n(t)[n(\tw)-n(t)].
\label{Delta_C0_simple}
\ee
This is plotted against $\tw$ or, more precisely, $\nu\w$.  The
theoretical estimate~(\ref{Delta_C0_simple}) rationalizes the
simulation data quite well, but there clearly are small corrections
for nonzero $c$. These are most noticeable when $t$ and $\tw$ are
close together inside the same plateau of the dynamics;
Eq.~(\ref{Delta_C0_simple}) then predicts a very small correlation
function difference because $n(\tw)\approx n(t)$. The actual values
are somewhat larger because there is an additional quasi-equilibrium
contribution in this regime.  This correction can be most easily
determined via FDT from the local susceptibility $\Chi_0$, to which we
turn next.

\begin{figure}
\hspace*{1in} \includegraphics[width=4.5in]{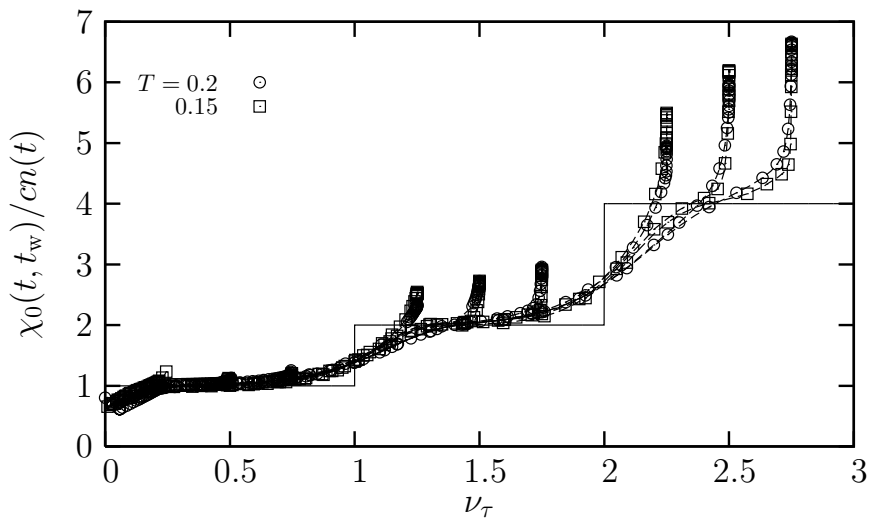}
\caption{Local susceptibility $\Chi_0(t,\tw)$ in the quasi-equilibrium
  regime, scaled by $cn(t)$ and plotted against
  $\nu_\tau=T\ln(t-\tw)$. The steps show the theoretical
  prediction~(\ref{Chi0_quasi_eq}) for $c\to 0$. Squares correspond
  to $T=0.15$, circles to $T=0.2$. For both temperatures we
show data for $\nu_t=0.25$, 0.5, 0.75, 1.25, 1.5, 1.75, 
2.25, 2.5 and 2.75 (from left to right).
\label{fig:Chi0_quasi_eq}
}
\end{figure}

To understand the effect of a field $h_i$ applied at time $\tw$ on the
evolution of spin $n_i$, let us take the history $n_{i-1}(t')$ of its
facilitating left neighbour as given in the time interval
$t'=\tw\ldots t$. Spin $n_i$ then behaves like an isolated spin except
for the fact that it cannot change state whenever $n_{i-1}(t')=0$: the
time interval available for it to relax from the state $n_i(\tw)$ is
reduced from $t-\tw$ to $\int_{\tw}^t dt'\,n_{i-1}(t')$. Since with
Glauber rates~(\ref{Glauber_rates}) the relaxation time of an isolated
spin is unity, we thus have
\be
\langle n_i(t)\rangle = c' + [n_i(\tw)-c']\exp\left(-\int_{\tw}^t
  dt'\,n_{i-1}(t')\right).
\label{ni_t_average}
\ee
Here, $c'$ is the perturbed up-spin density at site $i$.  To linear
order in the field strength $h_i$ it can be expanded as
\be
c' = \frac{1}{1+e^{\beta(1-h_i)}} = c + \beta h_i c(1-c) + \ldots 
\label{c_prime}
\ee
Inserting into~(\ref{ni_t_average}) and differentiating w.r.t.\ $\beta
h_i$, the initial value $n_i(\tw)$ drops out. After averaging over
the history of $n_{i-1}$ one thus gets for the local susceptibility the
general expression
\be
\fl \Chi_0(t,\tw) = c(1-c)\left[ 1 - \langle \exp\left(-\int_{\tw}^t
  dt'\,n_{i-1}(t')\right)\rangle \right] \equiv c(1-c)[1-r(t,\tw)].
\label{Chi0_clock}
\ee
For low $c$ the factor $1-c$ can be dropped, and it remains to
understand the behaviour of the ``relaxation integral''
$r(t,\tw)$. The average over the dynamics in its definition can, for a
large system, be replaced by an average over sites $j=i-1$:
\be
r(t,\tw) = \frac{1}{N}\sum_j \exp\left(-\int_{\tw}^t dt'\,n_j(t')\right).
\ee
Sites where a spin has been up for a total time larger than unity make
a negligible contribution, so $r(t,\tw)$ is to a good approximation
the fraction of sites with spins that have remained persistently down
between $\tw$ and $t$; in other words, $\Chi_0(t,\tw)/c$ is
the fraction of sites that have been up at some point
between $\tw$ and $t$. From this we can predict its dependence on
$t-\tw$ in the quasi-equilibrium regime, where $\tw$ and $t$
remain in the same plateau of the dynamics so that no domains
disappear and the up-spins defining them can be regarded as fixed. As
long as $\tau=t-\tw\gg 1$ but $\tau \ll c^{-1}$, only these fixed
up-spins contribute to $\Chi_0$; since these also determine the
overall up-spin density, one predicts $\Chi_0/c=n(\tw)$. Once $\tau$
increases past $c^{-1}$, the right neighbour of each fixed up-spin
will have had time to flip up once or more, so that
$\Chi_0/c=2n(\tw)$. This process then proceeds: for $\tau\gg c^{-k}$,
the relation~(\ref{East_barriers}) between energy barriers and
distances implies that the $2^k-1$ spins to the right of each fixed
up-spin will have been up at some point. (The ``$-1$'' arises because
we are considering flipping spins {\em up}; the energy barrier for
flipping a spin at distance $d$ {\em down}, as written
in~(\ref{East_barriers}), is the same as that for flipping a spin at
distance $d-1$ {\em up}.) Adding the contribution from the fixed up-spin
back on gives overall
\be
\Chi_0(t,\tw)=c\,n(\tw)2^{k_\tau}=c\,n(t)2^{k_\tau}.
\label{Chi0_quasi_eq}
\ee
(Either $n(\tw)$ or $n(t)$ can be used in the prefactor since by
assumption we remain in the same plateau.)  This quasi-equilibrium
prediction is checked against simulations in
Fig.~\ref{fig:Chi0_quasi_eq}, with good agreement. Our use of the term
``quasi-equilibrium'' is justified by the fact that exactly the same
step-like structure in $\Chi_0$ would be observed in equilibrium,
except for the replacement of the density of ``fixed'' up-spins,
$n(t)$, by the appropriate equilibrium value $c$. The simple
scaling~(\ref{Chi0_quasi_eq}) of the step heights with powers of two
does not seem to have been noticed in previous analyses of the
equilibrium dynamics~\cite{east,EisJaec93,SolEva2}.

The quasi-equilibrium regime lasts while $t$ and $\tw$ remain in the
same plateau of the dynamics, i.e. $k_t=k\w$. In the limit of
low $c$ we have $k_t=\max(k\w,k_\tau)$, so an equivalent condition
for quasi-equilibrium is $k_\tau\leq k\w$. When $t$ is held fixed as
in Fig.~\ref{fig:Chi0_quasi_eq}, the limit $k_\tau=k\w$ corresponds to
$\tau\approx\tw\approx t/2$ and so, again for low $c$, $k_\tau=k_t$. We thus
expect to see $k_t$ steps in $\Chi_0$ within the quasi-equilibrium
regime, as confirmed by the data in Fig.~\ref{fig:Chi0_quasi_eq}.

\begin{figure}
\hspace*{1in} \includegraphics[width=4.5in]{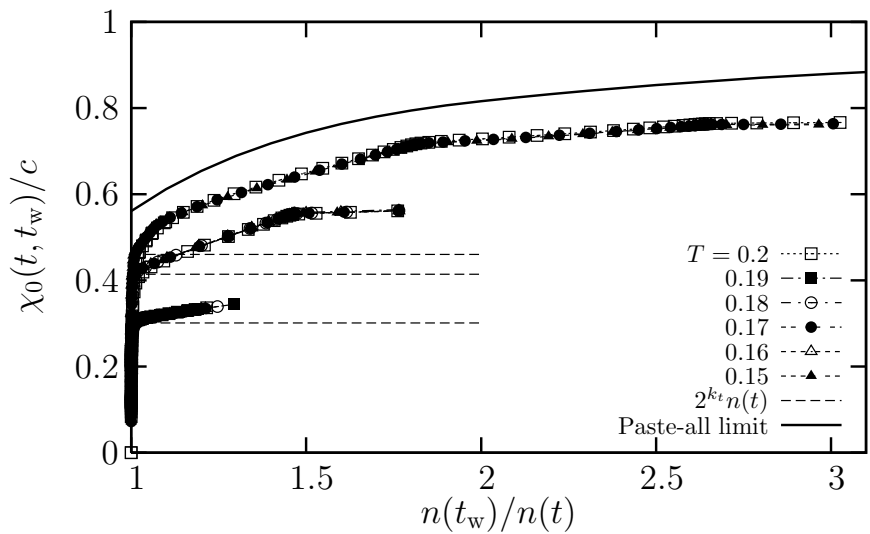}
\caption{Local susceptibility $\Chi_0$ in units of $c$, plotted
  against $\theta=n(\tw)/n(t)$ to show the out-of-equilibrium regime. 
  From bottom to top, we have $\nu_t=0.5$, 1.5 and 2.75.
  Horizontal
  lines indicate the predicted  height of the 
  quasi-equilibrium regime
  for the three cases. Solid
  line: theoretical prediction for the paste-all limit
  ($\nu_t\gg 1$). 
\label{fig:Chi0_paste_all}
}
\end{figure}
Beyond the quasi-equilibrium regime we can continue to identify $\Chi_0/c$
with the fraction of the chain swept by up-spins, but this now becomes
more difficult to calculate as domains and the corresponding ``swept
out'' areas merge. Within the paste-all regime, however, it is a
relatively simple matter to get accurate predictions by simulating the
paste-all dynamics and keeping track of the swept out areas within
each domain. The clock variable for this simulation is
$\dmin(t)/\dmin(\tw)$. Since in the paste-all regime the up-spin
density scales with the inverse of $\dmin$, we can equivalently write
\be
\Chi_0(t,\tw)/c={\cal F}(\theta), \qquad \theta = n(\tw)/n(t).
\label{Chi0_paste_all}
\ee
We explain in~\ref{sec:persistence} how the initial part of the
scaling function can be calculated analytically, with the result
\be
{\cal F}(\theta) =
e^{-\gamma}\left(1+\ln\theta-\frac{1}{2}\ln^2\theta\right),
\quad \mbox{for\ \ }\theta\leq 2.
\label{Chi0_scaling_analytical}
\ee
This is in very good agreement with the numerical values extracted
from the paste-all simulations, as demonstrated in more detail in
Fig.~\ref{fig:Xindep_test} below. We show in
Fig.~\ref{fig:Chi0_paste_all} the overall prediction for $\Chi_0/c$
versus $n(\tw)/n(t)$ and compare with numerical data. Even though the simulations are performed within the
first few plateaus, i.e.\ far from the paste-all regime,
the data for increasing $\nu_t$ clearly do approach the
theoretical prediction. The initial discontinuity of the theory
represents the contribution from the initial quasi-equilibrium regime,
which in the limit of low $c$ shrinks to the point $n(\tw)/n(t)=1$.
The value of $\Chi_0/c$ at this point is $\exp(-\gamma)\approx 0.56$:
each domain contains a swept-out equilibration zone of length
$\dmin(\tw)$, and the density of domains is
$n(\tw)=\exp(-\gamma)/\dmin(\tw)$, giving a swept out fraction
$\dmin(\tw)n(\tw)=\exp(-\gamma)$ of the chain. This is entirely
consistent with extrapolating to the same point from the
quasi-equilibrium regime: as discussed above, the latter ends where
$k_\tau=k\w$. At this point, from~(\ref{Chi0_quasi_eq}),
$\Chi_0/c=n(\tw)2^{k\w}$ which for large $k\w$ approaches
$\exp(-\gamma)$ from~(\ref{n_t_large_k}).

Before combining correlation and susceptibility data to get the FD
plots we return to the quasi-equilibrium corrections to the local
correlation function. To determine these we use the following
relation between local susceptibility and correlation:
\be
\Chi_0(t,\tw) = n(t)[1-n(\tw)]-C_0(t,\tw)+c[n(\tw)-n(t)].
\label{local_relation}
\ee
This is derived in~\ref{sec:local_relation}; remarkably, it is {\em
  exact} and applies not just to the East model but in fact to all spin
models with directed constraints. We used it to obtain the numerical data
for the susceptibility in the simulations, but also checked it against
direct susceptibility measurements. From the theoretical point of view
we note that the first two terms on the r.h.s.\
of~(\ref{local_relation}) cancel to leading order because
of~(\ref{C0_simple}); this of course must be so because
from~(\ref{Chi0_clock}) the susceptibility never becomes larger than
$c$. Because the $O(c)$-corrections to the correlation function are
difficult to estimate directly, Eq.~(\ref{local_relation}) would not
be useful to deduce accurate estimates for $\Chi_0$. We can, however,
turn it around to get an expression for the correlation function
(difference):
\be
\Delta C_0(t,\tw) = n(t)[n(\tw)-n(t)] + \Chi_0(t,\tw)-c[n(\tw)-n(t)].
\ee
The last two contributions on the r.h.s.\ are the exact corrections to the
irreversible coarsening estimate~(\ref{Delta_C0_simple}). They will
be significant only in the quasi-equilibrium regime $n(\tw)\approx
n(t)$, where furthermore the last term is negligible compared to the
second one. The remaining dominant contribution to $\Delta C_0(t,\tw)$
is simply $\Chi_0(t,\tw)$, as expected from FDT for dynamics in the
quasi-equilibrium regime. Using~(\ref{Chi0_quasi_eq}) gives explicitly
\be
\Delta C_0(t,\tw) \approx n(t)[n(\tw)-n(t)] + cn(t) 2^{k_\tau}.
\label{Delta_C0_improved}
\ee
This improved prediction is included in Fig.~\ref{fig:C0} above and
now shows very good agreement with the numerical data. As expected,
the correction term only affects the outcome when $t$ and $\tw$ are
close together and the correlation function is still close to its
equal-time value; elsewhere it is negligible.

\begin{figure}
\hspace*{1in} \includegraphics[width=4.5in]{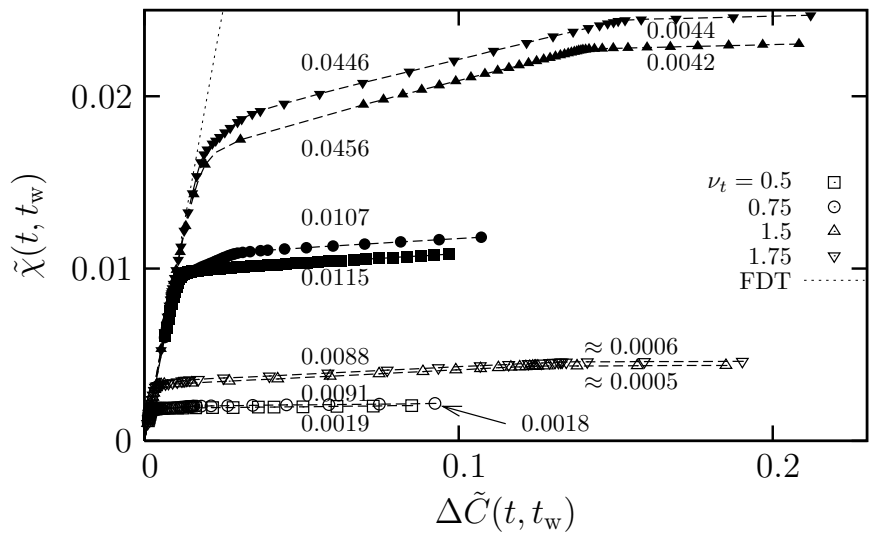}
\caption{Normalized FD plot for the local observable for $T=0.2$
  (full symbols) and $T=0.15$ (open symbols). The numbers 
  indicate the value
  of the fluctuation-dissipation ratio $X_0$ in the various
  straight-line segments. For $T\to 0$ one expects the plot to depend
  only on $k_t=\lfloor \nu_t\rfloor$, while for nonzero $T$ the sharp
  change at integer values of $\nu_t$ will be replaced by a smooth
  crossover. This is visible in the additional short linear segments
  just beyond the departure from quasi-equilibrium (dotted
  line) in the plots for $T=0.2$ and $\nu_t=0.75$, $1.75$.
\label{fig:FDlocal_raw}
}
\end{figure}
We can now combine the above results to obtain the FD plot for
the local observable in the East model. The raw results, with both axes
scaled by $C_0(t,t)$ to produce the normalized two-time correlation
$\tilde C_0(t,\tw)=C_0(t,\tw)/C_0(t,t)$ and susceptibility
$\tilde\Chi_0(t,\tw)=\Chi(t,\tw)/C_0(t,t)$, are shown in
Fig.~\ref{fig:FDlocal_raw}, for fixed $t$ with $\tw$ varying along the
curves. The data show clearly the initial quasi-equilibrium regime.
The curves depart from this where $\tw$ becomes small enough to have
left the plateau that time $t$ is in.  The kinks in this
out-of-equilibrium part of the curves, where points accumulate,
correspond to $\tw$ being well within a plateau; the next line
segment begins when this plateau is being left and $\tw$ moves
(backwards) through the previous stage of the dynamics. The segments
between the kinks appear straight to a good approximation, but close
inspection of the data for larger $\nu_t$ (as included in
Fig.~\ref{fig:Chi0_paste_all}, and Fig.~\ref{fig:FDT_inco} below)
does show a small amount of curvature, as observed also in the
not dissimilar triangular plaquette model~\cite{Robs_plaquette_FDT}.
The slopes of the segments, which give the local FDR
$X_0(t,\tw)$, are indicated in the plot. They are significantly
smaller than unity but do not, in this raw form, follow a clear
pattern.

\begin{figure}
\hspace*{1in} \includegraphics[width=4.5in]{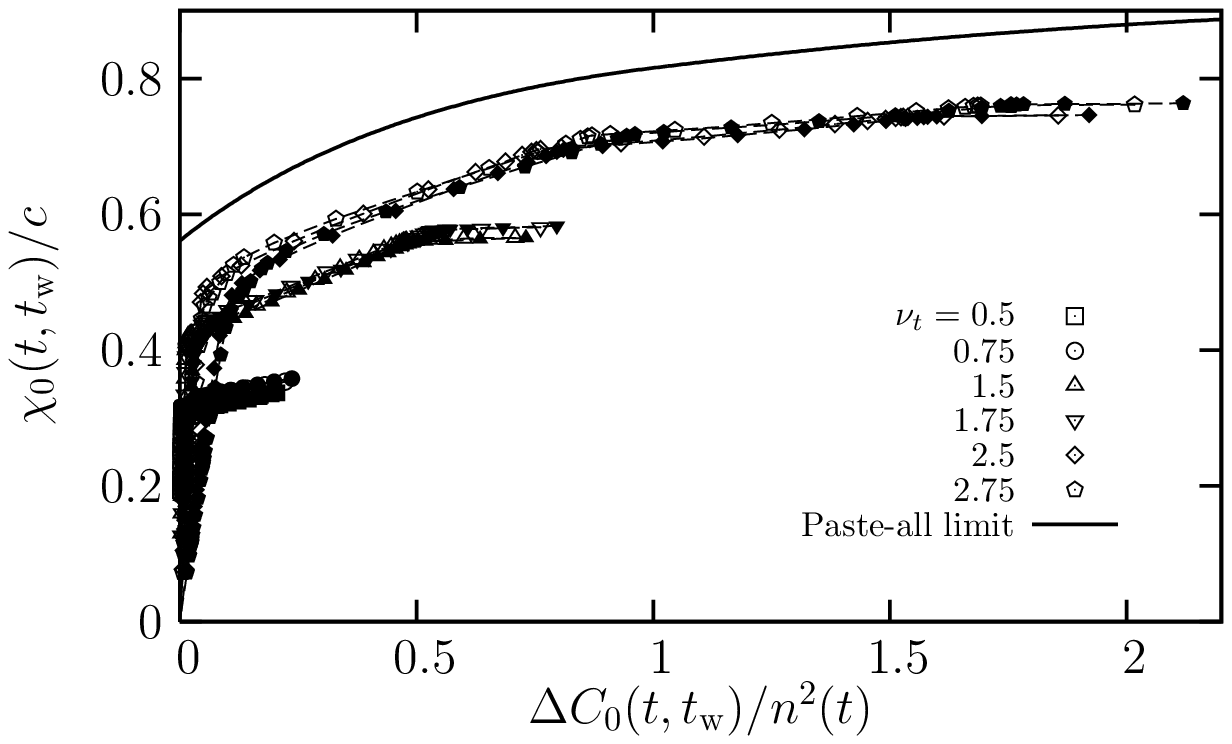} \\
\hspace*{1in} \includegraphics[width=4.5in]{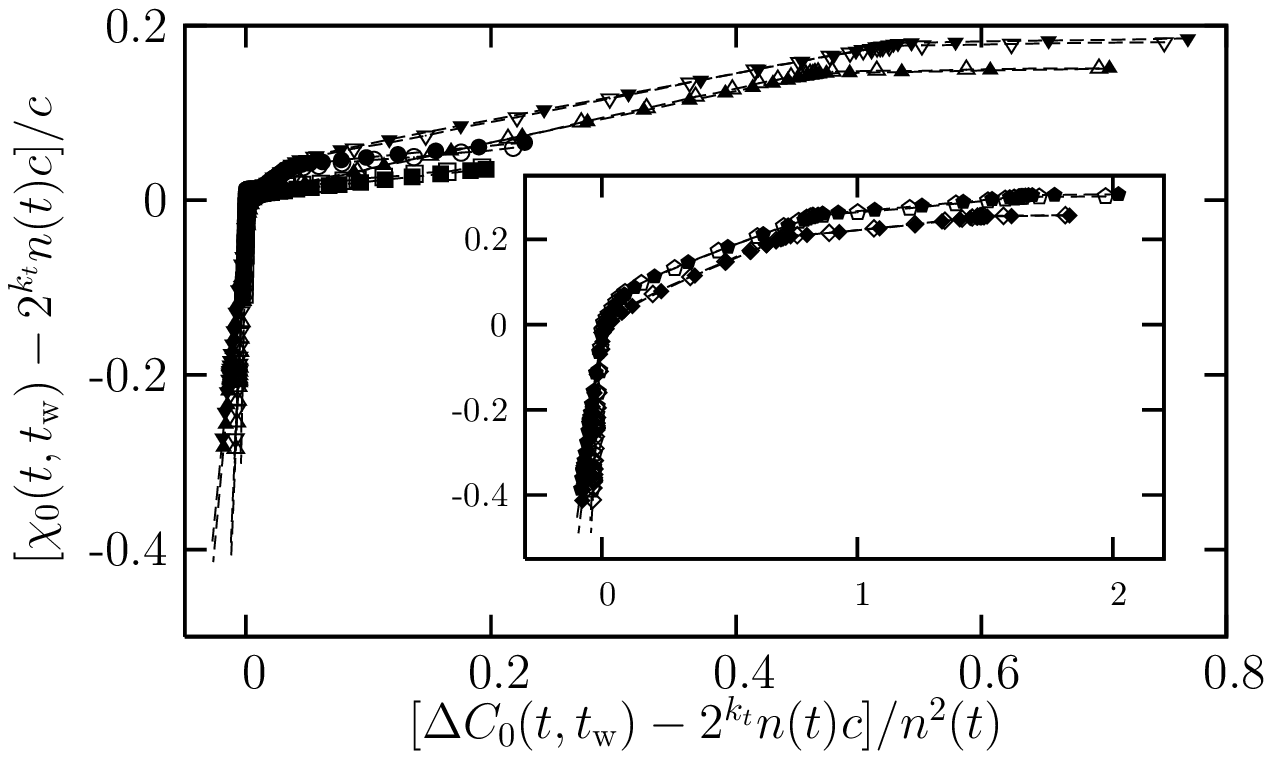} 
\caption{(Top) Scaled local FD plot for $T=0.15$ (open symbols) and $T=0.17$
  (full symbols) for different values of $\nu_t$. (Bottom) Same 
  with quasi-equilibrium correction subtracted; the data for
  $\nu_t=2.5, 2.75$ are shown separately in the inset for better visibility.
\label{fig:FDT_inco}
}
\end{figure}
To determine a more appropriate scaling for the FD plot we recall
from~(\ref{Chi0_paste_all}) that $\Chi_0/c$ is a scaling function of
$n(\tw)/n(t)$, at least in the paste-all limit. For the correlation
function difference we see from~(\ref{Delta_C0_simple}) that $\Delta
C_0(t,\tw)/n^2(t)=n(\tw)/n(t) - 1$ is a function of the same scaling
variable. (Since we are concerned with understanding the scaling in
the non-equilibrium regime, we ignore here the small quasi-equilibrium
corrections.) These considerations motivate one to plot, as we do in
Fig.~\ref{fig:FDT_inco} (top), $\Chi_0(t,\tw)/c$ vs $\Delta
C_0(t,\tw)/n^2(t)$. The curves for different temperatures now collapse
reasonably well, except for small shifts caused by the presence of the
initial quasi-equilibrium regime. If these are removed by subtracting
from both $\Chi_0$ and $\Delta C_0$ the relevant contribution
$cn(t)2^{k_t}$, as done in Fig.~\ref{fig:FDT_inco} (bottom), the
collapse is really quite good.

Looking more quantitatively at the local FDR and comparing with the FA
case, the most striking feature is that for the East model $X_0$ is
never negative, even though the dynamics is activated. This is not
just because any negative values are hard to see on the scale of the
FD plot, as was the case for the FA model at late times. In fact,
from~(\ref{Chi0_clock}) one has
\be
\fl R_0(t,\tw) = -\frac{\partial}{\partial\tw} \Chi_0(t,\tw) 
= c(1-c) \left\langle n_{i-1}(\tw)\exp\left(-\int_{\tw}^t
  dt'\,n_{i-1}(t')\right) \right\rangle,
\ee
and because this local impulse response is always positive, the same
is true of the local FDR. The
magnitude of $X_0$, on the other hand, is very small: the scaling of
Fig.~\ref{fig:FDT_inco} shows that $X_0$ is roughly of the order of
$c/n^2(t)$ for small $c$. To be precise, the expected scaling outside the
trivial quasi-equilibrium regime is
\be
X_0(t,\tw)=\frac{c}{n^2(t)}S_{k_t,\nu\w}.
\ee
This contains the overall prefactor from the scaling of the FD plot.
The remaining coefficient $S_{k_t,\nu\w}$ depends only on the plateau
$k_t$ in which $t$ is located, and the stage $\nu\w$ of the dynamics
that $\tw$ is traversing. Note here that the FDR is defined only for
integer $\nu\w$ in the limit $c\to 0$: for all other values, $\tw$
lies within a plateau and $C_0$ and $\Delta\Chi_0$ are both constant.
Table~\ref{table:S} shows the values of $S$ determined from our
numerical data. They show a relatively weak dependence on temperature,
consistent with the expected approach to nonzero limits for
$T\to 0$; as temperature decreases, also the expected independence from
the non-integer part $a_t$ of $\nu_t$ becomes manifest.

\begin{table}
\begin{center}
\begin{tabular}{|r|r||r|r|}
  \hline
 $S_{k_t,\nu\w}$ & $T$ & $\nu\w=0$ & $\nu\w=1$\\
  \hline\hline
  \multirow{2}{1.3cm}{$k_t=0$}
  & $0.2$  & 0.155 & ---\\  \cline{2-4}
  & $0.17$ & 0.149 & ---\\  \cline{2-4}
  & $0.15$ & 0.135 & ---\\
  \hline
  \multirow{2}{1.3cm}{$k_t=1$}
  & $0.2$  & 0.0269 & 0.295 \\  \cline{2-4}
  & $0.17$ & 0.0235 & 0.301 \\  \cline{2-4}
  & $0.15$ & 0.0166 & 0.306 \\  
  \hline
\end{tabular}

\begin{tabular}{|r|r||r|r|}
  \hline
 $S_{k_t,\nu\w}$ & $T$ & $\nu\w=0$ & $\nu\w=1$\\
  \hline\hline
  \multirow{2}{1.3cm}{$k_t=0$}
  & $0.2$  & 0.134 & ---\\  \cline{2-4}
  & $0.17$ & 0.132 & ---\\  \cline{2-4}
  & $0.15$ & 0.122 & ---\\
  \hline
  \multirow{2}{1.3cm}{$k_t=1$}
  & $0.2$  & 0.0257 & 0.263\\  \cline{2-4}
  & $0.17$ & 0.0203 & 0.266\\  \cline{2-4}
  & $0.15$ & 0.0154 & 0.276\\  
  \hline
\end{tabular}
\end{center}
\caption{Numerical values of $S_{k_t,\nu\w}$ for $a_t=0.5$ (top) and
  $a_t=0.75$ (bottom).
\label{table:S}
}

\end{table}

\begin{figure}
\hspace*{1in} \includegraphics[width=4.5in,clip]{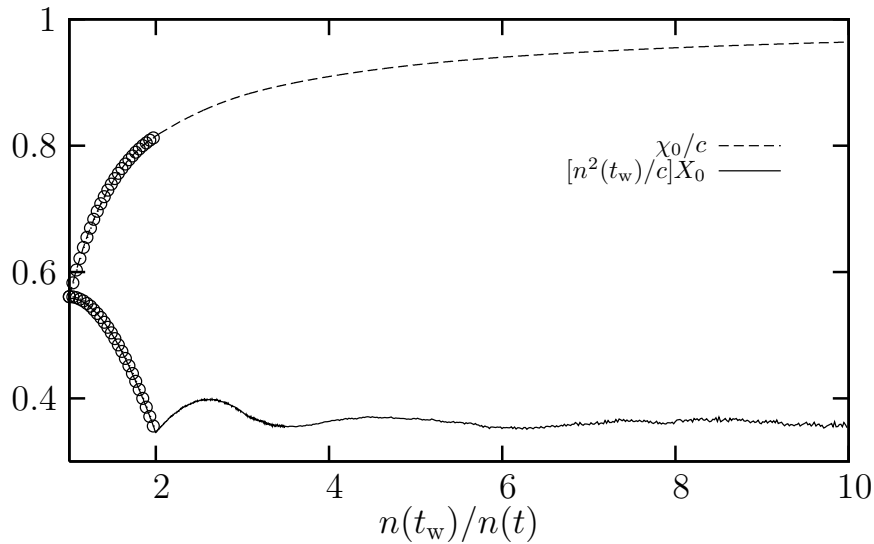}
  \caption{Paste-all prediction for $\Chi_0(t,\tw)/c$ (dashed line)
    and $[n^2(\tw)/c]X_0(t,\tw)$ (solid line), plotted against
    $\theta=n(\tw)/n(t)$. The lines show the results of numerical
    simulations of the paste-all dynamics; the
    circles give the closed-form prediction for $\theta\leq 2$.
\label{fig:Xindep_test}
}
\end{figure}

In our simulations we are necessarily restricted to exploring only the
first few plateaus. Theoretically, we can also study the interesting
paste-all limit where domain sizes are much larger than unity but
still small compared to the average equilibrium domain length $1/c$;
this could be viewed as the true asymptotic non-equilibrium regime.
We have included in Fig.~\ref{fig:FDT_inco} (top) the
paste-all prediction for the scaled FD plot. It is obtained by
combining the predicted scaling~(\ref{Chi0_paste_all}) with
$\Delta C_0(t,\tw)/n^2(t) = n(\tw)/n(t)-1=\theta-1$. Beyond the
quasi-equilibrium regime the limiting scaled FD plot is a smooth curve
(although from the theoretical analysis in~\ref{sec:persistence} one
expects discontinuities in higher-order derivatives at integer values
of $\theta$). Within a mean-field picture, this would be interpreted
as arising from an infinite hierarchy of relaxation timescales, each
``responsible'' for an infinitesimal segment of the FD
plot, as found for instance in the 
infinite range Sherrington-Kirkpatrick spin glass model~\cite{CugKur2}.  
Remarkably, even though the East model
with its purely local and directed interactions is very far from mean
field, this is exactly the scenario here. Along the $x$-axis of the plot
we have $\theta-1=n(\tw)/n(t)-1=\dmin(t)/\dmin(\tw)-1$; two different points
$\theta$ and $\theta'>\theta$ then correspond to waiting times with,
from~(\ref{t_paste_all}), a ratio $\tw'/\tw=(\theta'/\theta)^{1/(T\ln 
  2)}$ that diverges exponentially for $T\to 0$.

We note finally that for the closely related triangular plaquette
model, approximate relations for correlation and response functions
have recently been derived by assuming that relaxations in the
different stages of the dynamics are independent of each
other~\cite{Robs_plaquette_FDT}. The resulting factorization of the
local correlation functions does hold in the East model for small $c$,
as already discussed in~\cite{Robs_plaquette_FDT}. The local FDR was
also predicted to be independent of the later time $t$. To check the
quality of this approximation at least in the paste-all regime, we
differentiate~(\ref{Chi0_paste_all}) w.r.t.\
$\Delta C_0(t,\tw)/n^2(t)=\theta-1$ to get $X_0(t,\tw)=[c/n^2(t)]{\cal
  F}'(n(\tw)/n(t))$. Independence of $t$ would then require that
$[n^2(\tw)/c]X_0(t,\tw) = [n(\tw)/n(t)]^2{\cal F}'(n(\tw)/n(t))$ be
independent of $n(t)$, i.e.\ constant. As Fig.~\ref{fig:Xindep_test}
shows, this is a reasonable approximation when $n(\tw)/n(t)$ is large,
but for smaller values there are deviations which are in fact
oscillatory in $n(\tw)/n(t)$.
In the earlier, irreversible coarsening stages of the dynamics,
$t$-independence is also not a very accurate approximation. To check
this one can consider, instead of $S_{k_t,\nu\w}$ from
Table~\ref{table:S}, the combination $X_0/c=S_{k_t,\nu\w}n^2(t)$.
Fixing for example $\nu\w=0$ and comparing $k_t=0$ and $k_t=1$ one finds values
which, while closer than those of $S_{k_t,\nu\w}$ itself, still differ
by a factor around three.

\subsection{Global observable}
\label{sec:East_global}

\begin{figure}
\hspace*{1in} \includegraphics[width=4.5in]{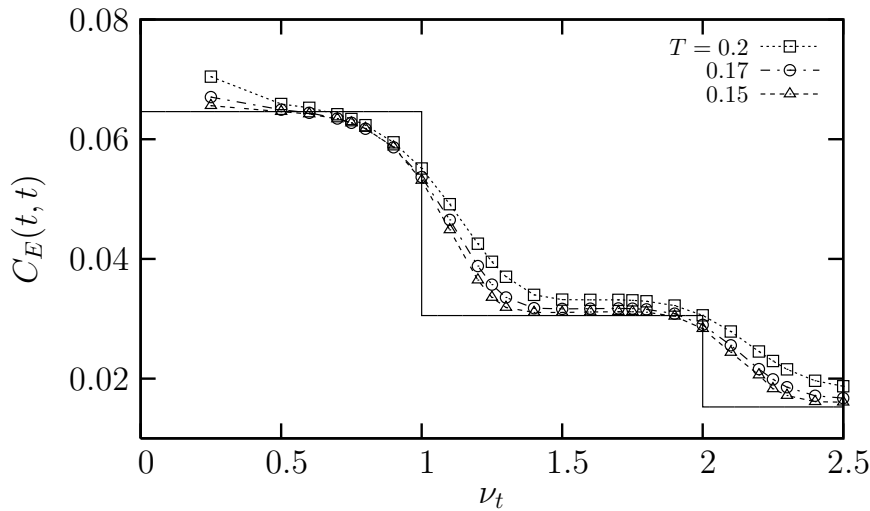}
\caption{Energy variance $C_E(t,t)$ versus $\nu_t$ for three different
  temperatures. The solid line shows the theoretical prediction for
  the limit $T\to 0$.
\label{fig:C_E_tt}
}
\end{figure}
We now switch from the local observable $n_i$ to its global analogue
$E=\sum_i n_i$; as in the FA case, this is just the energy function of
the system. The associated correlation and response are defined in
(\ref{equ:CEdef}, \ref{equ:REdef}).

We start in Fig.~\ref{fig:C_E_tt} with simulation results for the
equal-time correlation $C_E(t,t)$. Plotting against $\nu_t$ reveals
that this has a plateau structure similar to the defect density
$n(t)$, and for $c\to 0$ one expects sharp transitions
to occur at integer values of $\nu_t$. This can be
confirmed by a calculation within the irreversible coarsening regime,
as outlined in~\ref{sec:C_E_calc}. It produces values in very good
agreement with the simulation data, as indicated graphically in
Fig.~\ref{fig:C_E_tt}. The theory shows that, not unexpectedly, energy
fluctuations are related to the variance of the domain size
distribution, via
\be
C_E(t,t) = \frac{\overline{d^2}-\bar{d}^2}{\bar{d}^3}.
\label{C_E_dbar}
\ee

\begin{table}
\begin{center}
\begin{tabular}{|r|r|r|r|r|r|}
\hline
$C_E(t,\tw)$ & & $k\w=-1$ & $k\w=0$ & $k\w=1$ & $k\w=2$\\
\hline
\multirow{4}{1.5cm}{$k_t=0$}
 & $T=0.2$     & 0.0739  & 0.0659 & --- & ---\\  \cline{2-6}
 & $T=0.17$    & 0.0744  & 0.0649 & --- & ---\\  \cline{2-6}
 & $T=0.15$    & 0.0748  & 0.0646 & --- & ---\\  \cline{2-6}
 & Theory      & 0.0758  & 0.0646 & --- & ---\\
\hline
\multirow{4}{1.5cm}{$k_t=1$}
 & $T=0.2$   & 0.0254 & 0.0141 & 0.0332 & --- \\  \cline{2-6}
 & $T=0.17$  & 0.0254 & 0.0138 & 0.0316 & ---\\  \cline{2-6}
 & $T=0.15$  & 0.0256 & 0.0137 & 0.0310 & --- \\  \cline{2-6}
 & Theory    & 0.0261 & 0.0137 & 0.0305 & --- \\
\hline
\multirow{4}{1.5cm}{$k_t=2$}
 & $T=0.2$   & 0.00395 & 0.00240 & 0.000580 & 0.0188\\  \cline{2-6}
 & $T=0.17$  & 0.00376 & 0.00226 & 0.000148 & 0.0167\\  \cline{2-6}
 & $T=0.15$  & 0.00367 & 0.00232 &-0.000110 & 0.0160\\  \cline{2-6}
 & Theory    & 0.00385 & 0.00232 &-0.000233 & 0.0153 \\
\hline
\end{tabular}
\end{center}

\caption{Numerical values of energy correlation $C_E(t,\tw)$ by
  plateau of the dynamics, labelled by $k\w$ and $k_t\geq k\w$. Data
  for three temperatures are shown alongside the theoretical
  prediction for $T\to 0$. In this limit, $k\w=-1$ corresponds to
  $\tw=0$; the simulation data were in fact taken at a nonzero $\tw$
  corresponding for each $T$ to $\nu\w=-0.8$, hence the slightly more
  noticeable deviation from theory in the first column.
\label{table:C_E_ttw}
}
\end{table}
The same calculational approach can in fact also be used to predict
the two-time energy correlations $C_E(t,\tw)$, which have a plateau
structure in both $t$ and $\tw$. Table~\ref{table:C_E_ttw} summarizes
the predictions as a function of $k\w$ and $k_t$: good convergence of
the simulation results to the theory is observed as $c$ is reduced.
The only exception is $(k\w=1,k_t=2)$, where the predicted {\em
  negative} correlation is too small to be accurately determined from
the simulation data. That such negative correlations could arise can
be motivated by considering e.g.\ configurations with atypically
many short domains at time $\tw$, i.e.\ a high value of the energy; in
the next stage of the dynamics these will have to merge and so can
lead to unusually many large domains at a later time $t$.

\begin{figure}
\hspace*{1in} \includegraphics[width=4.5in]{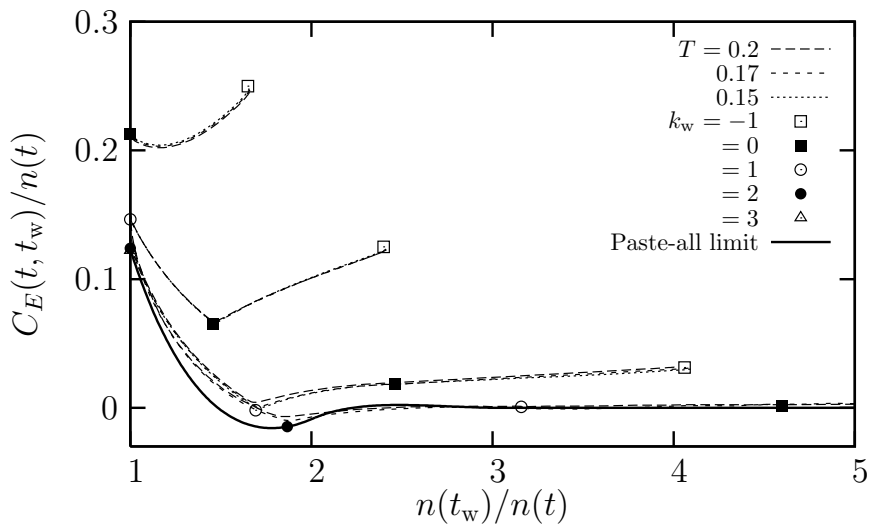}
\caption{$C_E(t,\tw)/n(t)$ from simulations, plotted against
  $n(\tw)/n(t)$. The various dashed lines show simulation data for
  $\nu_t=0.5$, 1.5, 2.5 and 3.5 (top to bottom). Results are displayed
  for temperatures $T=0.15$ (except for $\nu_t=3.5$), 0.17 and 0.2 but
  the variation with temperature is almost invisible on the
  scale of the graph. As $\nu_t$
  increases, convergence towards the curve predicted in the paste-all
  regime (solid line) is observed. For finite $\nu_t$, i.e.\ in the
  irreversible coarsening regime, only the
  plateau values can be predicted theoretically (see
  Table~\ref{table:C_E_ttw}); these are shown by the symbols
  and agree closely with the simulation results.
\label{fig:C_E_ttw}
}
\end{figure}
One may wonder whether negative energy-energy correlations are only a
quirk of the first few stages of the dynamics. To check this we
calculate in~\ref{sec:C_E_calc} the behaviour in the paste-all limit,
where $C_E(t,\tw)/n(t)$ becomes a scaling function ${\cal
G}(\theta)$ of $\theta=n(\tw)/n(t)$. In terms of the functions
$f_l(x)$ defined in~(\ref{fl_def}) this scaling function reads
\be
{\cal G}(\theta) =
e^{-\gamma}(\theta+1)-\left(1
+\sum_{l=1}^\infty \frac{1}{l!}\int_1^\theta d\theta'\,f_l(\theta')\right).
\label{G_theta}
\ee
It is plotted in Fig.~\ref{fig:C_E_ttw} and shows that (weak) negative
correlations do persist even in the paste-all limit.  The figure also
shows graphically the simulation data, with symbols indicating the
predicted plateau values from Table~\ref{table:C_E_ttw}. In spite of
the relatively low values of $\nu_t$ reached, the data are clearly
already moving towards the paste-all limit.

The excursion of $C_E(t,\tw)$ to negative values as illustrated in
Fig.~\ref{fig:C_E_ttw} implies in particular that the two-time energy
correlations are not monotonic in $\tw$ (at fixed $t$) as one would
usually expect. An additional source of non-monotonicity is the
behaviour around $\nu\w=0$, where in both simulations and theory we
find that $C_E$ starts to rise again as $\tw$ decreases. We attribute
this unusual behaviour to the relatively large value of the equal-time
correlation $C_E(\tw,\tw)=1/4$ in the initial plateau ($k\w=-1$).

\begin{figure}
  \hspace*{1in} \includegraphics[width=4.5in]{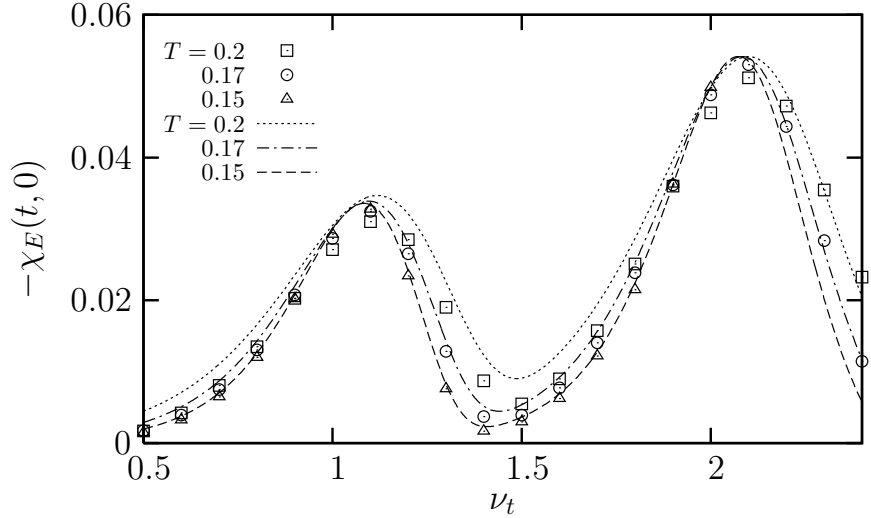}
  \caption{$-\Chi_E(t,0)$, plotted against
    $\nu_t$ for three temperatures. Symbols: simulation data; lines:
    theoretical prediction, Eq.~(\ref{Chi_E_estimate}).
  \label{fig:Chi_E_direct}
  }
\end{figure}

\begin{figure}
  \hspace*{1in} \includegraphics[width=4.5in]{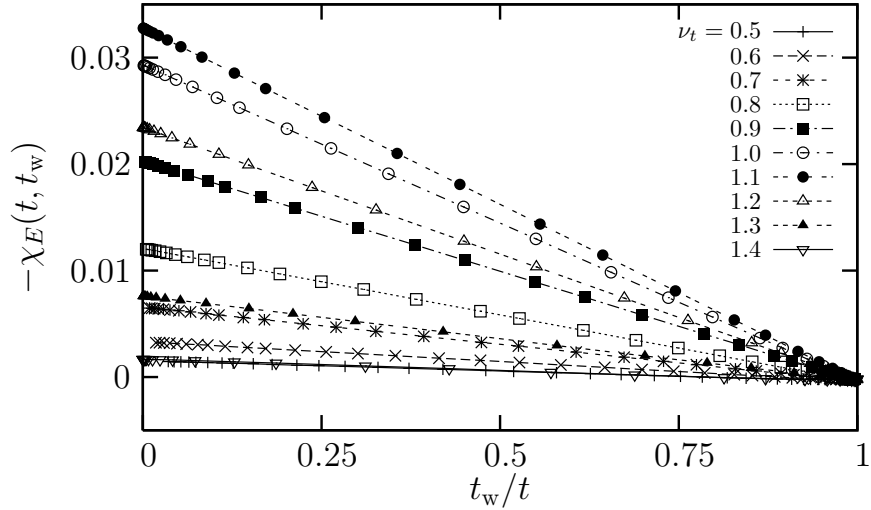}
  \caption{Negative energy susceptibility $-\chi_E(t,\tw)$ 
for $T=0.15$ and several $\nu_t$,
  plotted against $\tw/t$. The proportionality to $(1-\tw/t)$ 
expected from theory, Eq.~(\ref{Chi_E_estimate}), is well verified.
  \label{fig:chi_linear_dependence}
}
\end{figure}
Next we look at the energy susceptibility $\Chi_E(t,\tw)$. By
definition this is the response of the defect density $n(t)$ when the
equilibrium up-spin concentration is changed at time $\tw$ from $c$ to
$c'$, given in~(\ref{c_prime}). To gain some qualitative insight we
exploit the plateau structure of $n(t)$ in the unperturbed dynamics.
With successive relaxations taking place on timescales $\sim c^{-k}$,
$k=0,1,2,\ldots$ we can write (see~\ref{sec:nt})
\be
\ln\left(\frac{n(t)}{n(0)}\right) = -\sum_{k=0}^\infty g_k(c^k t),
\label{n_t_plateaus}
\ee
where the functions $g_k(\cdot)$ describe the relaxation within stage
$k$ and are independent of $c$ to leading order. Each $g_k(\zeta)$
increases from zero at $\zeta=0$ and exponentially approaches a finite
limit for large $\zeta$. For the first few stages one can calculate
explicitly (see~\ref{sec:nt})
\bea
g_0(\zeta)&=&\frac{1}{2}\left(1-e^{-\zeta}\right),
\label{g0}
\\
g_1(\zeta)&=&\frac{3}{8}\left(1-e^{-\zeta/2}\right),
\label{g1}
\\
g_2(\zeta)&=&\frac{7}{24}\left(1-e^{-2\zeta/3}\right) + 
\frac{15}{64}\left(1-e^{-\zeta/4}\right).
\label{g2}
\eea
Now when the field is switched on, the effective time interval
appearing in the argument of these functions will be replaced by
$c^k\tw+(c')^k(t-\tw)$. (For $\tw\neq 0$ this will not necessarily 
be exact except
when the $g_k$'s are single exponentials, but should give a reasonable
approximation nonetheless.) Using~(\ref{c_prime}) and differentiating
w.r.t.\ $\beta h$ then produces the following estimate for the energy
susceptibility:
\be
\Chi_E(t,\tw) = -(1-\tw/t)n(t)\sum_{k\geq 0} kc^kt g_k'(c^k t).
\label{Chi_E_estimate}
\ee
Each of the terms in the sum produces a ``bump'' in $\Chi_E(t,0)$ at
$t\sim c^{-k}$ with a height of order unity; the dependence on $\tw$
is only through the simple factor $1-\tw/t$. Simulation results are in
very good agreement with this prediction: see
Fig.~\ref{fig:Chi_E_direct}. We plot $-\Chi_E(t,0)$ because the
susceptibility is predicted to be {\em negative} -- recall that the
$g_k(\cdot)$ are increasing functions -- in a clear signature of
activated dynamics. Also the predicted linear dependence on $\tw$ is
well verified by our data, as shown in
Fig.~\ref{fig:chi_linear_dependence}.

\begin{figure}
  \hspace*{1in} \includegraphics[width=4.5in]{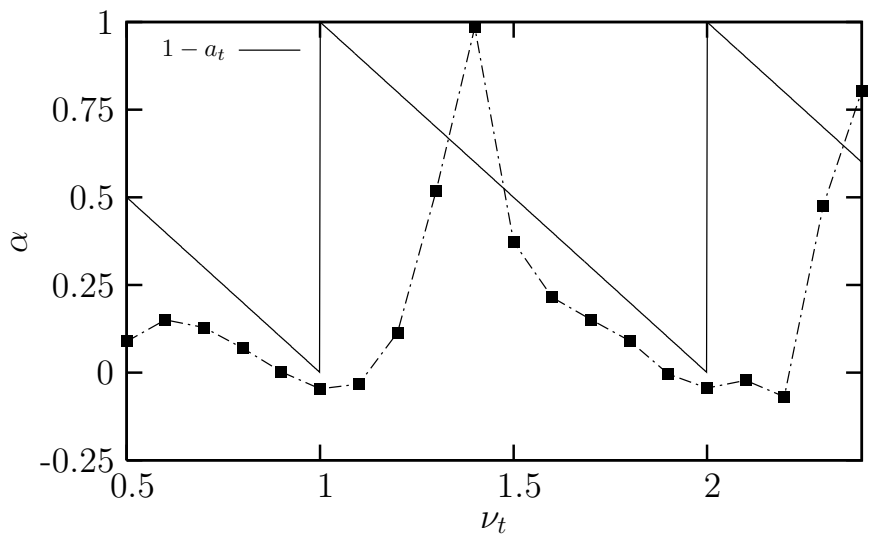}
  \caption{The exponent $\alpha$ as obtained by fitting the
  $c$-dependence of $\Chi_E(t,0)$ for fixed $\nu_t$, compared with the
  theoretical prediction $\Chi_E(t,0)\sim c^\alpha$ with $\alpha=1-a_t$.
\label{fig:Chi_E_alpha}
}
\end{figure}
The bumps correspond to the various stages of the dynamics; indeed,
within the plateaus between these stages the defect density is
independent of $c$ for small $c$ and so one expects the susceptibility
to vanish for $c\to 0$. More precisely, if we write $T\ln t = k_t +
a_t$ as usual with $0<a_t<1$, we have the scaling
\be
-\Chi_E(t,0)\sim c^\alpha, \qquad \alpha = 1-a_t.
\label{Chi_E_alpha}
\ee
To see this, note that all terms with $k\leq k_t$ in the
sum~(\ref{Chi_E_estimate}) are exponentially suppressed because the
argument of $g_k(\cdot)$ is large. In the remaining terms, on the
other hand, the function argument vanishes for $c\to 0$; since the
derivatives $g_k'(0)$ are constants of order unity, each term scales
as $c^k t$. The dominant contribution thus comes from $k=k_t+1$,
giving $-\Chi_E(t,0)\sim c^{k_t+1} c^{-k_t-a_t} = c^{1-a_t}$, as
claimed.  Intuitively, the dominant contribution to $-\Chi_E$ is from
the bump on whose uphill flank $t$ is situated. The
scaling~(\ref{Chi_E_alpha}) is checked against numerical fits of the
$c$-dependence in Fig.~\ref{fig:Chi_E_alpha}. Qualitatively we clearly
observe the 
expected non-monotonic dependence of the exponent $\alpha$ on $\nu_t$.
Quantitatively there are deviations which arise from the
fact that we only fit across three different values of $c$, and that
(compare Fig.~\ref{fig:Chi_E_direct}) we are not yet in the asymptotic
low-$c$ regime where the different bumps become well separated in time.

We note briefly that in the paste-all regime, where the dynamics no
longer separates into discrete stages, one can argue as above that the
unperturbed defect density decays as $n(t)\sim t^{-T\ln 2}$.
Perturbing $T=\ln(1/c)$ to $T'=\ln(1/c')=T+T^2(\beta h)+\ldots$ then shows
that
\be
-\Chi_E(t,0) \sim T^2(\ln t) t^{-T\ln 2} = T\nu_t 2^{-\nu_t} \propto
T\nu_t n(t).
\ee
At constant $\nu_t$ this is proportional to $T=\ln(1/c)$: this
logarithmic scaling is effectively a ``smoothed'' version across bumps
of the dependence on $c^{\alpha}$ with $\alpha=1-a_t\in[0,1]$ in the
irreversible coarsening regime. 

\begin{figure}
\hspace*{1in} \includegraphics[width=2.5in]{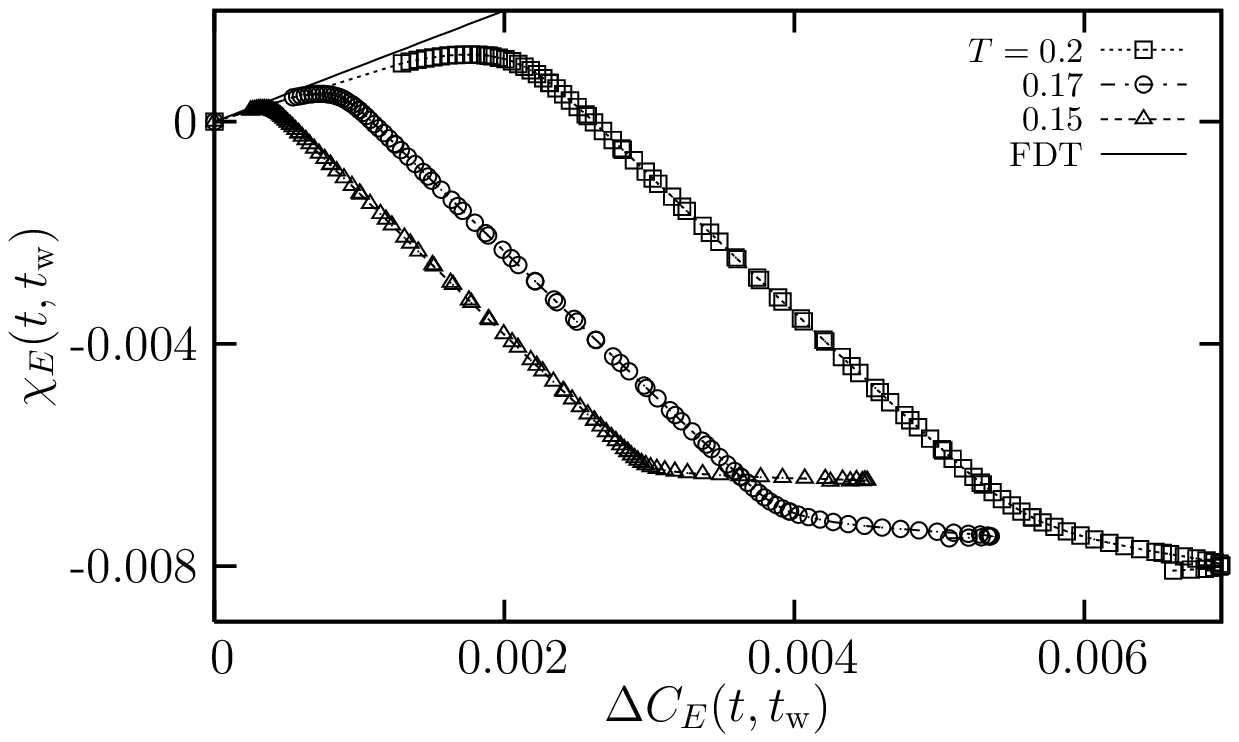}
\includegraphics[width=2.5in]{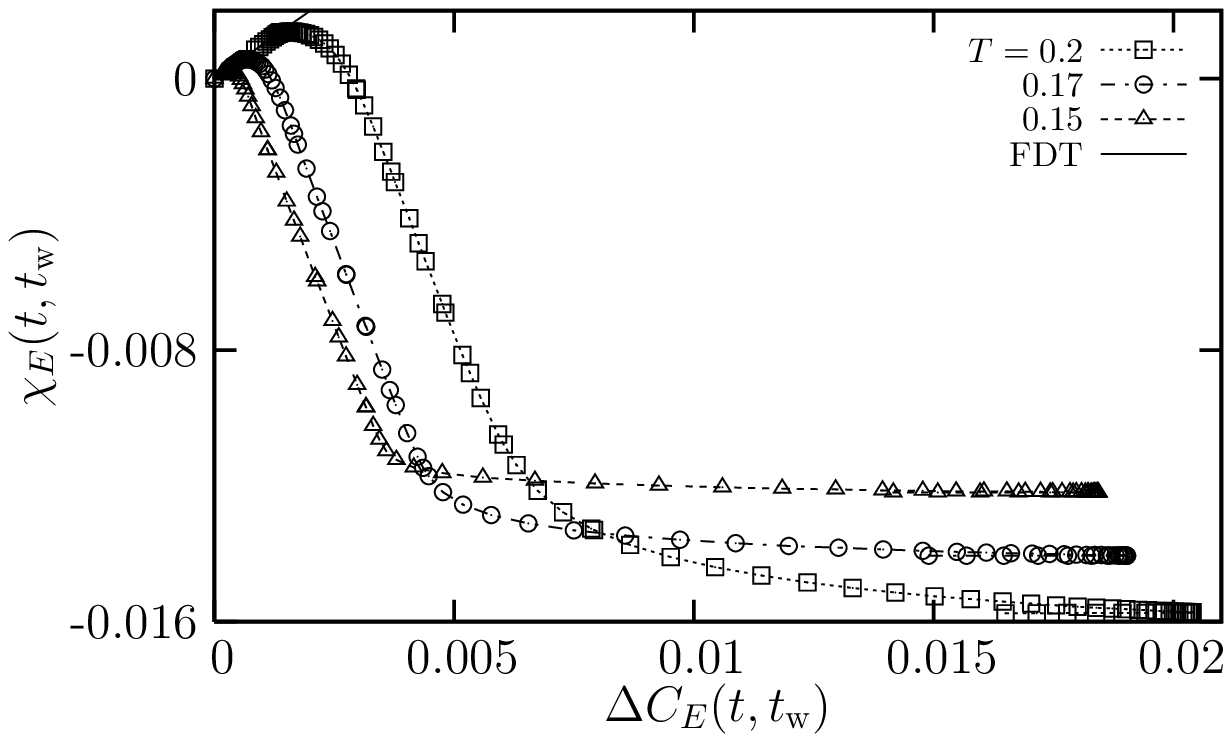} \\
\hspace*{1in} \includegraphics[width=2.5in]{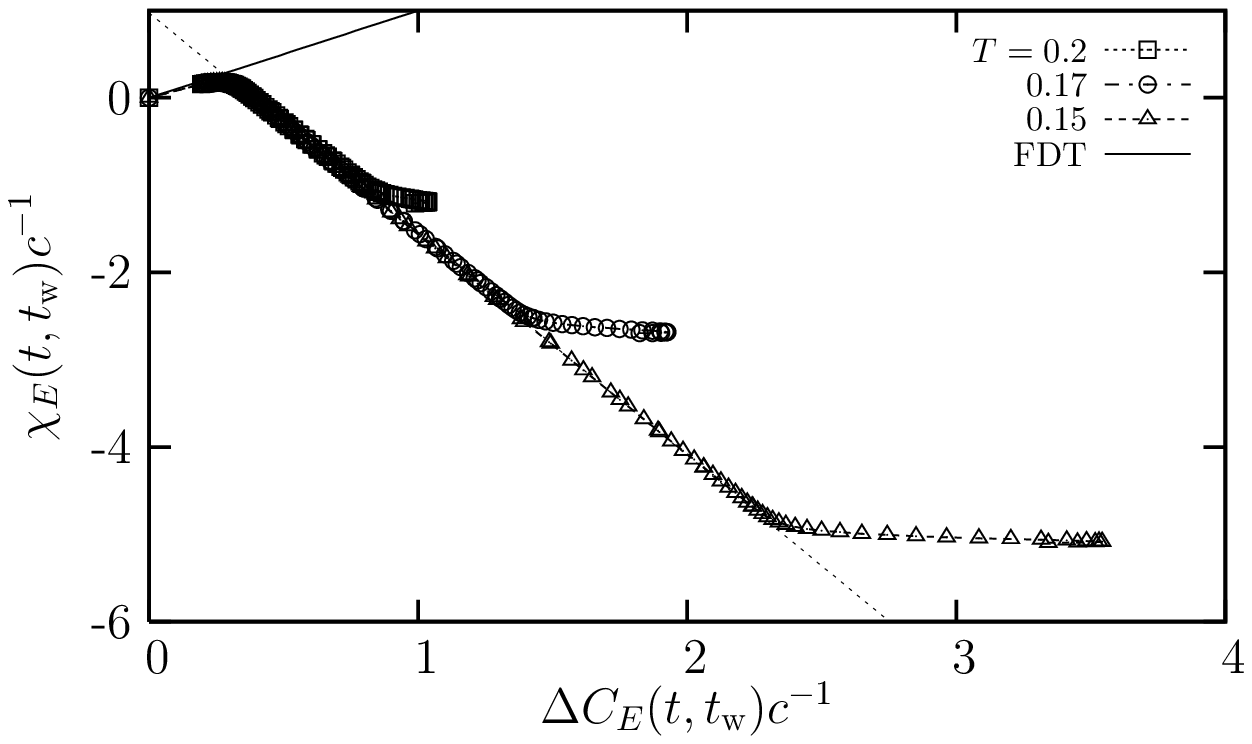}
\includegraphics[width=2.5in]{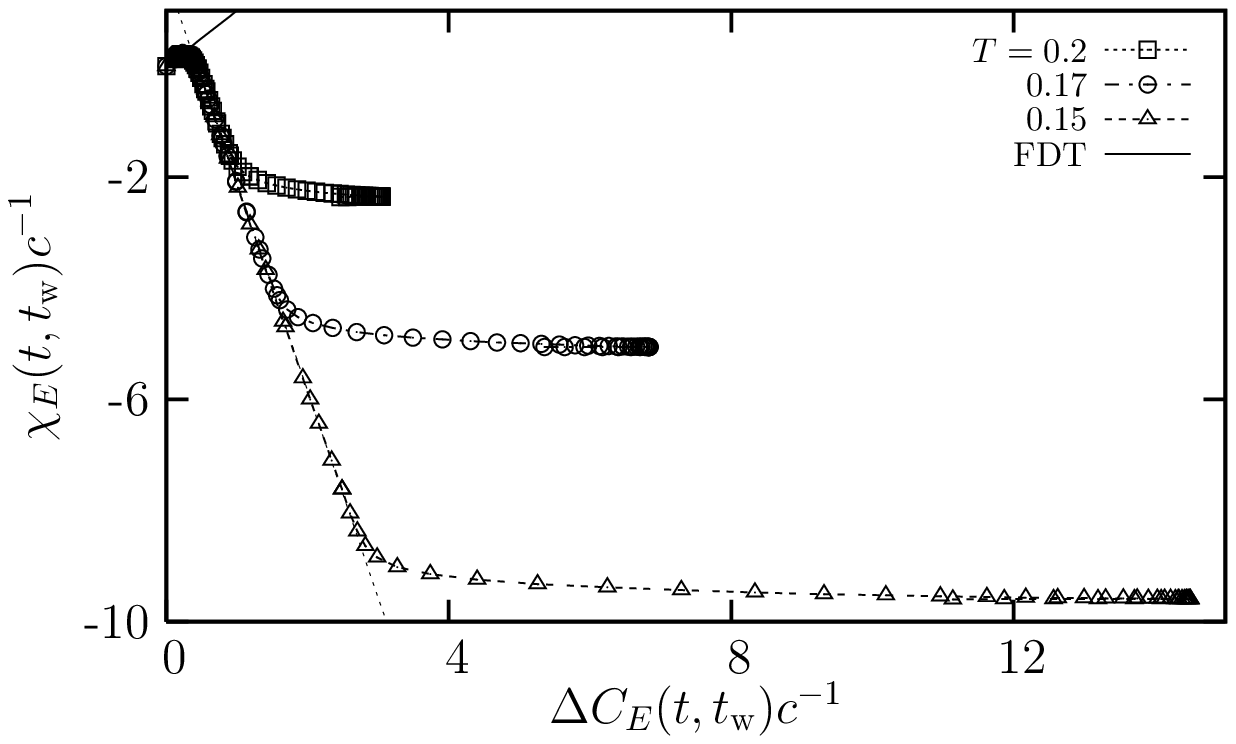}
\caption{Top: Energy FD plots for $\nu_t=0.7$ (left) and
  $1.7$ (right) for three different temperatures.
Bottom: Same plots with each axis rescaled by $1/c$;
fits to the linear parts give slopes of
$-2.55(2)$ and $-3.77(5)$, respectively.
\label{fig:FD_E_raw}
}
\end{figure}
We can now combine $\Chi_E$ and $C_E$ into FD plots.
Sample results from the numerical
simulations are shown in Fig.~\ref{fig:FD_E_raw}.
Three regimes can be discerned. Initially, for $\tw$
close to $t$, $\Chi_E(t,\tw)$ is positive and of order $c$. This is a
quasi-equilibrium response which we ignored in our estimates above.
There follows a section where the predicted negative activated
response becomes dominant; $|\Chi_E(t,\tw)|$ increases here as $\tw$
decreases, as expected from the dependence on $1-\tw/t$. For small
$c$, this factor can get close to unity before $\tw$ leaves the
plateau that $t$ is in. Only when this happens does $C_E(t,\tw)$ start
to change significantly, and we enter the third regime where
$\Chi_E(t,\tw)$ is essentially constant and the FD plot therefore
horizontal. (Within this regime the FD plot then eventually reverses
direction in the peculiar initial stage where $\tw$ is of order unity, 
when $C_E(t,\tw)$ starts to increase again.)

Our above theoretical estimates do not allow us to predict the precise
behaviour of the FD plot around the initial quasi-equilibrium regime.
Nevertheless, the fact that $\Chi_E$ reaches values of order $c$
suggests scaling both axes of the plot with $c^{-1}$;
Fig.~\ref{fig:FD_E_raw} indicates that a limit plot is then
approached as $c$ gets small at constant $\nu_t$. This describes the
initial quasi-equilibrium regime and the crossover to negative values
of $\Chi_E$. In the second regime that follows, the numerically
obtained FD plots are straight with negative slopes of order unity. To
rationalize this, consider $t$ and $\tw$ within the same plateau. Both
are then small compared to the timescale governing the following stage
of the dynamics ($k=k_t+1$), and we can linearize the evolution of the
$C_E(t,\tw)$ in both times.  This then leads to a linear dependence on
$t-\tw$ of $\Delta C_E(t,\tw)=C_E(t,t)-C_E(t,\tw)$. Combined with the
similarly linear relation $\Chi_E \propto (t-\tw)/t$, the FD plot should
then be a straight line as observed numerically. Working out the
relevant prefactors of $t-\tw$ for $\Delta C_E$ and $\Chi_E$ one can predict
for the energy-FDR in this regime as (we omit the details)
\be
X_E = -(k_t+1)\left[\frac{\overline{\Gamma d}}{\overline\Gamma \,\overline d}  
-\left(\frac{\overline{d^2}}{\bar{d}^2}-1\right)\right]^{-1}.
\label{X_E}
\ee
Here the averages are over the domain size distribution, $\Gamma(d)$
is the relaxation rate of domains of size $d$ (see~\ref{sec:nt}) and
all quantities are evaluated at the beginning of the current plateau.
The numerical value of~(\ref{X_E}) can be found relatively easily for
the first few plateaus. For $k_t=0$, there is only one nonzero rate,
$\Gamma(2)$. It follows that $\overline{\Gamma
  d}/\overline{\Gamma}=2$, and evaluating the remaining moments
$\bar{d}$ and $\overline{d^2}$ of $P_0(d)$ gives the prediction
$X_E=-2.54\ldots$, in full agreement with the value $X_E=-2.55(2)$
obtained from the simulations shown in Fig.~\ref{fig:FD_E_raw}. For
$k_t=1$ one predicts similarly, by using that $\Gamma(3)=2c^2/3$ and
$\Gamma(4)=c^2/4$ (see~\ref{sec:nt}), the value $X_E=-3.79\ldots$,
again in agreement with the simulation estimate $X_E=-3.77(5)$. In
the paste-all limit, where $\Gamma(d)$ is significant only for the
shortest domains, one has $\overline{\Gamma d}/\overline{\Gamma}
=\dmin$ so that the energy FDR $X_E=-(k_t+1)/[1-\exp(-\gamma)]$ grows
linearly with the index of the stage of the dynamics. Intuitively this
dependence arises because timescales grow as $c^{-k_t-1}$ and as $k_t$
grows so does the perturbation arising from the change in $c$. 

So far we have discussed the initial quasi-equilibrium regime of the
energy FD plot and the straight line section with negative FDR $X_E$
that follows. Because the ``height'' of the FD plot scales as
$\Chi_E(t,0)\sim c^{1-a_t}$ and $X_E$ is of order unity there, this
section only extends by a small amount $\sim c^{1-a_t}$ of the same
order along the $C_E$-axis. In the remaining ``third'' section, $X_E$
vanishes for $c\to 0$ as explained above. More precisely, this section
can be divided further into subsections where the FDR scales as
$X_E\sim c$, $X_E\sim c^2$ etc (up to $c^{k_t+1}$) as $\tw$ moves
backwards through the various plateaus of the dynamics. To see this,
recall that $X_E=-(\partial\Chi_E/\partial\tw)/(\partial
C_E/\partial\tw)$. The numerator equals
$-\partial\Chi_E/\partial\tw=\Chi_E(t,0)/t$
from~(\ref{Chi_E_estimate}) which scales as
$c^{1-a^t+(k_t+a_t)}=c^{1+k_t}$. For the correlation function we do
not have an explicit result that interpolates between plateaus but it
seems reasonable to assume a smooth dependence on $n(\tw)$. This then
gives the estimate $\partial C_E/\partial \tw\sim
(\partial/\partial\tw)n(\tw)$ which from~(\ref{n_t_plateaus}) will
scale as $c^{k\w+1}$ to leading order. Putting this together gives
$X_E\sim c^{k_t-k\w}$ and hence scalings $\sim c, c^2, \ldots,
c^{k_1+1}$ as claimed. The case where $\tw$ is still in the same
plateau as $t$ ($k\w=k_t$) is included and leads to the FDRs of order
unity discussed above.

\subsection{Non-local observables}
\label{sec:East_nonlocal}

In the previous subsections we found that both for local and global
observables the correlation functions have values of order unity and
exhibit a plateau structure that reflects the splitting of the
dynamics into discrete stages (in the irreversible coarsening regime).
The susceptibilities of local and global observables differ more
strongly: for the local
case, $\Chi_0$ is always positive and of order $c$, and the local FDR
is positive and of the order of $c/n^2(\tw)$ (cf.\ the
discussion at the end of Sec.~\ref{sec:East_local}). The global
susceptibility, on the other hand, is negative save for an initial
quasi-equilibrium part of $O(c)$, and the associated FDR $X_E$ is
negative with values of order unity before dropping to zero -- more
precisely $O(c)$, $O(c^2)$, \ldots -- in the
final segment of the FD plot. The question naturally arises of how
these results relate to each other.

\begin{figure}
\hspace*{1in} \includegraphics[width=4.5in]{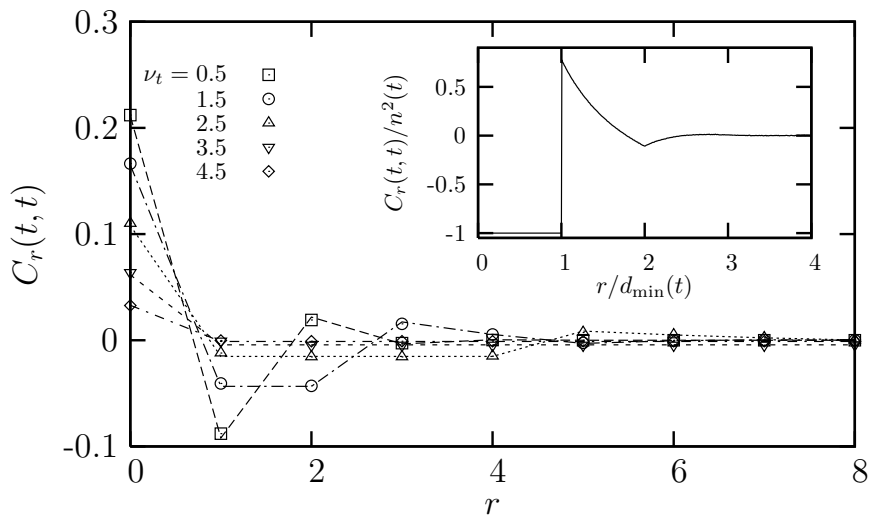}
\caption{Non-local equal-time correlations $C_r(t,t)$ from 
simulations (symbols)
  for $T=0.15$ and different values of $\nu_t$; these
  compare well with the irreversible coarsening predictions shown 
as dashed lines. Inset: 
  paste-all prediction for the same 
quantity, from Eq.~(\ref{Cr_tt_paste_all}).
\label{fig:C_r_tt}
}
\end{figure}

Based on experience with the FA model we initially experimented with
Fourier component observables. These, however, lead to FD plots that
are very difficult to interpret; also, the limit of short wavevectors
does not give direct access to the quantities for local observables.
Instead we consider observables
defined by random staggered fields $\epsilon_i$ with Gaussian
correlations $\langle \epsilon_i \epsilon_{i+r} \rangle =
\exp(-r^2/2\ell^2)$ in space. These have correlation and
susceptibility
\bea
C_\ell(t,\tw) &=& \sum_r e^{-r^2/2\ell^2} C_r(t,\tw), \\
\Chi_\ell(t,\tw) &=& \sum_r e^{-r^2/2\ell^2} \Chi_r(t,\tw).
\label{C_Chi_l_def}
\eea
As $\ell$ is increased from 0 to $\infty$ we can thus directly
interpolate between local and global observables. Before looking at
this $\ell$-dependence we turn to the non-local functions $C_r$ and
$\Chi_r$ as these are the building blocks from which $C_\ell$ and
$\Chi_\ell$ are constructed.

\begin{figure}
\hspace*{1in} \includegraphics[width=4.5in]{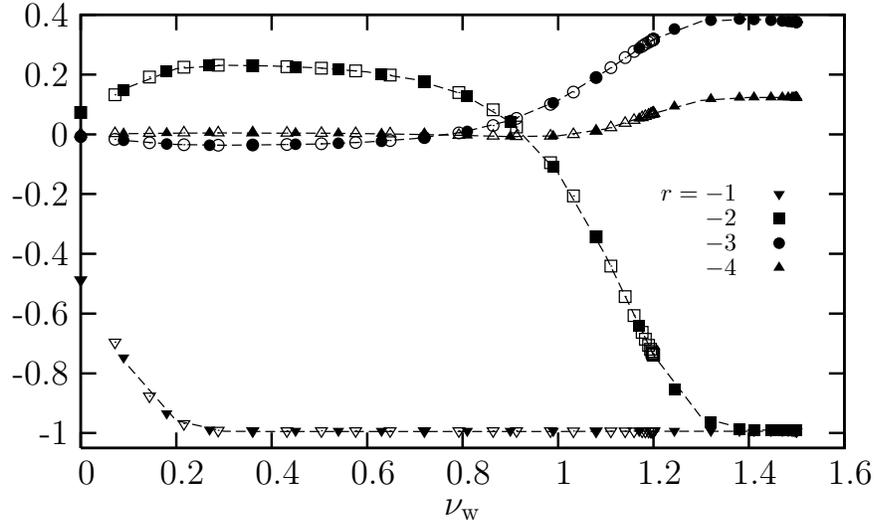}
\caption{Non-local correlation function $C_r(t,\tw)/[n(t)n(\tw)]$ for
  various negative $r$, plotted against $\nu\w$. Symbols: simulation
  data for $T=0.15$, for fixed $\nu_t=1.2$
  (open symbols) and $\nu_t=1.5$ (full symbols).
  Dashed lines: theoretical prediction in terms of equal-time
  correlations, $C_r(\tw,\tw)/n^2(\tw)$, from eq.~(\ref{Cr_ttw_negative_r}).
\label{fig:C_rneg_ttw}
}
\end{figure}

The equal-time correlations $C_r(t,t)$ are simple to predict in the
irreversible coarsening regime. Because the dynamics is of an
independent interval type, domains from the domain size distribution
$P_k(d)$ for the current plateau $k=k_t$ are arranged in random order.
Summing the probability of having an up-spin one, two (etc) domains
along one gets
\bea
C_r(t,t) &=& \langle n_i(t)n_{i+r}(t) \rangle -n^2(t) \\
&=& n(t)\left[ -n(t) + \delta_{r,0} + P_k(r) + \sum_{d} P_k(d)P_k(r-d) + \ldots\right].
\label{Cr_tt}
\eea
This is simple to evaluate by numerical convolution given that we know
$P_k(d)$; only the first few terms are needed since the $m$-th order
convolution term is nonzero only for $r\geq m\dmin(t)=m(2^k+1)$. The
result is shown in Fig.~\ref{fig:C_r_tt} for the first few plateaus
and is in very good agreement with numerical data. The negative
initial section for $0<r<\dmin(t)$ arises because no up-spins exist in
this distance range. This strong short-range repulsion also leads to
oscillations in the correlation for larger values of $r$, including
weak anti-correlations at some distances. In the paste-all limit one
can exploit that the infinite sum of convolutions in~(\ref{Cr_tt})
becomes a simple geometric series for the Laplace transforms. The
result is proportional to $\exp({\rm Ei}(s))$, and inverting the
transform gives in terms of the functions~(\ref{fl_def})
\be
\frac{C_r(t,t)}{n^2(t)} = {\cal H}(r/\dmin(t),1), \qquad
{\cal H}(x,1) = -1 + e^\gamma\sum_{l\geq 1}\frac{1}{l!}f_l(x).
\label{Cr_tt_paste_all}
\ee
A plot of this is included in Fig.~\ref{fig:C_r_tt} and clearly shows
that small negative values occur e.g.\ around $x=2$, outside the main
``repulsion zone'' $0<x<1$.

The two-time correlations
\be
C_r(t,\tw) = \langle n_{i+r}(t) n_i(\tw)\rangle - n(t)n(\tw)
\ee
are more difficult but can still be obtained relatively simply for
$r<0$. The survival of an up-spin at site $i+r$ from time
$\tw$ to time $t$ depends only on the arrangement of domain lengths to
the {\em left} of this site; in particular, it does {\em not} depend
on whether an up-spin is present at site $i$ (to the right of $i+r$) at time
$\tw$. Since the overall survival probability of an up-spin is just
$n(t)/n(\tw)$, this implies $\langle n_{i+r}(t)n_i(\tw)\rangle =
[n(t)/n(\tw)] \langle n_{i+r}(\tw)n_i(\tw)\rangle$ or, in terms of
$C_r(t,\tw)$,
\be
\frac{C_r(t,\tw)}{n(t)n(\tw)} = \frac{C_r(\tw,\tw)}{n^2(\tw)}, \qquad
{\rm for} \ r<0.
\label{Cr_ttw_negative_r}
\ee
This prediction is very well verified by our simulation results
in Fig.~\ref{fig:C_rneg_ttw}, which confirm for various negative $r$
that the left hand side of Eq.~(\ref{Cr_ttw_negative_r}) is both
independent of $t$ and varies with $\tw$ in the matter predicted.

\begin{figure}
\hspace*{1in} \includegraphics[width=4.5in,clip]{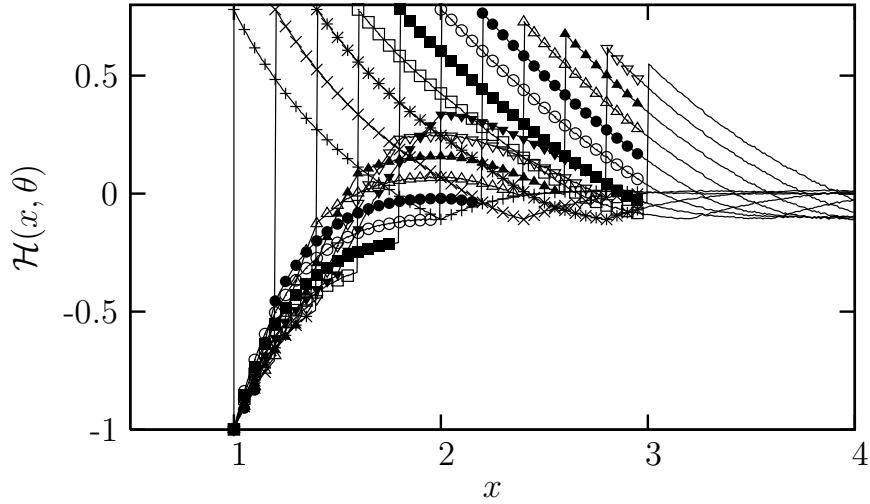}
\caption{Scaling function ${\cal H}(x,\theta)$ for the non-local
  correlation function in the paste-all regime. Curves are for
  $\theta=1,1.2,\ldots,3$ from left to right. Lines: Simulations of
  paste-all dynamics; symbols: analytical prediction~(\ref{H_scaling}).
\label{fig:Cr_ttw_paste_all}
}
\end{figure}
\begin{figure}
\hspace*{1in} \includegraphics[width=4.5in]{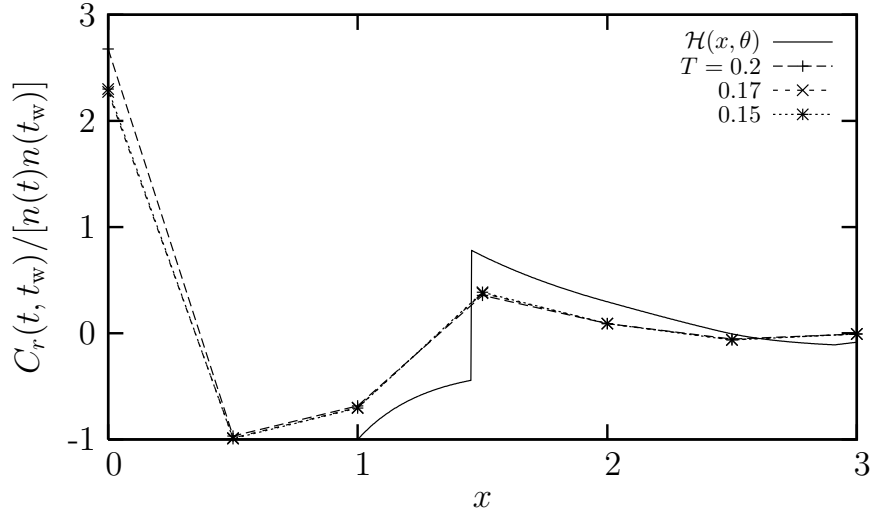}
\caption{Non-local correlation function $C_r(t,\tw)$, as obtained from
  simulations at $\nu_t=1.5$ and
  $\nu\w=0.5$ and for the three temperatures indicated. The axes are
  as in Fig.~\ref{fig:Cr_ttw_paste_all}, showing the
  correlations scaled by $n(t)n(\tw)$ against 
  the scaled distance $x=r/\dmin(\tw)$.
The solid line graphs the paste-all prediction for the limit of large
$\nu\w$ and $\nu_t$.
\label{fig:crttw_scaling}
}
\end{figure}

For positive values of $r$, no similar simple argument
applies. However, one can still calculate $C_r(t,\tw)$ within the
paste-all regime; this involves keeping track of which domain lengths
from time $\tw$ have been swallowed up into the larger domains at time
$t$ (see~\ref{sec:C_r_calc}). One finds the following scaling:
\be
\frac{C_r(t,\tw)}{n(t)n(\tw)} = {\cal H}(x,\theta), \quad
x=\frac{r}{\dmin(\tw)}, \quad \theta=\frac{n(\tw)}{n(t)}.
\label{C_r_ttw_scaling}
\ee
The scaling function ${\cal H}$ can be calculated explicitly in the
range $0\leq x\leq 3$ and $1\leq \theta\leq 3$, with the result:
\bea
{\cal H}(x,\theta) &=& - 1 +
e^\gamma\frac{\theta}{x}\Biggl\{1-\frac{\Theta(\theta-x)}{x}
\nonumber\\
& &+\Theta(x-2)\left[-\frac{x-1}{x}\ln(x-1)+1-\frac{2}{x}\right]
\nonumber\\
& &+\Theta(x-\theta-1)
\left[\frac{x-1}{x}\ln\left(\frac{(x-\theta)(x-1)}{\theta}\right)
  - \frac{x-\theta-1}{x-\theta}\right]
\nonumber\\
& &+\Theta(x-2\theta)
\left[\frac{1}{x}\ln\left(\frac{x-\theta}{\theta}\right) 
+ \frac{2}{x} - \frac{1}{x-\theta}\right]
\nonumber\\
& &+
\Theta(\theta-x-1)\frac{1}{x}\ln\left(\frac{\theta}{(x+1)(\theta-x)}\right)
\nonumber\\
& &+\Theta(\theta-2x)\frac{1}{x}\ln\left(\frac{2(\theta-x)}{\theta}\right)
\Biggr\}.
\label{H_scaling}
\eea
(Notice that for negative distances, the scaling is independent of $t$
from~(\ref{Cr_ttw_negative_r}):
${\cal H}(-x,\theta)={\cal H}(-x,1)$, and due to the spatial symmetry
of the equal-time correlation this is identical to ${\cal H}(x,1)$
from~(\ref{Cr_tt_paste_all}).) We plot ${\cal H}(x,\theta)$ for
positive $x$ in Fig.~\ref{fig:Cr_ttw_paste_all} for a range of
values of $\theta$, comparing also with data from direct simulations
of the paste-all dynamics. The two-time correlations are seen to
have an extremely rich spatial structure, which arises from the
interplay of the different lengthscales $\dmin(\tw)$ and $\dmin(t)$.
The strict exclusion zone for $r<\dmin(\tw)$ ($x<1$) remains at the
later time $t$. Larger distances $r$ {\em can} separate spins at the two
different times $\tw$ and $t$
(${\cal H}>-1$), but correlations only become positive beyond the
lengthscale $r=\dmin(t)$ ($x=\theta$). Simulation data in the first
plateaus (Fig.~\ref{fig:crttw_scaling}) qualitatively follow these
predicted trends; later plateaus would evidently need to be considered
to see more quantitative agreement with the paste-all limit of large
$k\w$ and $k_t$.

\begin{figure}
\hspace*{1in} \includegraphics[width=4.5in,clip]{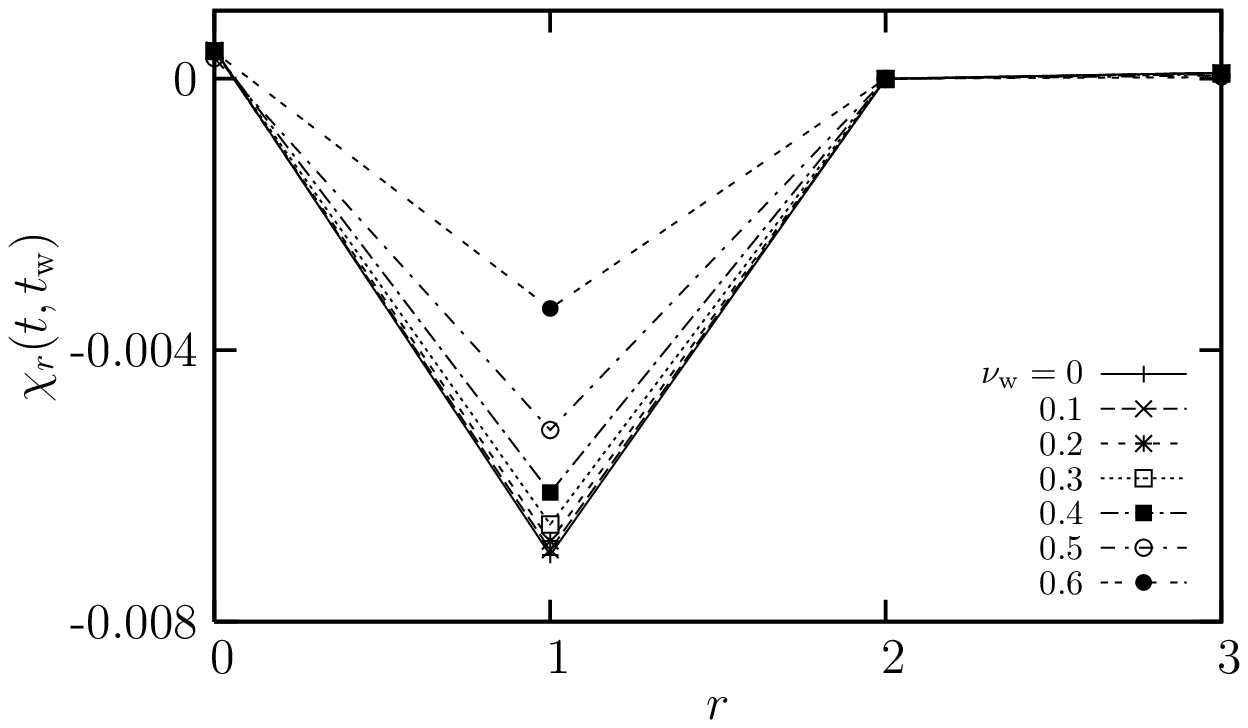} \\
\hspace*{1in} \includegraphics[width=4.5in,clip]{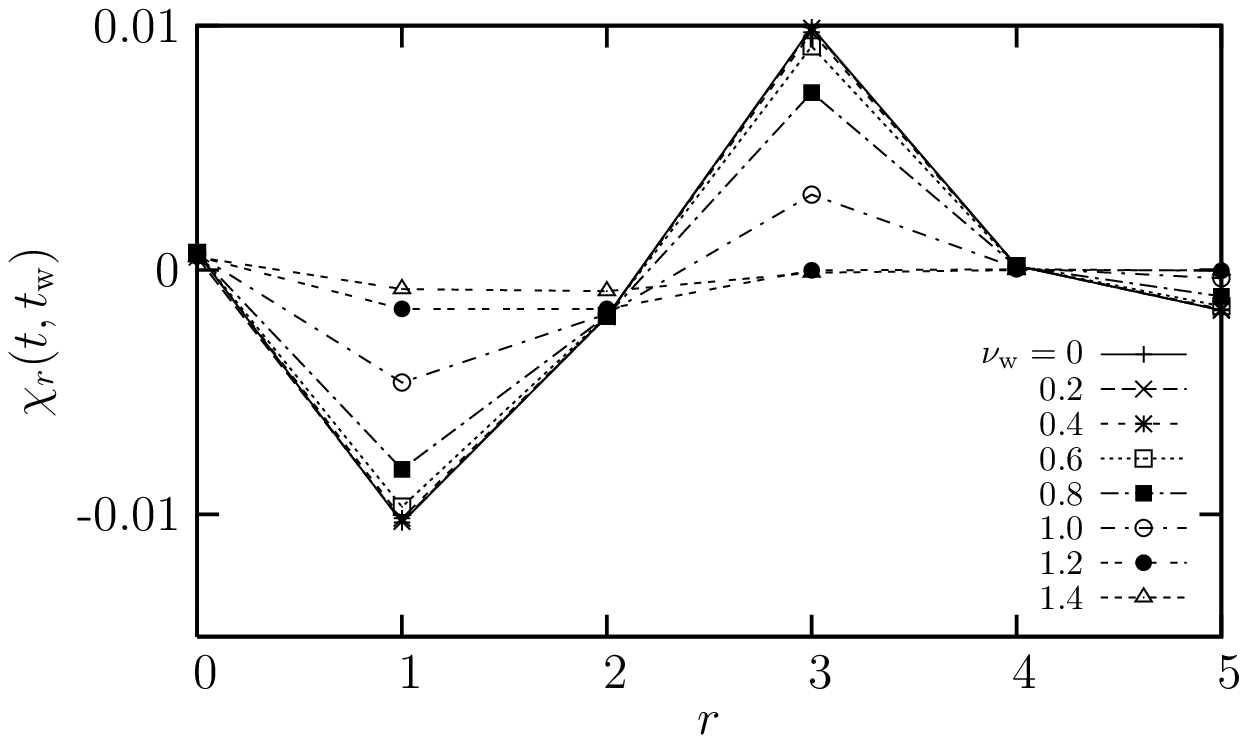}
\caption{Non-local susceptibility from simulations at $T=0.15$ and for
  $\nu_t=0.7$ (top) and $\nu_t=1.5$ (bottom). We plot $\Chi_r(t,\tw)$
  against $r$ for a range of waiting times as indicated by the legend.
\label{fig:chi_r}
}
\end{figure}

\begin{figure}
\hspace*{1in} \includegraphics[width=4.5in,clip]{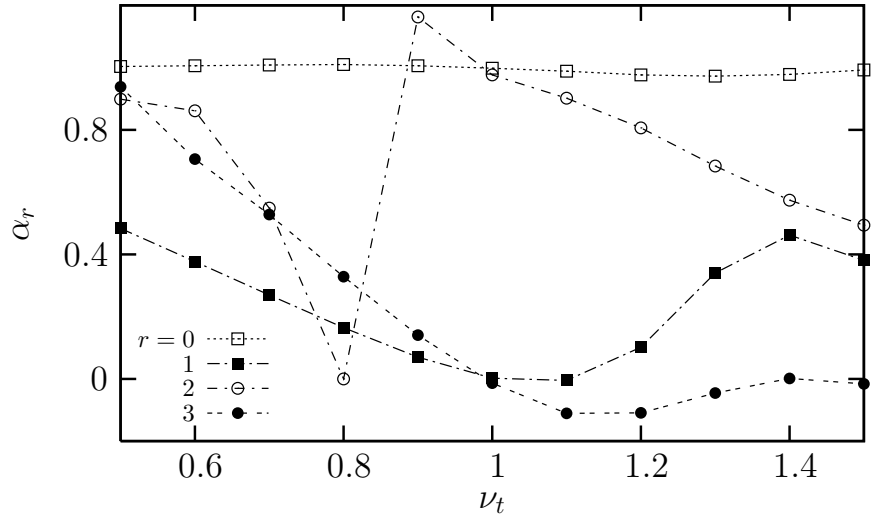}
\caption{Scaling exponent $\alpha_r$ obtained from fits of simulation
  data to $\Chi_r(t,0)\sim c^{\alpha_r}$ at fixed $\nu_t$. The
  exponent is plotted against $\nu_t$ for $r$ as indicated by the
  legend.
\label{fig:expo_nu_r}
}
\end{figure}

Next we consider the non-local susceptibility $\Chi_r(t,\tw)$. Here
again the case of negative $r$ is simpler: $\Chi_r(t,\tw)$ then has to
vanish. This is because the evolution of spin $n_{i+r}(t)$ is only
affected by the spins to its left; in particular, it remains
unperturbed by a field applied at site $i>i+r$. For positive $r$, on
the other hand, $\Chi_r(t,\tw)$ will be nonzero but there are
competing effects governing its sign and magnitude, making theoretical
analysis difficult. We show selected simulation data for $\Chi_r(t,\tw)$
in Fig.~\ref{fig:chi_r}, and the exponent $\alpha_r$ extracted by fits
to the expected scaling $\Chi_r(t,0)\sim c^{\alpha_r}$ at fixed
$\nu_t$ in Fig.~\ref{fig:expo_nu_r}.

To develop some intuition, consider $r=1$ and a local spin
configuration $(n_{-1},n_0,n_1)=(1,0,1)$ at time $\tw=0$ when the
field is applied to site $0$. This will increase the chances of $n_0$
flipping up, after a time $\sim c^{-1}$, and hence increase the
probability of $n_1$ flipping down. The resulting contribution to
$\Chi_1(t,0)$ is a ``bump'' around $t=c^{-1}$ with negative sign and
amplitude of order unity. If the next up-spin to the left of $n_0$ is
further away, a similar effect occurs but with a longer delay because
an up-spin front needs to propagate to $n_{-1}$. Overall this should
give rise to a series of negative bumps in $\Chi_1(t,0)$ centred around $t=c^{-1}$, $c^{-2}$
etc, similarly to $\Chi_E(t,0)$. At constant $\nu_t$ the scaling with
$c$ is then $\sim c^{1-a_t}$ as argued after~(\ref{Chi_E_alpha}). This
is in reasonable agreement with the data (Fig.~\ref{fig:expo_nu_r}).
For $r=2$, an argument based on initial configurations such as
$(n_{-1},n_0,n_1,n_2)=(1,0,0,1)$ suggests a similar structure but with
the bump around $t=c^{-1}$ absent, giving the scaling $\Chi_2(t,0)\sim
c^{2-\nu_t}$ for $\nu_t<2$ (and $\sim c^{1-a_t}$ thereafter). This is
again in general accord with the numerical trends seen in
Fig.~\ref{fig:expo_nu_r}. Closer inspection reveals, however, that
$\Chi_2(t,0)$ is in fact {\em positive} below $\nu_t\approx 0.8$. This sign
change, which also makes the measurements of the exponent $\alpha_r$ in
this region rather unreliable, is not
accounted for by our naive argument. For $r=3$ this issue becomes more
pronounced still; in fact,
$\Chi_3(t,0)$ is positive throughout the range that we can
explore (Fig.~\ref{fig:chi_r}). The dominant
contribution for $t<c^{-3}$ is in this case an effect {\em across}
domains: starting from $(n_{-1},n_0,n_1,n_2,n_3)=(1,0,1,0,1)$, a
field at site 0 will speed up the disappearance of the domain formed
by the first three spins.  This causes $n_3$ to survive for {\em
  longer}, giving a positive contribution to $\Chi_3(t,0)$. The
timescale $t\sim c^{-1}$ on which this term becomes significant is set
by the rate for the first domain to disappear. One would expect, then,
the scaling $\Chi_3(t,0)\sim c^{1-a_t}$ for $\nu_t<1$. In fact,
the relevant scaling function -- which for short chains can be
calculated explicitly -- turns out to start off quadratically for
small values of its argument, so that instead $\Chi_3(t,0)\sim
c^{2(1-a_t)}$, in good agreement with our numerics
(Fig.~\ref{fig:expo_nu_r}). It is not clear to us at present how to
systematically account for effects of this type to get an overall
prediction for the low-$c$ scaling of $\Chi_r(t,0)$. Also the
inclusion of the full $\tw$-dependence of $\Chi_r(t,\tw)$ is not
trivial: for $k_t=0$ (Fig.~\ref{fig:chi_r} top) a simple
proportionality to $(1-\tw/t)$ can be checked to describe the data
well, whereas for $k_t=1$ (Fig.~\ref{fig:chi_r} bottom) this is no
longer the case.

\begin{figure}
\hspace*{1in} \includegraphics[width=4.5in,clip]{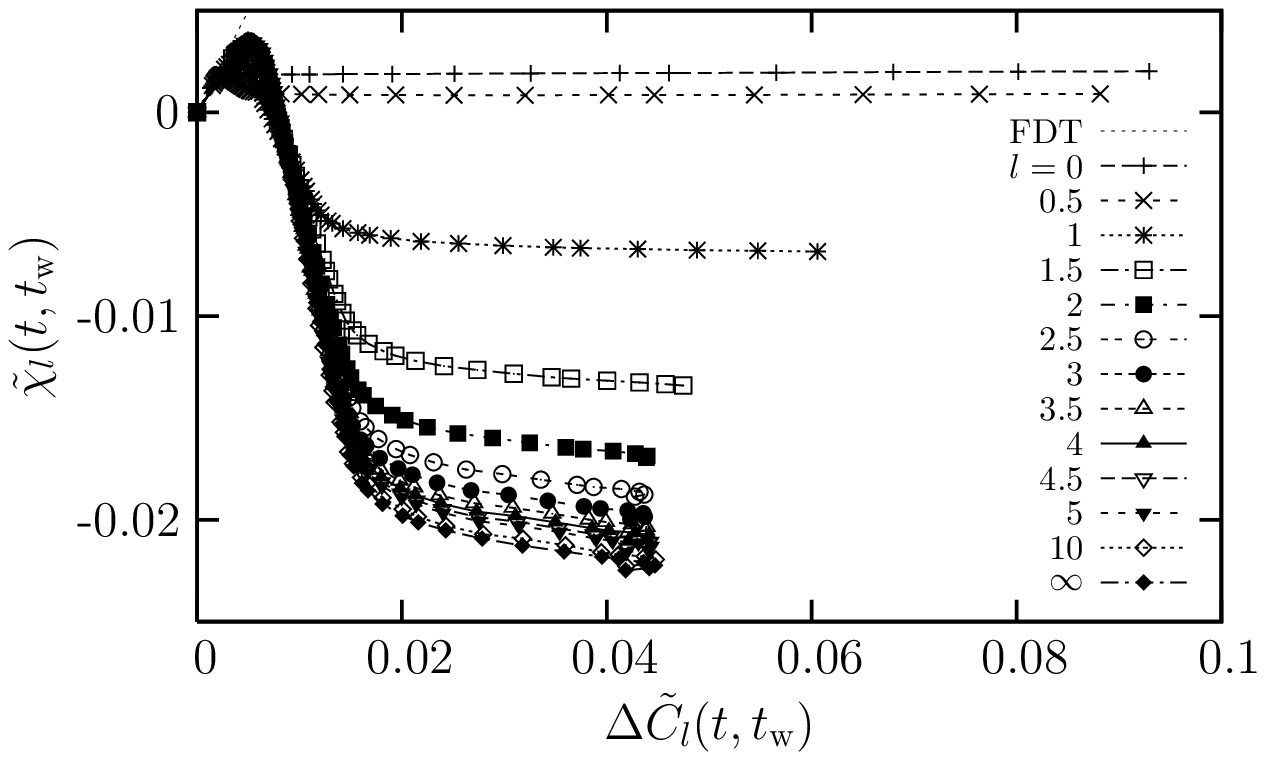} \\
\hspace*{1in} \includegraphics[width=4.5in,clip]{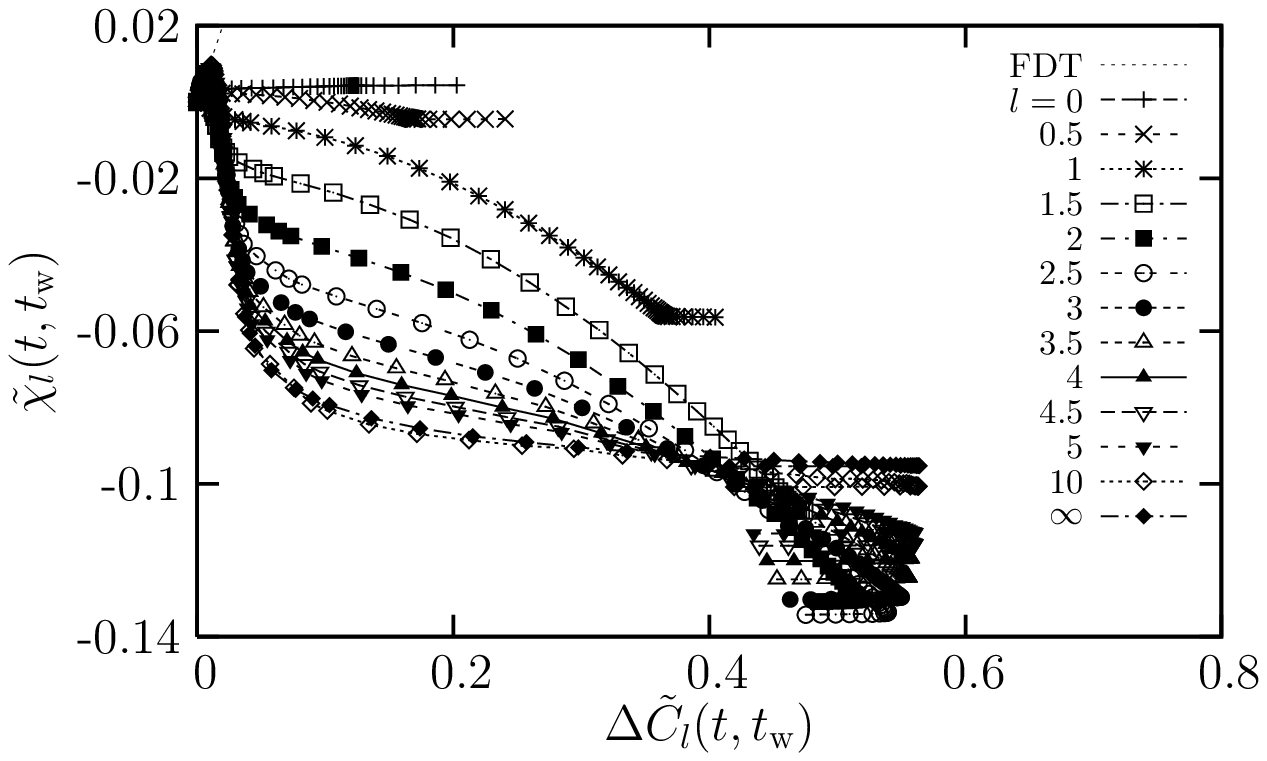}
\caption{Normalized FD plot 
  for observables defined by Gaussian staggered fields,
  for $T=0.15$ and $\nu_t=0.5$ (top), $\nu_t=1.5$ (bottom). The
  lengthscale $\ell$ of the field correlation is given in the
  legend. \label{fig:FDT_gaussian}
}
\end{figure}

Finally we show in Fig.~\ref{fig:FDT_gaussian} the FD plots for the
$\ell$-dependent correlation and response defined
in~(\ref{C_Chi_l_def}). These interpolate between the local ($\ell=0$)
and global limits ($\ell\to\infty$) as they must. For the times
considered here the dynamical lengthscales (typical domain sizes) are
short enough that the global behaviour is approached already for
moderate values of $\ell$. The approach to the local limit $\ell=0$,
on the other hand, must be expected to become singular for small $c$.
This can be seen as follows. Consider $\Chi_E(t,\tw)=\sum_r
\Chi_r(t,\tw)$. We know that for $\tw=0$ and fixed $\nu_t$ this
quantity scales as $c^{1-a_t}$ for $c\to 0$. There must therefore be
one or several values of $r$ for which $\Chi_r(t,0)$ scales in the
same way. But then these values will dominate the sum over $r$ in
$\Chi_\ell(t,0)$; in particular, they will always dominate over the
local term ($r=0$) with its $O(c)$ contribution. Thus, for any nonzero
$\ell$ we expect $\Chi_\ell(t,0)\sim c^{1-a_t}$, whereas at $\ell=0$
itself the scaling is $\sim c$. Assuming that similar arguments apply
also to the two-time quantities $\Chi_\ell(t,\tw)$, one expects the FD
plots for sufficiently small $c$ to switch effectively discontinuously
from the local FD relation for $\ell=0$ to behaviour dominated by
non-local effects for $\ell>0$.

\section{Discussion and Conclusions}
\label{sec:conclusion}

We have studied two paradigmatic kinetically constrained models (KCMs)
of glassy dynamics: the FA (Fredrickson-Andersen) and East models.
Deploying a variety of analytical techniques and comparing with
detailed numerical simulations, we have analysed in particular the
correlation and response functions during the aging after a quench to
low temperature, along with the resulting fluctuation-dissipation
ratios (FDRs). Local as well as global observables were considered, and
Fourier mode and Gaussian staggered field observables
respectively allowed us to interpolate between these two limiting
cases.

In the FA model, with its effective dynamics of diffusing and
coagulating defects, a clear physical scenario emerges: the asymptotic
FDR $X^\infty$ for well-separated times $t\gg\tw$ is {\em negative}.
Its precise value depends on whether one is above or below the
critical dimension $\dc=2$, with $X^\infty=-3$ for $d>2$ and
$X^\infty=-3\pi/(6\pi-16)$ in $d=1$. The underlying physics is,
however, the same: the negative FDR arises from the activated nature
of the aging dynamics. Where a temperature increase would, in
equilibrium, increase the defect density $n$, here its main effect is
to speed up the decrease of $n(t)$ and so reduce its value. The
asymptotic FDR can be determined from observables of any wavelength or
characteristic lengthscale, although for local observables this is
very awkward because the interesting aging effects are buried
underneath a dominant quasi-equilibrium signal. Much better suited is
the global observable, i.e.\ the energy or total number of defects,
which produces a fluctuation-dissipation (FD) plot that is close (in
$d=1$) or exactly equal (in $d>2$) to a straight line of slope
$X^\infty$.

The East model has a more strongly cooperative behaviour than the
FA model, and a correspondingly more subtle pattern of violations of
the fluctuation-dissipation theorem (FDT). For the local observable,
the FDR $X_0(t,\tw)$ is always positive but small, of order
$c/n^2(\tw)$, where $c$ is the (small) equilibrium defect
concentration. In the paste-all limit of large domain sizes, which
corresponds to a long-time limit taken within the aging regime, we find an
intriguing similarity with mean-field predictions 
for spin glass dynamics: a continuous
hierarchy of relaxation timescales leads to a curved FD plot that is
effectively composed of a sequence of infinitesimal straight line
segments.

For global observables, on the other hand, also the East model
displays the negative FDRs characteristic of activated aging dynamics.
For times $t$ and $\tw$ in the same ``plateau'' of the dynamics, the
FDR has (negative) values of order unity which we can predict
theoretically; as $\tw$ becomes smaller it then drops to $O(c)$,
$O(c^2)$ and so on. The contrast to the local observable can be traced
to the different scaling with $c$, and different signs, of the
distance dependent susceptibilities $\Chi_r$. Observables probing
intermediate lengthscales $\ell$ can interpolate between the local and
global cases, although we argued that for $c\to 0$ the non-local
effects would dominate for any $\ell>0$.

The apparent decoupling of the local response from activation effects
in the East model can be understood as follows. The response of a spin
to a local field is governed by the history of its facilitating
neighbour on the left. The field does not affect this ``clock''
because of the directed nature of the East model, and so one does not
get any speed-up effects: the susceptibility only reflects the direct
influence of the field and is positive. For spins at some distance
$r>0$ from the site where the field is applied, one has the opposite
scenario. These spins have a pure ``speed-up'' response: the spin at
the field site will be up more often, thus accelerating the dynamics
of all spins on the right.

We already alluded in the introduction (Sec.~\ref{sec:summary}) to the
wider implications of our results. The non-trivial FDRs we have found
in the aging regime reflect the growth of a purely dynamic lengthscale
and cannot be related to the (trivial) equilibrium properties of KCMs.
For most observables the FDRs are negative, precluding an
interpretation in terms of an effective temperature. The negative sign
nevertheless has a clear physical interpretation as arising from the
activated nature of the aging dynamics. In the FA case, also the
actual value of the asymptotic FDR (in the sense of widely separated
times $t\gg\tw$) is robust among observables probing the entire range
of lengthscales, from purely local behaviour to system-spanning global
observables. In the East model the situation is more subtle, with the
FD behaviour of local observables decoupled from activation effects.
The latter do show up, however, in appropriate non-local and global
observables. Because the East model has a hierarchy of relaxation
timescales, the (negative) value of the FDR then also varies depending
on which stage of the dynamics is being considered.

\begin{figure}
\hspace*{1in} \includegraphics[width=4.5in,clip]{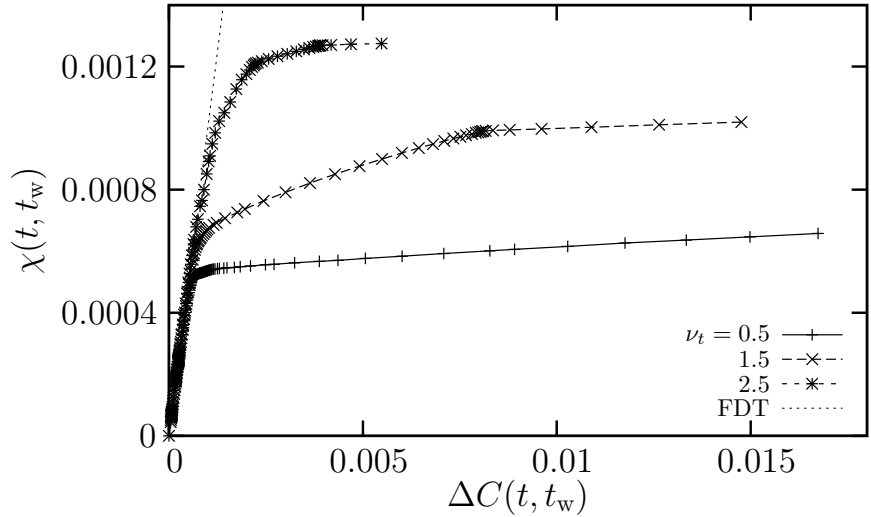}
\caption{FD plot for the local observable in the three-dimensional
  analogue of the East model, for $T=0.15$ and $\nu_t=0.5$, 1.5, 2.5
  (from bottom to top). As in $d=1$,
  the FD plot consists of approximately straight segments whose number
  increases with $t$ (see Fig.~\ref{fig:FDlocal_raw}).
\label{fig:NEF}
}
\end{figure}
To what extent will our results apply to other KCMs? One suspects that
the strictly directed KCMs might form a somewhat separate group with
regard to FD behaviour, because the effects of a local field can never
propagate back to the site where it was applied. Qualitatively similar
effects to those seen here for the East model would therefore be
expected in, for example, the analogous three-dimensional model
studied in Ref.~\cite{BerGar05}. The local FD plot shown in
Fig.~\ref{fig:NEF} demonstrates that this is indeed the case.
In $d=3$ there is again a hierarchy of timescales in the
non-equilibrium dynamics, with the
defect density relaxing in steps as in
Fig.~\ref{fig:nt}. This leads to FD plots that are (approximately)
piecewise linear, with a number of segments depending on the 
position of $t$ in the hierarchy (compare
Fig.~\ref{fig:FDlocal_raw} for $d=1$). The two-dimensional triangular
plaquette model~\cite{Robs_plaquette_FDT} also shows some broad
similarities to the East model, but with additional subtleties because
one now has to 
choose between studying the dynamics of the defects or of the
underlying spin system. When the direction of facilitation is itself
treated as a dynamical variable~\cite{arrow_model}, the model becomes
isotropic overall and one might expect the connection between local
and global observables to be restored as for the FA model. It would 
therefore be very interesting to study the aging dynamics
of such a model.

A significant open question concerns the FD behaviour of undirected
but highly cooperative KCMs, for example the FA model in $d=2$ with
two rather than one up-spin neighbours required to facilitate a move.
Numerical studies on similar models with conserved
dynamics~\cite{Kurchan97,SellittoEPJB} indicate a mean-field-like
scenario without any apparent activation effects. It seems clear,
however, that such effects should be present, certainly for global
observables: the relaxation to lower defect densities must, as for our
simpler one-spin facilitated FA models, proceed more quickly for
higher $c$ and result in a negative susceptibility. Here and also more
widely in experimental studies we hope that our study will stimulate
the search for well-defined negative FDRs. Arising as they do from
activated dynamics, they should be essentially ubiquitous among
systems exhibiting glassy dynamics.

\ack

LB, JPG and PS are grateful to the Newton Institute, Cambridge, for
providing a stimulating environment in which to discuss some of the
central questions of this paper. JPG acknowledges financial support
from EPSRC (grants GR/R83712/01 and GR/S54074/01) and the University
of Nottingham (grant FEF 3024). 
We thank Robert Jack for useful discussions and collaborations 
on the subject of this article.

\appendix

\section{Functions used in Section~\ref{sec:FA1d}}

Two functions appearing throughout Sec.~\ref{sec:FA1d} are the modified Bessel function
$I_n(t)$ of integer order $n$ \cite{MaySol07,Mathbook}, 
\begin{equation} 
  I_n(t) = \int_0^{2\pi} \frac{\rmd\varphi}{2\pi} \cos(n \varphi) \, \rme^{t \cos(\varphi)}, 
  \label{equ:I} 
\end{equation} 
and the function $H_n(t)$, 
\begin{equation}
  H_n(t) = \frac{1}{2} \int_0^t \rmd t' \, \rme^{-t'} \left[ I_{n-1}(t') - I_{n+1}(t') \right].
  \label{equ:H}
\end{equation}
Elementary properties that can be read off from these equations are $I_n(0) = \delta_{n,0}$, $I_{-n}(t) 
= I_n(t)$, $H_0(t) = 0$ and $H_{-n}(t) = -H_n(t)$. In our analysis we make extensive use 
of the asymptotic scalings of these functions. At fixed order $n$ and in the 
limit $t \to \infty$, 
\begin{equation}
  \rme^{-t} I_n(t) \sim \frac{1}{\sqrt{2 \pi t}} 
  \quad \mbox{and} \quad 
  H_n(t) \sim 1, 
  \label{equ:IHx}
\end{equation}
whereas for $t,n \to \infty$ simultaneously with $n^2/t$ fixed, 
\begin{equation}
  \rme^{-t} I_n(t) \sim \frac{1}{\sqrt{2 \pi t}} \, \rme^{-n^2/(2t)} 
  \quad \mbox{and} \quad 
  H_n(t) \sim \Erfc \left( \frac{n}{\sqrt{2t}} \right). 
  \label{equ:IHxn}
\end{equation}
Here $\Phi(z) = (2/\sqrt{\pi}) \int_z^\infty \rmd u \, \rme^{-u^2}$ denotes the complementary 
error function as in the main text. The following functional relations are also of great value for our analysis, 
\begin{eqnarray} 
  \partial_t  I_n(t) & = & \frac{1}{2} \left[ I_{n-1}(t) + I_{n+1}(t) \right], 
  \label{equ:Id} \\
  \frac{n}{t} I_n(t) & = & \frac{1}{2} \left[ I_{n-1}(t) - I_{n+1}(t) \right]. 
  \label{equ:Ir} 
\end{eqnarray} 
To calculate Fourier transforms we use the identities ($t \geq \tw \geq 0$) 
\begin{eqnarray}
\fl \sum_{n = -\infty}^\infty \rme^{-\rmi q n} I_{n+i}(t) I_{n+j}(t) = 
  \rme^{\rmi \frac{1}{2} (i+j) q} I_{i-j}\left(2 t \cos \frac{q}{2}\right), 
  \label{equ:S1} \\ 
\fl \sum_{n = -\infty}^\infty \rme^{-\rmi q n} I_n(t-\tw) [I_{n-m}+I_{n+m}](t+\tw) 
  \nonumber \\ 
  = 2 T_m\,\left( \frac{t \cos(q/2)^2 + \tw \sin(q/2)^2}{A} \right) I_m(2A), 
  \label{equ:S2} \\
\fl \sum_{n=-\infty}^\infty \rme^{-\rmi q n} [I_{n-1} - I_{n+1}](t-\tw) [I_{n-1} - I_{n+1}](t+\tw)
  \nonumber \\ 
  = 2 \left[ \cos(q) I_0(2A) - \frac{t^2 \cos(q/2)^2 - \tw^2 \sin(q/2)^2}{A^2} I_2(2A) \right], 
  \label{equ:S3}
\end{eqnarray}
where $A = \sqrt{t^2 \cos(q/2)^2 + {\tw}^2 \sin(q/2)^2}$ and $T_n(x) = 
\cos(n \arccos x)$ is the
Chebyshev polynomial of degree $n$. Equations~(\ref{equ:S1}) and (\ref{equ:S2}) follow 
from the integral representation Eq.~(\ref{equ:I}) and trigonometric identities. The sum 
Eq.~(\ref{equ:S3}) can be reduced to Eq.~(\ref{equ:S2}) by using Eq.~(\ref{equ:Ir}) 
and expressing the resulting factor $n^2$ as a second derivative with respect to $q$.

\section{Initial stages of irreversible coarsening dynamics}
\label{sec:nt}

In this section we give further details of the solution of the
irreversible coarsening dynamics, leading in particular to the
explicit expressions for the scaling functions~(\ref{g0}--\ref{g2})
for stages $k=0,1,2$ of the dynamics.

The dynamical equations~(\ref{Pd_dynamics}) for stage $k$ can be solved in terms of the generating functions~\cite{SolEva1}
\be
G(z,t)=\sum_d P(d,t)z^d, \qquad
H(z,t)=\sum_{2^{k-1}<d\leq 2^k} P(d,t)z^d,
\ee
where $H$ is the analogue of $G$ restricted to the active
domains. Bearing in mind that~(\ref{Pd_dynamics}) applies only for the
inactive domain lengths $d$, one has
\be
\frac{\partial}{\partial
t}[G(z,t)-H(z,t)]=G(z,t)\left(-\frac{\partial}{\partial t}\right)H(z,t),
\ee
and hence
\be
G(z,t) = 1 + [G(z,0)-1]\exp[H(z,0)-H(z,t)].
\label{G_z_t}
\ee
In the limit $t\to\infty$, where all active domains have disappeared
($H\to 0$), this gives a relation between the generating functions in
plateaus $k-1$ and $k$:
\be
G_k(z) = 1 + [G_{k-1}(z)-1]\exp[H_{k-1}(z)].
\ee
If we define $h_k(d)$ as the inverse transform of $\exp[H_k(z)]-1$,
i.e.\ $\exp[H_k(z)]-1 = \sum_d h_k(d) z^d$, this gives the following
recursion for the domain length distributions
\be
P_k(d) = P_{k-1}(d) - h_{k-1}(d) + (P_{k-1}\ast h_{k-1})(d).
\label{P_d_recursion}
\ee
Starting from the initial condition $P_{-1}(d) = 2^{-d}$ one finds
easily the particular values that we will need later:
\be
P_{-1}(1)=\frac{1}{2}, \quad
P_0(2)=\frac{3}{8}, \quad
P_1(3)=\frac{7}{24}, \quad
P_2(4)=\frac{15}{64}.
\label{P_d_explicit}
\ee
The overall distributions, which were obtained by evaluating the
recursion numerically, are graphed in Fig.~\ref{fig:Pd_East}.

To get a prediction for the evolution of the defect density $n(t)$
during stage $k$, we exploit that $1/n(t)=\bar{d}(t) =
(\partial/\partial z)G(z=1,t)$. From~(\ref{G_z_t}) this gives
\be
n(t) = n(0)\exp[H(1,t)-H(1,0)],
\label{nt_per_stage}
\ee
and we need the evolution of $H(1,t)$. Since active domains cannot be
recreated during the coarsening process, we can write
\be
H(1,t) = \sum_{2^{k-1}<d\leq 2^k} P(d,t) = \sum_{2^{k-1}<d\leq 2^k}
P(d,0) S(d,t),
\label{H1}
\ee
where $S(d,t)$ is the survival probability of a domain of length
$d$. In stage $k=0$ this is very simple: the disappearance of a domain
of length $d=1$ corresponds to a mobile up-spin flipping down, so
$S(1,t)=\exp[-(1-c)t]\approx \exp(-t)$ for $c\to 0$. Also
$P_{-1}(1)=1/2$
at the beginning of stage 0, so $H(1,t)=\exp(-t)/2$. Inserting
into~(\ref{nt_per_stage}) directly leads to the expression~(\ref{g0})
for $g_0(\zeta)$. For later stages and correspondingly larger $d$ one can
proceed using a ``leaky'' Markov chain that exits as soon as spin
$n_d$ flips down. Taking $k=1$ as an example, active domains have
length $d=2$. The initial state of the domain is 101 (i.e.\ $n_0=1$,
$n_1=0$, $n_2=1$). The only other possible state that can be reached
before spin $n_2$ flips down is 111. The leaky transition matrix
between these two states is
\be
W = \left(\begin{array}{cc}
-c & 1-c\\
c  & -2(1-c)
\end{array}
\right).
\ee
The first column represents the transition at rate $c$ from 101 to
111; the second column has the reverse transition, with rate $1-c$, and in
addition the rate $1-c$ for exiting via 111 $\to$ 110. The survival
probability is just the probability of remaining in the chain,
starting in state 101:
\be
S(2,t)=(1,1)e^{Wt}\left(\begin{array}{c}1\\0\end{array}\right)
= \left(\frac{1}{2}+\frac{2-c}{4D}\right)e^{\lambda_1 t}
+\left(\frac{1}{2}-\frac{2-c}{4D}\right)e^{\lambda_2 t},
\ee
with $D=\sqrt{1-2c+5c^2/4}$ and the eigenvalues
$\lambda_{1,2}=-1+c/2\pm D$. To get $g_1(\zeta)$ we need to consider the
survival probability on the timescale $t=\zeta/c$, taking the limit $c\to
0$ at fixed $\zeta$. Since $\lambda_2$ stays of order unity, the
second exponential then disappears. In the first one,
$\lambda_1=-c/2$ to leading order while the prefactor tends to 1 so
that overall $S(2,t=\zeta/c)\to \exp(-\zeta/2)$. Inserting
into~(\ref{H1}) and using~(\ref{P_d_explicit}) for the prefactor
$P_0(2)$ then gives the result~(\ref{g1}) for $g_1(\zeta)$. One
proceeds similarly for $g_2(\zeta)$ to derive~(\ref{g2}), the main
difference being that there are now two active domain lengths, $d=3$
and $d=4$. The required leaky transition matrices are of size $4\times 4$ and
$8\times 8$, respectively, and the survival probability is evaluated
in the regime $t=\zeta/c^2$. By taking derivatives of the survival
functions at $\zeta=0$ one further obtains the rates $\Gamma(d)$
required for the prediction of the negative energy FDR~(\ref{X_E}), as
\be
\Gamma(1)=1, \quad
\Gamma(2)=\frac{c}{2}, \quad
\Gamma(3)=\frac{2c^2}{3}, \quad
\Gamma(4)=\frac{c^2}{4}\, .
\ee

\section{Exact relation between local correlation and response}
\label{sec:local_relation}

We outline the derivation of the exact relation~(\ref{local_relation})
between the local correlation $C_0$ and susceptibility $\Chi_0$ in the
East model. For the latter we have the expression~(\ref{Chi0_clock}),
$\Chi_0(t,\tw) = c(1-c)[1-r(t,\tw)]$, in terms of the relaxation
integral
\be
r(t,\tw) = \langle \hat{r}(t,\tw)\rangle, \qquad
\hat{r}(t,\tw) = \exp\left(-\int_{\tw}^t dt'\,n_{i-1}(t')\right).
\ee
For the correlation function we need to evaluate $\langle
n_i(t)n_i(\tw)\rangle$. As in the derivation of $\Chi_0$ we first take
the history of spin $n_{i-1}$ from time 0 to $t$ as fixed. If also
$n_i(\tw)$ is given, the average value of $n_i(t)$ is
$c+[n_i(\tw)-c]\hat{r}(t,\tw)$, using the same argument as for
$\Chi_0$. Thus
\be
\langle n_i(t)n_i(\tw)\rangle_{n_i(\tw),n_{i-1}(0\ldots t)} = 
n_i(\tw)[c+(1-c)\hat{r}(t,\tw)],
\ee
where the subscript indicates the fixed quantities. Now we average
over $n_i(\tw)$, which has evolved from the initial average defect
density $n(0)$ at time 0 to a later value governed by the relaxation
integral $\hat{r}(\tw,0)$:
\bea
\langle n_i(t)n_i(\tw)\rangle_{n_{i-1}(0\ldots t)} &=& 
\{c+[n(0)-c]\hat{r}(\tw,0)\}[c+(1-c)\hat{r}(t,\tw)] \\
&=& c^2+c(1-c)\hat{r}(t,\tw) + c[n(0)-c]\hat{r}(\tw,0)
\nonumber\\
& &{}+
(1-c)[n(0)-c]\hat{r}(t,0),
\label{nt_ntw_aux}
\eea
where we have exploited that $\hat{r}(t,\tw)\hat{r}(\tw,0)=\hat{r}(t,0)$.
Still at fixed history of $n_{i-1}$, the same arguments as above give the
average densities at times $\tw$ and $t$ as
\bea
\langle n_i(\tw)\rangle_{n_{i-1}(0\ldots t)} &=& c+[n(0)-c]\hat{r}(\tw,0),\\
\langle n_i(t)\rangle_{n_{i-1}(0\ldots t)} &=& c+[n(0)-c]\hat{r}(t,0).
\eea
These can be used to eliminate the occurrences of $\hat{r}(\tw,0)$ and
$\hat{r}(t,0)$ from~(\ref{nt_ntw_aux}):
\bea
\langle n_i(t)n_i(\tw)\rangle_{n_{i-1}(0\ldots t)} &=&
c^2+c(1-c)\hat{r}(t,\tw) + c[\langle n_i(\tw)\rangle_{n_{i-1}(0\ldots t)}-c]
\nonumber
\\
& &{} + (1-c)[\langle n_i(t)\rangle_{n_{i-1}(0\ldots t)}-c].
\eea
Averaging over the history of $n_{i-1}$ gives then
\be
\langle n_i(t)n_i(\tw)\rangle = 
-c(1-c)[1-r(t,\tw)] + cn(\tw) + (1-c)n(t),
\ee
and so finally for the autocorrelation function $C_0(t,\tw)=\langle
n_i(t)n_i(\tw)\rangle - n(t)n(\tw)$:
\be
C_0(t,\tw) = -c(1-c)[1- r(t,\tw)] + cn(\tw) + (1-c)n(t) -
n(t)n(\tw).
\ee
The first term on the r.h.s.\ is $-\Chi_0(t,\tw)$, and rearranging
slightly gives the promised result~(\ref{local_relation}). It is clear
from the derivation that this exact relation will hold for all
kinetically constrained spin models with directed constraints, as long
as we replace $n_{i-1}(t')$ by the appropriate facilitation factor
$f_i(t')$ for spin $n_i$. The key ingredient is that $n_i$ does not
affect the evolution of this facilitation factor; this is why we can
first fix the facilitation history and average over the dynamics of
$n_i$.

\section{Energy correlations}
\label{sec:C_E_calc}

Here we sketch how the energy correlations for the East model can be
calculated within the irreversible coarsening regime. The energy
$E=N/\bar{d}$ is inversely proportional to the average domain length.
Since the latter has fluctuations $\delta\bar{d}$ of over $N^{-1/2}$
we can write
\be
C_E(t,\tw) = \frac{N\langle \delta\bar{d}(t)\delta\bar{d}(\tw)\rangle}
{\bar{d}^2(t)\bar{d}^2(\tw)}.
\label{C_E_starting_point}
\ee
The fluctuations $\delta\bar{d}$
arise from the corresponding fluctuations $\delta P(d)$ in the domain
size distribution. At equal times, $\tw=t$, the latter are easy to
obtain because of the independent interval nature of the dynamics: the
actual arrangement of domain lengths at some given time can be
obtained by repeatedly sampling domain lengths from the (average)
domain length distribution $P(d)$ and lining them up along the chain
until the total length $N$ is reached. By a relatively simple
combinatorial calculation one then finds for large $N$ the following
expression for the covariance of the $\delta P(d)$:
\be
N\langle \delta P(d)\delta P(d')\rangle = \bar{d}
\left[P(d)\delta_{d,d'} - P(d)P(d')\right].
\label{delta_P_d}
\ee
Summing over $d$ and $d'$, this implies $\langle [\sum_d \delta
P(d)]^2\rangle = 0$ as it must be because the distribution is always
normalized. By multiplying with $dd'$ first and then summing, on the
other hand, one obtains the variance of $\delta\bar{d}$ as
$\bar{d}(\overline{d^2} - \bar{d}^2)$ and hence the
expression~(\ref{C_E_dbar}) for the equal-time energy correlations.

For the two-time correlations one exploits that the
recursion~(\ref{P_d_recursion}) for the domain size distribution in
the plateaus of the dynamics holds whatever the initial shape of this
distribution. We can apply it, in particular, to a domain size
distribution $P_{k\w}(d)+\delta P_{k\w}(d)$ in plateau $k\w$ perturbed by a
small fluctuation. Linearization in the small quantities $\delta
P_{k\w}(d)\sim N^{-1/2}$ then tells us how this fluctuation propagates
into the later plateau $k_t$. We can write the outcome in the generic form
\be
\delta P_{k_t}(d) = \sum_{d'} M_{k_tk\w}(d,d') \delta P_{k\w}(d'),
\ee
with an appropriate (infinite) matrix $M_{k_tk\w}$ depending on both
the initial and final plateau. Inserting
into~(\ref{C_E_starting_point}) gives then
\bea
\fl C_E(t,\tw) &=& \bar{d}_{k\w}^{-2}\bar{d}^{-2}_{k_t}
\sum_{d,d',d''} dd'' M_{{k_t}{k\w}}(d,d') 
N\langle \delta P_{k\w}(d') \delta P_{k\w}(d'')\rangle \\
\fl &=& \bar{d}_{k\w}^{-1}\bar{d}^{-2}_{k_t} 
\sum_{d,d',d''} dd'' M_{{k_t}{k\w}}(d,d') 
\left[P_{k\w}(d')\delta_{d',d''} -
P_{k\w}(d')P_{k\w}(d'')\right] \\
\fl &=& \bar{d}_{k\w}^{-1}\bar{d}^{-2}_{k_t} 
\sum_{d,d'} d M_{{k_t}{k\w}}(d,d') P_{k\w}(d')(d'-\bar{d}_{k\w}).
\label{C_E_ttw_general}
\eea
To evaluate the sum numerically without explicitly calculating and
storing $M_{k_tk\w}$, we perturb the domain size distribution in
plateau $k\w$ from $P_{k\w}(d)$ to
$P_{k\w}(d)[1+\epsilon(d-\bar{d}_{k\w})]$ with some small ``field''
$\epsilon$, and find the resulting small change (proportional to
$\epsilon$) in the average domain length in plateau $k_t$. Multiplying
by the prefactor $\bar{d}_{k\w}^{-1}\bar{d}^{-2}_{k_t}$ then gives the results
shown in Table~\ref{table:C_E_ttw}.

A closed form solution for the propagation of perturbations from $\tw$
to $t$ can be obtained in the paste-all regime of large domain
lengths. We work with the normalized lengths $x=d/\dmin(\tw)$ as
before; the clock variable is $\theta=\dmin(t)/\dmin(\tw)$ so that at
``time'' $\theta$ there are no domains of length $x<\theta$. The
distribution $P_\theta(x)$ of domain lengths at time $\theta$ obeys
the master equation
\be
\frac{\partial}{\partial\theta}P_\theta(x) =
-\delta(x-\theta)P_\theta(\theta) + P_\theta(x-\theta)P_\theta(\theta).
\label{paste_all_master}
\ee
The first term on the right captures the disappearance of domains of length
$x=\theta$; these merge with their right neighbours into larger
domains as represented by the second term. (Because $x\geq
\theta$ always, $P_\theta(x)$ has a step discontinuity at
$x=\theta$; the $P_\theta(\theta)$ in the master equation is to be
understood as the nonzero probability $P_\theta(x=\theta^+)$ to the
right of this discontinuity, i.e.\ the probability density of the
shortest domains present.) The initial condition $P_1(x)$ at
$\theta=1$ is the scaling distribution $\tilde{P}(x)$ from~(\ref{East_Px}).

To solve~(\ref{paste_all_master}) one goes to Laplace transforms
\be
\hat{P}_\theta(s)=\int_0^\infty dx\,e^{-sx}P_\theta(x),
\ee
to find
\be
\frac{\partial}{\partial\theta}\hat{P}_\theta(s) =
e^{-\theta s} P_\theta(\theta)[\hat{P}_\theta(s)-1].
\ee
Integrating w.r.t.\ $\theta$ gives explicitly
\be
\hat{P}_\theta(s) = 1+[\hat{P}_1(s)-1]\exp\left(\int_1^\theta d\theta'
\,e^{-\theta's} P_{\theta'}(\theta')\right).
\label{paste_all_master_LT}
\ee
The initial condition is the Laplace transform of $\tilde{P}(x)$,
given by $\hat{P}_1(s)=1-\exp[-\Ei(s)]$. Since the paste-all
limit is a scaling regime one expects
$P_\theta(x)=P_1(x/\theta)/\theta$ and in particular
$P_\theta(\theta)=P_1(1)/\theta=1/\theta$. This is indeed the correct
solution of~(\ref{paste_all_master_LT}) as the integral over $\theta'$
then becomes $\Ei(s)-\Ei(\theta s)$ so that
$\hat{P}_\theta(s) = 1-\exp[-\Ei(\theta s)] = \hat{P}_1(\theta s)$.

To apply the above description of the paste-all dynamics to the
calculation of the energy correlation function, we substitute
$P_\theta(x) \to P_\theta(x)+\delta P_\theta(x)$ everywhere and
linearize in the small perturbation $\delta P_\theta$. In the Laplace
transform version~(\ref{paste_all_master_LT}) this yields
\be
\fl
\delta \hat{P}_\theta(s) =
\delta\hat{P}_1(s)e^{\Ei(s)-\Ei(\theta s)} +
[\hat{P}_1(s)-1]e^{\Ei(s)-\Ei(\theta s)}
\int_1^\theta d\theta'\,e^{-\theta's}\delta P_{\theta'}(\theta'),
\ee
or
\be
e^{\Ei(\theta s)}\delta\hat{P}_\theta(s) =
e^{\Ei(s)}\delta\hat{P}_1(s) - 
\int_1^\theta d\theta'\,e^{-\theta's}\delta P_{\theta'}(\theta').
\label{delta_hat_P}
\ee
As explained above, we need to insert as the initial perturbation
$\delta P_1(x) = P_1(x)(x-\bar{x}) = P_1(x)(x-e^\gamma)$, i.e.\ 
\be
\delta \hat{P}_1(s) = -\frac{\partial}{\partial
  s}\hat{P}_1(s)-e^\gamma\hat{P}_1(s) 
= \frac{e^{-s}}{s}e^{-\Ei(s)} - e^\gamma[1-e^{-\Ei(s)}].
\ee
This gives for the Laplace transform~(\ref{delta_hat_P}) of the
perturbed domain size distribution at time $\theta$
\be
e^{\Ei(\theta s)}\delta\hat{P}_\theta(s) =
\frac{e^{-s}}{s} - e^\gamma[e^{\Ei(s)}-1]
- \int_1^\theta d\theta'\,e^{-\theta's}\delta P_{\theta'}(\theta').
\label{delta_almost}
\ee
It remains to determine $\delta P_{\theta'}(\theta')$ from the
condition that $\delta P_\theta(x)=0$ for $x<\theta$. On the l.h.s.\
of~(\ref{delta_almost}) we have the (Laplace transform of) a
convolution of $P_\theta(x)$ with another function; this convolution
then also vanishes for $x<\theta$. The same must therefore be true of
the r.h.s., which is the Laplace transform of
\be
\Theta(x-1)-e^\gamma\sum_{l\geq 1}\frac{1}{l!}f_l(x) -
\Theta(\theta-x)\delta P_x(x).
\ee
The condition that this must vanish for $x<\theta$ implies that in
this range
\be
\delta P_x(x) = \Theta(x-1)-e^\gamma\sum_{l\geq 1}\frac{1}{l!}f_l(x).
\label{delta_P}
\ee
Because we can make $\theta$ as large as desired, this expression must
then hold for all $x$. We can now reinsert this result
into~(\ref{delta_almost}) and take the limit $s\to 0$ to obtain the
perturbation $\delta\bar{x}_\theta$ in the average domain length, using that
$\delta\hat{P}_\theta(s) = -s\,\delta \bar{x}_\theta + O(s^2)$ and
$\Ei(s) = -\ln s - \gamma + s + O(s^2)$:
\be
- \frac{\delta\bar{x}_\theta}{e^\gamma\theta} = 
-2+e^\gamma - \int_1^\theta d\theta'\,\delta P_{\theta'}(\theta'),
\ee
or, after inserting~(\ref{delta_P}),
\be
\delta\bar{x}_\theta = e^\gamma\theta \left(
1-e^\gamma + \theta - 
e^\gamma\sum_{l\geq 1}\frac{1}{l!}\int_1^\theta d\theta'\,f_l(\theta')\right).
\ee
From~(\ref{C_E_ttw_general}) we can then finally write down the
two-time energy correlation. By considering the combination
$C_E(t,\tw)/n(t)$ we remove one 
factor of $\bar{d}_{k_t}^{-1}$ and also ensure that all factors of
$\dmin(\tw)$ arising from our length rescaling cancel from the
result. The remainder of the prefactor is $1/(e^\gamma
e^\gamma\theta)$ so that overall
\be
C_E(t,\tw)/n(t) = 
e^{-\gamma}(\theta+1) - 1 - \sum_{l\geq 1}\frac{1}{l!}\int_1^\theta
d\theta'\,f_l(\theta').
\ee
This is the scaling function ${\cal G}(\theta)$ given
in~(\ref{G_theta}); for $\theta\leq 2$ it can be written explicitly as
\be
{\cal G}(\theta) = e^{-\gamma}(\theta+1)-1-\ln\theta.
\ee
For $\theta=1$ in particular one obtains the scaling of the equal-time
correlations in the paste-all limit as
$C_E(t,t)/n(t)=2e^{-\gamma}-1=0.1229\ldots$ This value gives the
$y$-axis intercept of the paste-all curve in Fig.~\ref{fig:C_E_ttw}.

\section{Non-local correlations}
\label{sec:C_r_calc}

In this appendix we describe how the non-local two-time correlations in
the East model can be calculated in the paste-all regime. The aim is
to derive the scaling function ${\cal H}(x,\theta)$ defined
in~(\ref{C_r_ttw_scaling}), in the nontrivial region of positive $x$.

As explained in the main text, two-time spatial correlations require
one to keep track of which domains from time $\tw$ have been merged
into the domains at time $t$. We can then characterize a domain at $t$
by the (scaled) lengths of the $\tw$-domains it contains. If these
lengths are, in order, $x_1$, \ldots, $x_l$, we write the fraction of
such $t$-domains as $P^{(l)}_\theta(x_1,\ldots,x_l)$; here
$\theta=n(\tw)/n(t)=\dmin(t)/\dmin(\tw)$ as before. For this more
detailed description of the domains, the master
equation~(\ref{paste_all_master}) becomes
\bea
\fl \frac{\partial}{\partial\theta}P^{(l)}_\theta(x_1,\ldots,x_l) &=&
-\delta(x_1+\ldots+x_l-\theta)P^{(l)}_\theta(x_1,\ldots,x_l) 
\label{twotime_master}\\
\fl &&{}+ \sum_{l'=1}^{l-1}
\delta(x_1+\ldots+x_{l'}-\theta)
P^{(l')}_\theta(x_1,\ldots,x_{l'})
P^{(l-l')}_\theta(x_{l'+1},\ldots,x_l).
\nonumber
\eea
The first term represents the usual disappearance of domains of total
length $\theta$. The second one keeps track of the appearance of
new domains; as two domains merge, the lists of the lengths of the
$\tw$-domains which they contain are simply appended to each other.

It turns out that the above equations for the $P^{(l)}_\theta$ can be
solved explicitly, but it is useful to check first how they enter the
scaling function ${\cal H}(x,\theta)$. From the
definition~(\ref{C_r_ttw_scaling}), $\H+1$ is the density of
$\tw$-spins a distance $-x$ from a $t$-spin, divided by the overall
density of $\tw$-spins. The latter is $1/\bar{x}=e^{-\gamma}$ in the
paste-all regime, so $e^{-\gamma}(\H+1)$ for positive $x$ is the density
of $\tw$-spins that are a distance $x$ to the left of a $t$-spin. This
can be written as
\be
e^{-\gamma}(\H+1) = Q_\theta(x)+(Q_\theta\ast P_\theta)(x)+(Q_\theta
\ast P_\theta\ast P_\theta)(x)+\ldots
\label{H_general}
\ee
The terms in this series represent $\tw$-spins that are within (or on
the left boundary of) the first, second, third etc. $t$-domain to the left
of the $t$-spin considered, hence the appearance of the length
distribution $P_\theta(x)$ of $t$-domains. The factor $Q_\theta(x)$
records the positions of $\tw$-spins {\em within} $t$-domains:
\bea
Q_\theta(x) &=& \sum_{m=1}^\infty Q^{(m)}_\theta(x),
\\
Q^{(m)}_\theta(x) &=& \sum_{l=m}^\infty \int dx_1\cdots dx_l\,
P^{(l)}_\theta(x_1,\ldots,x_l)\delta(x_{l-m+1}+\ldots+x_l-x),
\eea
where $Q^{(m)}_\theta(x)$ accounts for the contributions from the
$m$-th $\tw$-spin within a $t$-domain, counting leftwards from the
$t$-spin on its right boundary. We see that we do not need {\em all}
details of the $P^{(l)}_\theta$; the $Q^{(m)}_\theta$ are sufficient.
Unfortunately, the latter do not obey closed equations because they
do not contain information about the total length of the
$t$-domain. We therefore generalize so that this is also kept track of
and define the quantities
\bea
\tilde Q^{(m)}_\theta(x,x\tot) &=& \sum_{l=m}^\infty \int dx_1\cdots dx_l\,
P^{(l)}_\theta(x_1,\ldots,x_l) \times
\nonumber\\
& &\times\delta(x_{l-m+1}+\ldots+x_l-x)\delta(x_1+\ldots+x_l-x\tot).
\eea
These give the fraction of $t$-domains of length $x\tot$ for which the
$m$ rightmost $\tw$-domains contained within add up to a length
$x$. Taking the simplest case $m=1$ as an example, one now derives
from~(\ref{twotime_master}) the evolution equation
\bea
\frac{\partial}{\partial\theta}\Qo(x,x\tot) &=&
-\delta(x\tot-\theta)\Qo(x,x\tot)
+ \frac{1}{\theta}\Qo(x,x\tot-\theta).
\label{tilde_Q_1}
\eea
The factor $1/\theta$ here is the total rate of disappearance of
domains; formally it is calculated from
\be
\sum_{l'=1}^{\infty}
\int dx_1\cdots dx_{l'}\,\delta(x_1+\ldots+x_{l'}-\theta)
P^{(l')}_\theta(x_1,\ldots,x_{l'}),
\ee
which just gives the density $P_\theta(\theta)=1/\theta$ of
$t$-domains of length $\theta$ (as enforced by the delta function) at
time $\theta$.

The initial condition for~(\ref{tilde_Q_1}) at $\theta=1$, i.e.\
$t=\tw$, is $\tilde Q^{(1)}_1(x,x\tot)=P_1(x)\delta(x-x\tot)=\tilde
P(x)\delta(x-x\tot)$ because there is then no distinction between
$\tw$-domains and $t$-domains. (It is only through this initial
condition that the $x$-dependence enters.) Going to Laplace transforms
$x\tot\to s$, integrating with respect to $\theta$ and transforming
back gives then
\bea
\int_0^{\infty}\frac{dx'}{\theta} \E\!\left(\frac{x\tot-x'}{\theta}\right)
\Qo(x,x') &=& \tilde P(x)\E(x\tot-x)
\label{Qop_aux}
\\
& &{} - \int_1^\theta \frac{d\theta'}
{\theta'}\E\!\left(\frac{x\tot-\theta'}{\theta'}\right)\Qop(x,\theta'),
\nonumber
\eea
where
\be
\E(x)=\delta(x)+\sum_{l\geq 1}\frac{1}{l!}f_l(x)
\label{E_def}
\ee
is the inverse Laplace transform of $\exp[\Ei(s)]$. We now need to
find $\Qop(x,\theta')$ from the condition that $\Qo(x,x')$
vanishes for $x'<\theta$. Since $\E(\cdot)$ is zero for negative
argument, the l.h.s.\ of~(\ref{Qop_aux}) vanishes for
$x\tot<\theta$ so that in this regime
\be
\tilde P(x)\E(x\tot-x) = \int_1^\theta \frac{d\theta'}
{\theta'}\E\!\left(\frac{x\tot-\theta'}{\theta'}\right)\Qop(x,\theta').
\ee
Again because of the vanishing of $\E(\cdot)$ for negative argument we
can restrict the integration to $\theta'\leq x\tot$. Also
$\Qop(x,\theta')$ must vanish for $\theta'<x$ -- the total length
cannot be smaller than the length of the rightmost $\tw$-domain within
-- so the lower integration limit can be raised from $\theta'=1$ to
$\theta'=x$. After relabelling $x\tot\to\theta$ our condition becomes
\be
\tilde P(x)\E(\theta-x) = \int_x^\theta \frac{d\theta'}
{\theta'}\E\!\left(\frac{\theta-\theta'}{\theta'}\right)\Qop(x,\theta').
\ee
It is from this expression that $\Qop(x,\theta')$ is to be found.  The simplest
regime is $\theta/x<2$. Then the argument of $\E(\cdot)$ on the right
is always less than 1 and so only the first term in~(\ref{E_def})
contributes, giving
\be
\Qo(x,\theta)=\tilde P(x)\E(\theta-x),
\qquad{\rm for}\ \theta/x<2.
\ee
Since $x\geq 1$ this solution applies, in particular, whenever
$\theta<2$, where it simplifies further to 
\be
\Qo(x,\theta)=\tilde P(x)\delta(\theta-x).
\label{Qo_theta_lt_2}
\ee
This makes sense: a $t$-domain of size $\theta<2$ cannot contain two
or more $\tw$-domains, so that necessarily
$\theta=x$. Inserting~(\ref{Qo_theta_lt_2}) into~(\ref{Qop_aux}) yields
\be
\fl \int_0^{\infty}\frac{dx'}{\theta} \E\!\left(\frac{x\tot-x'}{\theta}\right)
\Qo(x,x') = \tilde P(x)\left[\E(x\tot-x)
- \frac{\Theta(\theta-x)}{x} \E\!\left(\frac{x\tot-x}{x}\right)\right]
\ee
One can now integrate over $x\tot$, or equivalently go back to Laplace
transforms and take $s\to 0$, to find
\be
Q^{(1)}_\theta(x)=\int dx'\,\Qoo(x,x')=\theta \tilde
P(x)\left[1-\frac{\Theta(\theta-x)}{x}\right].
\ee
This expression is exact for $\theta<2$. If also $x<2$, then it is the
only contribution to~(\ref{H_general}) because all other terms involve
two or more $\tw$-domains whose combined length will be above 2. We
have thus obtained the first line of~(\ref{H_scaling}). To get the
full expression valid for $\theta<3$ and $x<3$ one has to extend the
above calculation for $Q^{(1)}_\theta(x)$ to the range $2<\theta<3$;
one also needs to calculate $Q^{(2)}_\theta(x)$.
We omit the details.

\section{Persistence function}
\label{sec:persistence}


Here we outline how to calculate the persistence function of
down-spins~(\ref{Chi0_scaling_analytical}) in the paste-all regime of
the East model. From the discussion before~(\ref{Chi0_paste_all}) this
governs the behaviour of the local susceptibility $\Chi_0$ in the East
model.

As discussed in Sec.~\ref{sec:East}, in the paste-all regime each
domain contains an equilibration zone on its left, of size
$\theta$. (We use the same definitions of $x$ and $\theta$ as in
previous sections.) Each equilibration zone has been fully ``swept'' by
up-spins, i.e.\ it contains no persistent down-spins.  However,
regions to the right of the current equilibration zone may already
have been swept previously. For example, when two domains merge, only
the equilibration zone of the original domain on the left remains, but
we have to keep track of the fact that the equilibration zone in the
domain on the right has already been swept by up-spins. We therefore
define $P_\theta(u,x)$ as the fraction of domains that have length $x$
and an unswept zone of length $u$ (on their right end). This has two
components:
\be
P_\theta(u,x) = P'_\theta(x)\delta(x-\theta-u) + P''_\theta(u,x).
\label{P_decomposition}
\ee
The first part represents domains where the swept area is exactly the
equilibration zone and so $u=x-\theta$; the second part describes
domains where the swept zone is larger and correspondingly
$u<x-\theta$. Initially, at $\theta=1$, only the first part is
present, and so $P'_1(x)=\tilde P(x)$ and $P''_1(u,x)=0$. This is
because we want to count persistence from time $\tw$, i.e.\ the swept
zones are reset to the current equilibration zones at that point. At some later
``time'' $\theta$, the fraction of the chain occupied by persistent
down-spins is $\langle u\rangle_\theta$ divided by the average domain
length $e^\gamma\theta$. The function ${\cal F}(\theta)$ is the
complement of this, i.e.\ the fraction of swept areas
\be
{\cal F}(\theta) = 1 - \frac{\langle u\rangle_\theta}{e^\gamma\theta}.
\label{F_general}
\ee

To calculate the required average $\langle u\rangle_\theta$ we start
from the evolution equations for $P'$ and $P''$,
\bea
\frac{\partial}{\partial\theta} P'_\theta(x) &=& 
-\delta(x-\theta)P'_\theta(x) + P''_\theta(x-\theta,x),
\label{Pp_evolution}
\\
\frac{\partial}{\partial\theta} P''_\theta(u,x) &=& 
-\delta(x-\theta-u)P''_\theta(x-\theta,x) + \frac{1}{\theta}
P_\theta(u,x-\theta).
\label{Pdp_evolution}
\eea
In~(\ref{Pp_evolution}) the first term describes disappearance of
domains of length $\theta$; note that only domains in the $P'$-part of
the distribution can disappear in this way because at disappearance
the equilibration zone covers the whole of the domain and is therefore
identical to the swept zone. The second term accounts for the fact
that when the equilibration zone (length $\theta$) grows long enough
to cover all of the existing swept zone (length
$x-u=x-(x-\theta)=\theta$), domains are counted in $P'$ rather than
$P''$. The first term in~(\ref{Pdp_evolution}) records the
corresponding loss of domains from $P''$. The last term, finally,
accounts for the creation of new domains as domains of length $\theta$
disappear and merge with their right neighbours. The factor $1/\theta$
is the total rate of disappearance of domains as in~(\ref{tilde_Q_1});
the length $u$ of the unswept zone remains unchanged during a merger while
the overall domain lengths $\theta$ and $x-\theta$ add.

The first evolution equation~(\ref{Pp_evolution}) can be integrated
directly to give
\be
P'_\theta(x) = \Theta(x-\theta)\left[\tilde P(x) + \int_1^\theta
d\theta'\, P''_{\theta'}(x-\theta',x)\right],
\ee
while for the second one we insert~(\ref{P_decomposition}) and then
proceed as in the derivation of~(\ref{Qop_aux}) to obtain
\bea
\int_0^{\infty}\frac{dx'}{\theta} \E\!\left(\frac{x-x'}{\theta}\right)
P''_\theta(u,x') &=& \int_1^\theta \frac{d\theta'}{\theta'}\Biggl[
-\E\!\left(\frac{x-\theta'-u}{\theta'}\right) P''_{\theta'}(u,\theta'+u)
\nonumber\\
& &{} + \frac{1}{\theta'}\,
\E\!\left(\frac{x-2\theta'-u}{\theta'}\right)
P'_{\theta'}(\theta'+u)\Biggr].
\eea
One sees that all quantities are determined once we know
$P''_{\theta'}(u,\theta'+u)$. Fortunately, this {\em vanishes} for
$\theta'<2$. The reason is that domains that have not yet merged have
identical swept and equilibration zones and so are counted in
$P'$. Domains in the $P''$-part of the distribution must then be the
product of a merger of at least two of the domains that were present
at time $\tw$. Their swept zone is, as a consequence, the result of a
merger of at least two equilibration zones and must have length at
least 2. It follows that $P''_{\theta'}(u,x)=0$ for $u>x-2$, and
substituting $x=\theta'+u$ with $\theta'<2$ proves our claim.

With the above simplification for $\theta<2$ we get
$P'_\theta(x)=\Theta(x-\theta)\tilde P(x)$ and
\be
\int_0^{\infty}\frac{dx'}{\theta} \E\!\left(\frac{x-x'}{\theta}\right)
P''_\theta(u,x') = \int_1^\theta \frac{d\theta'}{(\theta')^2}
\E\!\left(\frac{x-2\theta'-u}{\theta'}\right)
\tilde P(\theta'+u),
\ee
or, after integration over $x$,
\be
\int_0^\infty dx'\,P''_\theta(u,x') = \theta \int_1^\theta
\frac{d\theta'}{(\theta')^2} \tilde P(\theta'+u).
\ee
Putting everything together gives for the average of $u$
\bea
\langle u\rangle_\theta
&=& \int dx\,du\,u\,P_\theta(u,x)
\\
&=& \int dx\,(x-\theta) P'_\theta(x) + \int dx\,du\,u\,P''_\theta(u,x)
\\
&=& \int dx\,(x-\theta) \Theta(x-\theta) \tilde P(x) + \int du\,u\,
\theta \int_1^\theta \frac{d\theta'}{(\theta')^2} \tilde P(\theta'+u)
\\
&=& I(\theta) + \theta \int_1^\theta \frac{d\theta'}{(\theta')^2}
I(\theta'),
\label{u_av_almost}
\eea
where
\bea
I(\theta)&=&\int_\theta^\infty dx(x-\theta)\tilde P(x)
\\
&=&\int_1^\infty dx(x-\theta)\tilde P(x) - \int_1^\theta
dx(x-\theta)\tilde P(x)
\\
&=&e^\gamma-\theta - \int_1^\theta dx\,\left(1-\frac{\theta}{x}\right)
\\
&=&e^\gamma+1-2\theta+\theta\ln\theta.
\eea
Carrying out the remaining $\theta'$-integration
in~(\ref{u_av_almost}) yields
\be
\langle u\rangle_\theta =
(e^\gamma-1)\theta-\theta\ln\theta+\frac{\theta}{2}\ln^2\theta,
\ee
and, after inserting into~(\ref{F_general}), the
result~(\ref{Chi0_scaling_analytical}) from the main text. One can
push the calculation further, certainly up to $\theta<3$, but the
resulting expressions become very unwieldy and so are not given here.

\section*{References}

\bibliography{review_intro}

\end{document}